\documentclass[]{aa}  
\usepackage[dvipsnames,svgnames,x11names]{xcolor}

\usepackage{graphicx}
\usepackage{txfonts}
\usepackage{natbib}
\usepackage[colorlinks=true, linktocpage, linkcolor={blue!70!black}, citecolor={blue!70!black}, urlcolor={blue!70!black}]{hyperref}

\usepackage{ulem}
\newcommand{\change}[1]{{#1}}

\begin{document} 

\title{\change{Two decades of km-resolution satellite-based measurements of the precipitable water vapor above the Atacama Desert}}
\subtitle{A comparison of climate measurements for submillimeter astronomy} 
\titlerunning{A comparison of ground and satellite-based PWV data for the Atacama Desert}

\author{
    Pablo G\'omez Toribio\inst{1}\and
    Tony Mroczkowski\inst{2}\and
    Anna Cabr\'e\inst{3}\and
    Carlos De Breuck\inst{2}\and
    Ricardo Bustos\inst{4}\and
    Rodrigo Reeves\inst{5}
}

\authorrunning{G\'omez Toribio et al.}

\institute{
    Universitat Polit\`ecnica de Catalunya (UPC), Facultat de Matem\`atiques i Estad\'istica, Campus Diagonal Sud, Carrer de Pau Gargallo, 14, 08028, Barcelona, Catalonia, Spain\\\email{pablog.toribio@gmail.com}
    \and
    European Southern Observatory (ESO), Karl-Schwarzschild-Strasse 2, Garching 85748, Germany
    \and
    University of Pennsylvania, Department of Earth and Environmental Science, 240 South 33rd Street, Philadelphia, Pennsylvania 19104-6316, USA
    \and 
    Departamento de Ingeniería Eléctrica, Universidad Católica de la Santísima Concepción (UCSC), Alonso de Ribera 2850, Concepción, Chile
    \and
    CePIA, Departamento de Astronomía, Universidad de Concepción (UdeC), Casilla 160 C, Concepción, Chile
}

\date{Original form February 22, 2021; Revised March 7, 2022; accepted ???}

\abstract{The Atacama Desert has long been established as an excellent site for submillimeter observations.  Yet identifying potentially optimal locations for a new facility within this region can be costly, traditionally requiring long field campaigns in multiple locations that rely on the construction of weather stations and radiometer facilities to take data over sufficiently long timescales, often years.  Meanwhile, high-level remote sensing data products from satellites have generally only been available at $\gtrsim$25~km resolution, limiting their utility for astronomical site selection.}
{We aim to improve and expedite the process of site characterization and selection through the use of kilometer resolution satellite data, \change{using fiducial ground-based data to improve the accuracy of satellite precipitable water vapor (PWV) measurements in the extremely dry conditions found in the region around the Atacama Astronomical Park (Parque Astron\'omico de Atacama; hereafter AAP) in northern Chile.}}
{We analyze the daytime precipitable water vapor (PWV) values inferred using near-infrared measurements from the Moderate Resolution Imaging Spectrometer (MODIS) \change{installed on the} {\it Aqua} and {\it Terra} satellites, comparing the level-2 satellite products to those from existing ground-based measurements from the radiometer at the Atacama Pathfinder Experiment (APEX) site\change{, and also with the radiometer sited by the Universidad Católica de la Santísima Concepción (UCSC) and the Universidad de Concepción (UdeC) at the Atacama Cosmology Telescope (ACT) near Cerro Toco.} Since both the APEX and UdeC-UCSC radiometer data have been extensively tested and compared to atmospheric transmission models, particularly in low-PWV conditions of interest for astronomy, we use these data to re-calibrate the MODIS data for the entire region, reducing systematic errors to a level of \change{$\lesssim 5\%$, assuming the absolute calibration for the radiometer data is correct}.}
{After re-calibration, the satellite data allow mapping of the PWV across the region, and we identify several promising sites.  Our findings confirm previous results, but provide a more complete and higher resolution picture, filling in key spatial and temporal information often missing from dedicated field campaigns.  
We also examine the seasonal trends in the ground-based data from APEX and in the satellite for a large region encompassing the AAP, finding that both data sets provide tentative indications that \change{the median} PWV has increased \change{by $\approx 0.3$~mm} over the past two decades.
\change{Since the time span of our study is short, we compare our results to model predictions from the World Climate Research Programme (WCRP) Coupled Model Intercomparison Project Phase 6 (CMIP6), and find the trend is in general supported.}}
{We demonstrate a potentially powerful method for siting new facilities such as the Atacama Large Aperture Submillimeter Telescope and extensions to global very long baseline interferometry networks like the Event Horizon Telescope.  Over time, the ability to determine long term trends will improve as further satellite observations are accumulated and new instruments are deployed.}


\maketitle

\section{Introduction}\label{sec:intro}

The Atacama Desert has long been established as an excellent site for millimeter and submillimeter observations \citep[e.g.][]{Radford2008, Radford2016, Otarola2019, Cortes2020, Morris2022}, owing to its reputation for low precipitable water vapor (PWV)\footnote{We note that outside of the field of astronomy, precipitable water vapor is more commonly referred to as total precipitable water (TPW).  Since the audience for this work is largely astronomers, we adhere here to the nomenclature generally adopted by astronomers.} as one of the driest deserts on Earth.
It therefore serves as host to many of the preeminent millimeter and submillimeter astronomical observatories and experiments built in the last several decades, such as the Atacama Submillimeter Telescope Experiment \citep[ASTE;][]{Ezawa2004}, Atacama Cosmology Telescope \cite[ACT;][]{Fowler2007}, Atacama Pathfinder Experiment \citep[APEX;][]{Guesten2006}, Atacama Large Millimeter/Submillimeter Array \citep[ALMA;][]{Wootten2009}, CCAT-prime \citep[CCAT-p, also known as the Fred Young Submillimeter Telescope, or FYST;][]{Parshley2018}, Simons Observatory \citep[SO;][]{Ade2019}, and CMB-S4 \citep{CMB-S4_2019}, to list a few.
\change{In addition to impacting astronomical observations, water vapor is of interest for studies of the sustainability of life, particularly in extremely dry environments \cite[e.g.][]{McHugh2015}, and the transmission of solar radiation in the context of renewable energy \cite[e.g.][]{TheAtacamaSurfaceSolarMaximum}.  In this study, we are focused on the implications for submillimeter and millimeter-wave astronomy.}

Submillimeter atmospheric transmission can be characterized mainly by two components: a dry component dominated by telluric lines such as oxygen, and a wet component dominated by (liquid) water vapor in the atmosphere.  In large part, the dry component is a function of altitude, while contributions from the wet component can be reduced by choosing sites that minimize the PWV.  This leads astronomers to site their observatories at high and dry locations, though we note of course that considerations such as accessibility (road access, infrastructure, and legal restrictions\footnote{For example, as noted in e.g. \cite{DeBreuck2018} and \cite{Otarola2019}, both the hardships of working at high altitude and Chilean labor law make it challenging to build and operate a facility at more than 5500 meters above sea level.}), wind conditions, and temperature play a crucial role.

Long-term trends and interannual variability for these sites are difficult to establish due to the \change{sparsity} of weather stations in this region. The most detailed measurement campaigns, especially at high elevations, are generally those done as part of site testing \citep[e.g.][]{Otarola2019}. 

Recently, the work presented in \cite{Cantalloube2020} examined the impact of climate change and the El Niño \change{Southern Oscillation (ENSO) on the} ability to perform high resolution astronomical imaging using ground based observatories sited in locations traditionally exhibiting superb atmospheric transmission and seeing.
They used onsite observations of temperature for the period covering 2000-2020, and integrated water vapor (IWV)\footnote{We note that IWV, typically expressed in units of kg m$^{-2}$ and PWV, typically in units of mm, are related simply through the density of liquid water, and can be treated as essentially interchangeable.} for 2015-2020, at Paranal Observatory.   
\change{\cite{Cantalloube2020} also relied on the fifth European Centre for Medium-Range Weather Forecasts (ECMWF) Reanalysis \citep[ERA5;][]{Hersbach2020_ERA5}, which models approximately four decades of satellite data, and the ECMWF twentieth century reanalysis \citep[ERA-20C;][]{ERA20CAnAtmosphericReanalysisoftheTwentiethCentury}, which attempts to model historical climate records covering 1900-2010.  In order to assess the expected future trends and study long-term changes, they used climate projections from the Coupled Model Intercomparison Project Phase 6 \citep[CMIP6;][]{CMIP6}, which has a spatial resolution of 1 degree (100~km).}
They note that climate change will soon change the conditions for operations and observations, especially as temperature increases, though any given location can be dominated by its microclimate variability.  In terms of findings directly relevant to submillimeter astronomy, \cite{Cantalloube2020} also identify a potential trend for worsening atmospheric turbulence, which would affect phase coherence, and they report an increase in the number of days with low water vapor in recent years for the region surrounding Paranal, although they note that their time series is too short to determine the statistical significance.

At the same time, \cite{BOHM2020103192} studied the variability of the integrated water vapor column over the entire Atacama region, mainly using climate reanalysis data from \change{ERA-20C to determine trends over the entire 20th Century at 125~km resolution. They validate the ERA-20C data using satellite data from MODIS spanning 2000-2010, and use the former to show that the variability in the integrated water vapor increases in years dominated by La Niña, producing more extreme values as a result.}

Through this study, \change{we aim to establish the reliability of using MODIS NIR measurements to study spatial and temporal trends in PWV, which can strongly impact ground-based astronomical observations at nearly all wavelengths.
Our study focuses on the northern Atacama Desert in and around the Atacama Astronomical Park\footnote{\url{https://www.conicyt.cl/astronomia/sobre-el-parque/}} \citep[AAP, a.k.a. Parque Astron\'omico de Atacama; see][for further site details]{Bustos2014}.
In addition to the AAP itself, our region of interest includes the ALMA concession and surrounding regions nearby, such as the site hosting the Large Latin American Millimeter Array \citep[LLAMA;][]{Romero2020}.
Previous studies \citep[e.g.][]{Suen2014} have also relied on similar satellite data to perform studies of PWV in the context of submm astronomy.  However, several studies have demonstrated that MODIS NIR PWV measurements can be biased high with respect to data taken using ground-based GPS stations in dry regions \citep[e.g.][]{VAQUEROMARTINEZ2017, He2019, Khaniani2020} .  In this study we perform our own calibration correction, correcting the MODIS data retrievals to agree with those from the APEX radiometer. We then validate our recalibration correction by comparing the MODIS PWV with measurements from the radiometer installed by the Universidad Católica de la Santísima Concepción (UCSC) and the Universidad de Concepción (UdeC) at the Atacama Cosmology Telescope (ACT) near Cerro Toco.}

We note that the AAP surrounds the ALMA concession, and we therefore only show the outer border of the former in each map figure (see e.g. the shape region in Figure \ref{fig:topo}).

\section{Data}\label{sec:data}

\subsection{Elevation data}\label{sec:data:topography}

\begin{figure}
    \centering
    \includegraphics[clip,trim=0mm 3.5mm 3mm 2mm,width=0.99\columnwidth]{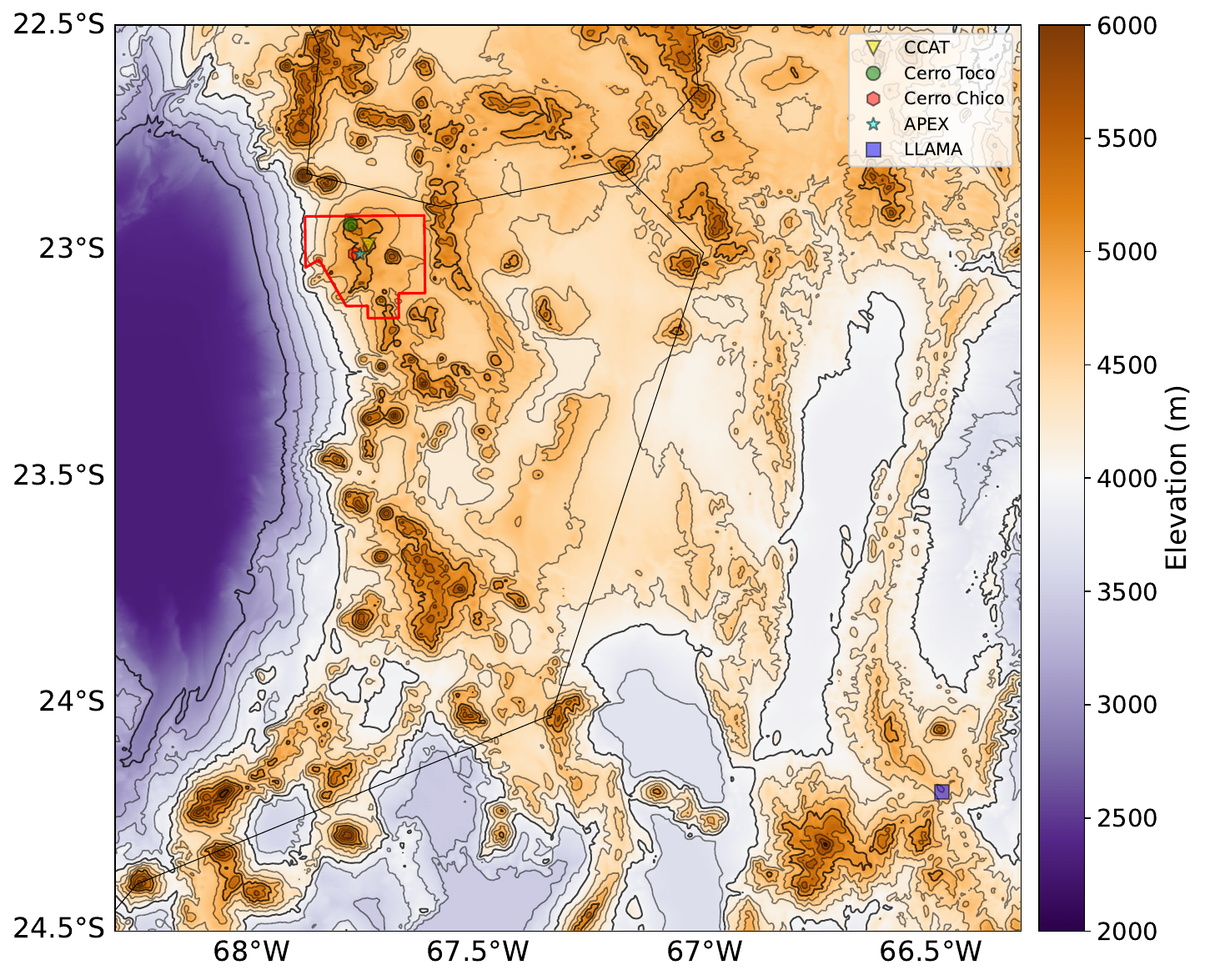}
    \caption{Topography of the region considered using the 2020 release from GEBCO \citep{geosciences8020063}, which relies on the SRTM15+ dataset \citep{tozer2019global}.  
    The locations of several noteworthy sites are indicated by markers, as labelled in the legend. The outer boundary of the AAP, which encloses the ALMA concession, is denoted by the red shape region.  The borders between Chile, Bolivia, and Argentina are shown in black. The elevation contours start at 3000 meters a.s.l., and are incremented in +250 meter steps.  Thicker contour lines appear at elevations of 3000, 4000, 5000, and 6000 meters a.s.l.}
    \label{fig:topo}
\end{figure} 

It is generally accepted that higher elevations generally imply a lower PWV, as is seen in models for the atmospheric transmission \citep[e.g.][]{paine_scott_2017_438726,Pardo2019}.  By way of comparison, we employ 15\arcsec\ resolution topographic data from the Shuttle Radar Topography Mission (SRTM15+; \citealt{tozer2019global}), which measured the surface rock elevation across the globe. The data are shown in Figure \ref{fig:topo} for the region of interest to the current study. The data were obtained using the 2020 General Bathymetric Chart of the Oceans data release (GEBCO; \citet{geosciences8020063}), which provides the data in a readily accessible netCDF format.\footnote{\url{https://www.gebco.net/data_and_products/gridded_bathymetry_data/}} 
The inclusion of these data allow us to focus on sites with elevations ranging from 5000 to 5500 meters above sea level (a.s.l.).

\subsection{Satellite near infra-red data}\label{sec:data:modis}

\begin{figure}
    \centering
    \includegraphics[clip,trim=0mm 3.5mm 2mm 2mm,width=0.99\columnwidth]{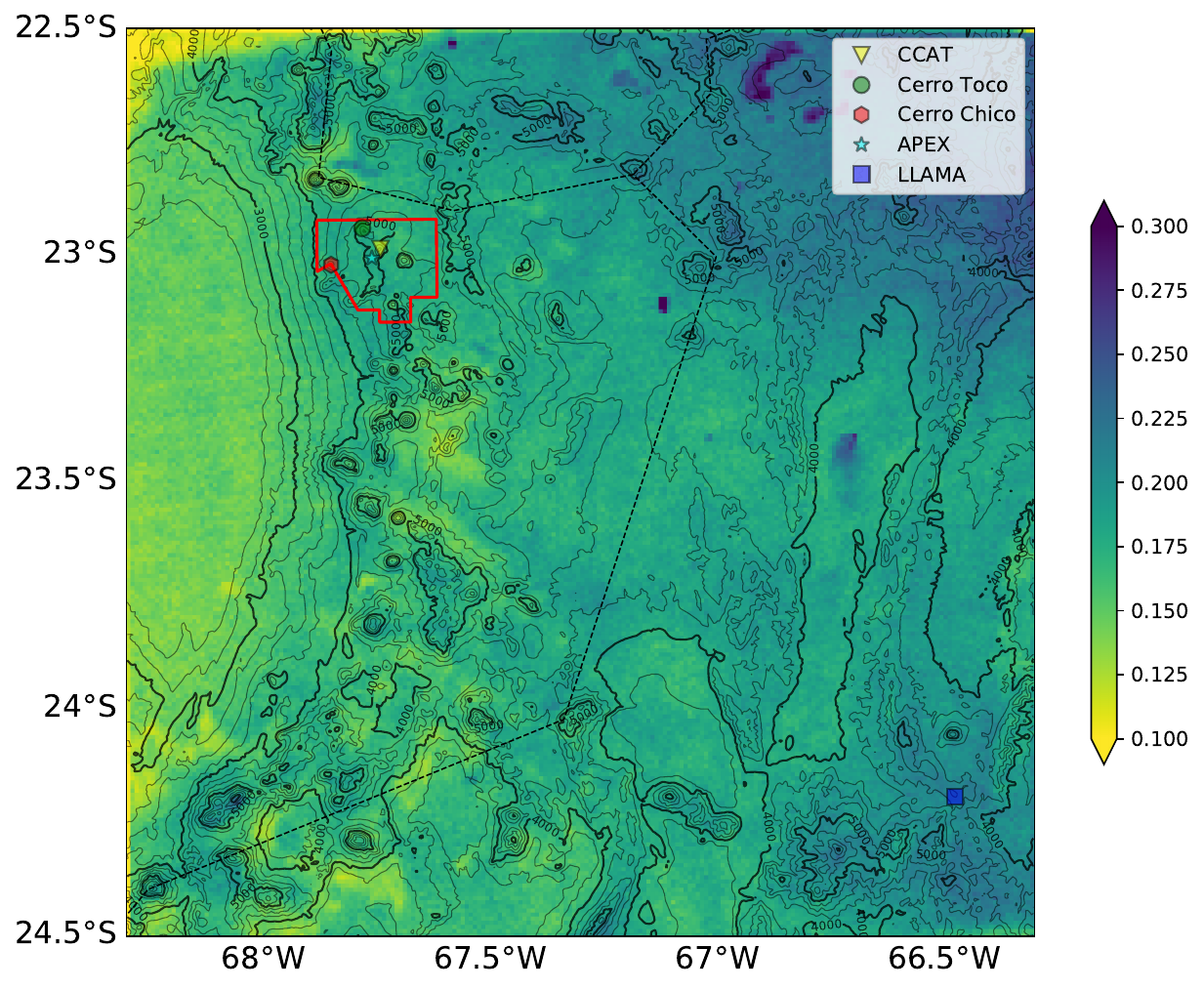}
    \caption{Fractional cloud coverage inferred from the flagging of MODIS data.
    The borders, AAP+ALMA boundary, elevation contours, and markers are the same as in Fig.~\ref{fig:topo}.}
    \label{fig:clouds}
\end{figure}

The National Aeronautics and Space Administration (NASA) launched \change{the {\it Terra} and {\it Aqua} satellites at the end of 1999 and mid-2002, respectively, with each carrying a Moderate Resolution Imaging Spectroradiometer (MODIS) instrument.}
In similar but offset orbits, each satellite provides two measurements per site per day for our region of interest using 36 near infrared (NIR) spectral bands spanning 0.4--1.4 $\mu$m and providing a resolution of approximately 250--1000 meters.
These level-2 measurements are used to infer the integrated column of water vapor using absorption, aerosol scattering, and surface reflection of solar radiation at a resulting resolution of 1~km for daytime data.  Hereafter, we equate the integrated column of water vapor with precipitable water vapor.
The PWV is inferred using the ratios of clear NIR windows at 0.865 and 1.24 $\mu$m, and lines due to absorption by atmospheric water at 0.905, 0.936, and 0.94 $\mu$m \citep[see][]{Gao1998,Gao2015_MOD,Gao2015_MYD}.\footnote{\url{https://modis.gsfc.nasa.gov/data/atbd/atbd_mod03.pdf}}
\change{As the NIR measurements depend on transmitted and reflected sunlight, they are necessarily only available during daytime.

As cloud cover can affect the ability to compute the NIR transmittance through the atmosphere and therefore affect completeness of the dataset, we exclude data flagged for cloud cover.  In Figure \ref{fig:clouds}, it can be seen that the cloud coverage fraction is typically $\approx$15-25\%, and is typically higher for locations farther to the east.
We note that some data flagged for cloud cover at a given location may be due to solar glint, rather than poor atmospheric transmission.  Therefore, the flagging does not necessarily imply that a given flagged measurement has high PWV.}

MODIS {\it Terra} and {\it Aqua} also provide mid-IR data at a more limited resolution of 5~km.  These data include nighttime measurements.  However, initial tests indicated large systematic offsets between the NIR and mid-IR measurements, confirming the results of \cite{Wang2017}.  
Since the \change{topography of the region varies greatly on scales $<5$~km} (see Section \ref{sec:data:topography}), we exclude the mid-IR data from our analysis, noting that -- as established in e.g.\ \cite{Suen2014} -- the PWV generally drops at night.  Thus, the daytime PWV we report on here can serve as a generally good indicator of the worst-case scenario conditions.  A future work will address the diurnal variability in PWV in more detail.

\subsection{Ground-based water vapor radiometer data}\label{sec:data:radiometers}

We use PWV data collected since May 2006 by the Atacama Pathfinder Experiment using their ground-based water vapor radiometer mounted inside the Cassegrain cabin of the telescope\footnote{See \url{http://archive.eso.org/eso/meteo_apex.html}.}. In short, the radiometer measures the irradiance in the 183~GHz water line of the atmosphere, which is sampled in three (up to 2012) or six (from 2012) bands. This is combined with a model of the atmosphere using local measurements of temperature, atmospheric pressure and humidity as additional input parameters. The PWV is then derived using the {\tt ATM model} \citep{Pardo2019} which is then used to infer the atmospheric transmission.  Since water dominates the transmission at this frequency, the signal provides information on the wet component of the atmosphere.    
The results in \citet{Cortes2020} indicate that the systematic uncertainty in the calibration of the APEX radiometer data should be $< 3\%$ (i.e. more than 97\% accurate) \change{if the atmospheric modelling is taken as fiducial}.  
Since we are mainly interested in using the APEX radiometer values to correct the values inferred from MODIS {\it Terra} and {\it Aqua} during conditions of low PWV, we only use values of PWV $< 3$~mm to re-calibrate (or correct) the MODIS {\it Terra} and {\it Aqua} data. 

\section{Methodology}\label{sec:methods}

\begin{figure*}
    \begin{minipage}[t]{0.215\textwidth}
        \centering
        \includegraphics[clip,trim=35mm 2.5mm 43mm 0mm,width=0.99\textwidth]{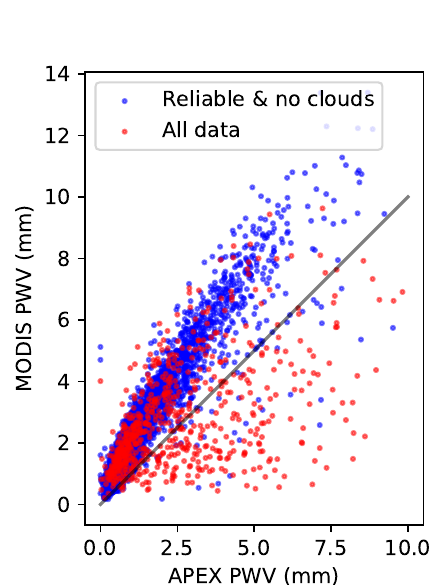}
    \end{minipage}
    \begin{minipage}[t]{0.785\textwidth}
    \centering
        \includegraphics[width=0.495\textwidth]{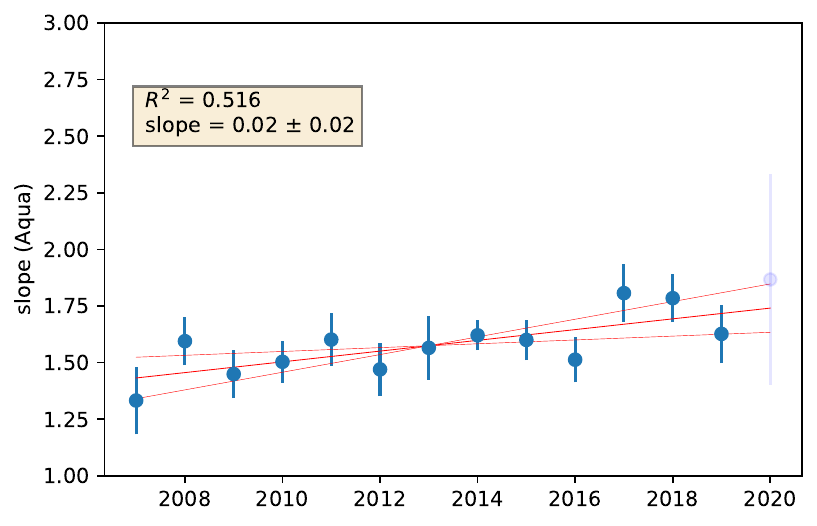}
        \includegraphics[width=0.495\textwidth]{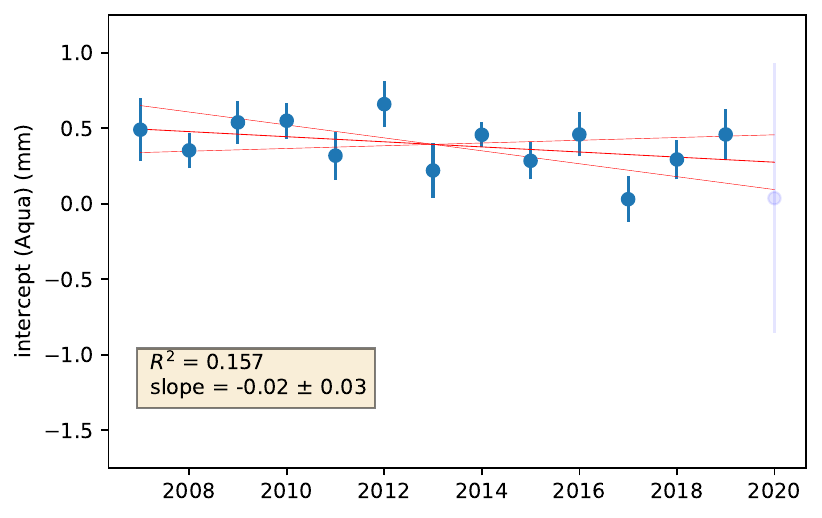}
        \includegraphics[width=0.495\textwidth]{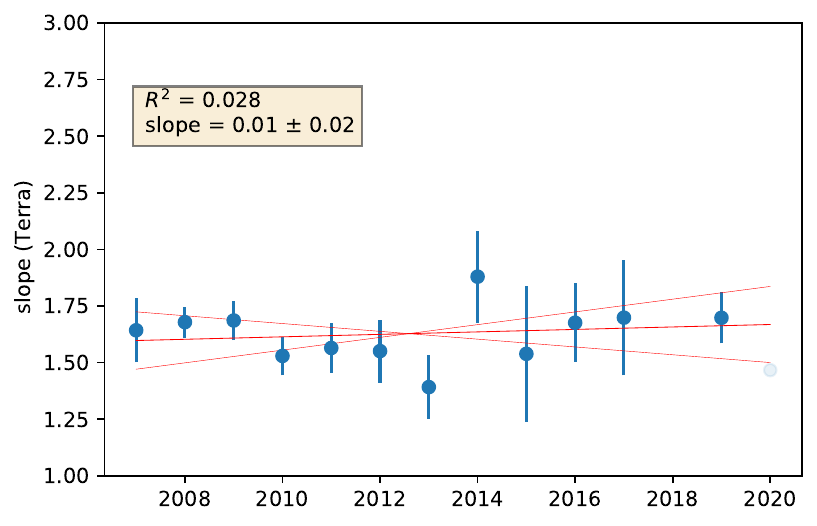}
        \includegraphics[width=0.495\textwidth]{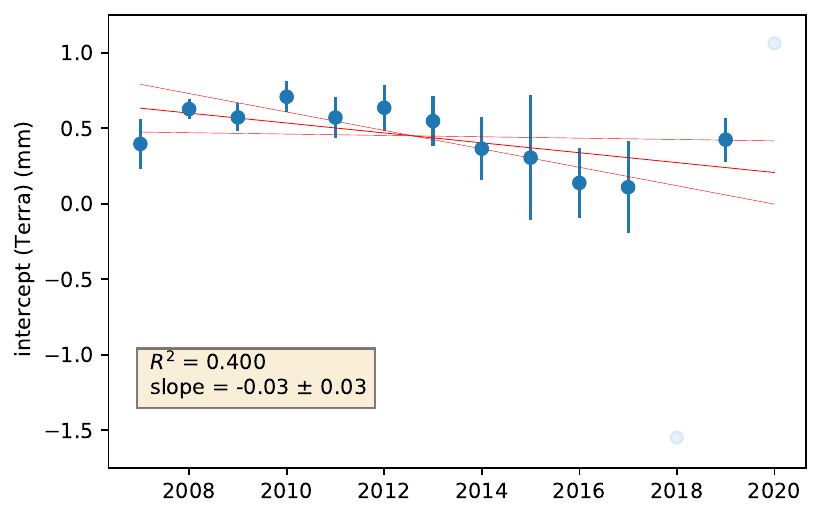}
    \end{minipage}
    \caption{
    {\bf Upper left:} Scatter in the values of PWV as measured by MODIS versus those from the APEX site radiometer.  MODIS data flagged by NASA as both `reliable' and `no clouds' are reported in blue, while the rest of the data are shown in red, and are excluded from the fitted relations to the right. A unity slope line with zero intercept is shown in black as a visual guide.
    {\bf Right four panels:}
    The fit slopes and intercepts used for re-calibration of the MODIS PWV data. The error bars denote the $95\%$ confidence interval for each value and the red lines denote the best fit and  $\approx$2-$\sigma$ deviations from this. 
    As the data reported include up to the beginning of August 2020, we exclude 2020 from the analysis but show the data point in faint blue for completeness.
    The {\bf upper panels} show the case of {\it Aqua}, for which a significant time dependence in the slope is clear and confirmed by the higher value for the linear regression coefficient of determination $R^2$.
    The {\bf lower panels} show the case of {\it Terra}, for which we do not see any significant time variation.
    }
    \label{fig:pwv_recal}
\end{figure*}

We obtain the MODIS NIR data from the Level-2 and Atmosphere Archive \& Distribution System (LAADS) website hosted by, and part of, the National Aeronautics and Space Administration (NASA).\footnote{\url{https://ladsweb.modaps.eosdis.nasa.gov/search/}}  
The {\it MOD05\_L2} and {\it MYD05\_L2} data correspond to {\it Terra} and {\it Aqua}, respectively. 
We select the 2$^\circ \times 2^\circ$ region spanning longitudes of 66.3$^\circ$ to 68.3$^\circ$ West, and latitudes of 22.5$^\circ$ to 24.5$^\circ$ South (i.e. the region -68.3$^\circ$,-22.5$^\circ$,-66.3$^\circ$,-24.5$^\circ$), which includes the entire area around the Parque Astron\'omico de Atacama (Atacama Astronomy Park, see region indicated in red in Figure \ref{fig:topo}) in Chile, which surround APEX, ALMA, and several other observatories. The 2$^\circ \times 2^\circ$ region was selected to include the future site of LLAMA nearby in Argentina.  

While the level-2 data are gridded at a resolution of 1~km, the grid itself is at an arbitrary orientation dependent upon the scan direction of the satellite during that pass. 
We therefore regridded it onto a new grid with pixels that are 1/110 degree ($\approx 1.01$~km) in extent. We note that this operation forced some measurements into the same pixel bin, while others have no data. Hence, for each scan, we first averaged the available data for each pixel and then performed a simple bilinear interpolation to fill in the gaps of those empty ones. To avoid having misleading data on those days when recorded data was scarce, we decided to apply this interpolation only to those pixels with at least 4 out of the 8 surrounding pixels with original data.

Using nearly 20 years of MODIS data, we compare the MODIS PWV values for the spatial bin containing the APEX site to the in-situ APEX radiometer measurements.  For this, we separately analyze the {\it Terra} and {\it Aqua} satellite data. After determining the scaling between the PWV measurements and any temporal evolution versus the APEX data, we apply these corrections to the original MODIS data. 

\change{PWV values inferred from the APEX radiometer measurements are recorded every minute. However, the MODIS data we obtained are stored in files containing values from a five minute interval.  While it is in principle possible to obtain more accurate information about the timestamp of the MODIS measurement, we instead facilitate comparison by simply averaging the radiometer measurements appropriate for the 5 minute interval of the MODIS file under consideration, as we are not concerned with higher time resolution than this.}
Using only the data flagged as useful, reliable and without clouds (see the MODIS atmosphere QA plan for Collection 061, page 18)\footnote{\url{https://modis-images.gsfc.nasa.gov/_docs/QA_Plan_C61_Master_2017_03_15.pdf}}, \change{we fit a simple, linear scaling relation between MODIS NIR measurements and the ground-based APEX radiometer measurements with an $R^2=0.784$.}  The MODIS data are shown compared to the APEX radiometer values in the upper left panel of Figure \ref{fig:pwv_recal}, with data flagged as useful shown in blue. 
As can be seen in the figure, the slope is not 1:1; rather the space-based MODIS NIR measurements of PWV are systematically higher than the ground-based radiometer ones. 
While it is clear the majority of the data points we excluded fall on the same scaling relation as those included (e.g. the cluster of points at MODIS PWV $<3$~mm and APEX PWV $<2.5$~mm in the leftmost panel of Fig.~\ref{fig:pwv_recal}), the data that were potentially affected by clouds or were otherwise `unreliable' exhibit systematically higher scatter.

Since the ground-based measurements have been shown to be accurate to within an uncertainty of $<$3\%, we derive and apply a simple slope-intercept scaling relation to correct this:
\begin{equation}
    \label{eq:correction}
    \widehat{\text{PWV}}_\text{c}(t) = \frac{\text{PWV}_\text{m}(t) - b(t)}{m(t)}
\end{equation}
Here $\text{PWV}_\text{c}(t)$ is the corrected value for the PWV, $\text{PWV}_\text{m}(t)$ is the measured value, $b(t)$ is the (possibly time-dependent) value for the intercept, and $m(t)$ is the (possibly time-dependent) value for the slope.

\begin{figure}
    \centering
    \includegraphics[clip,trim=0mm 2mm 2mm 2mm,width=0.99\columnwidth]{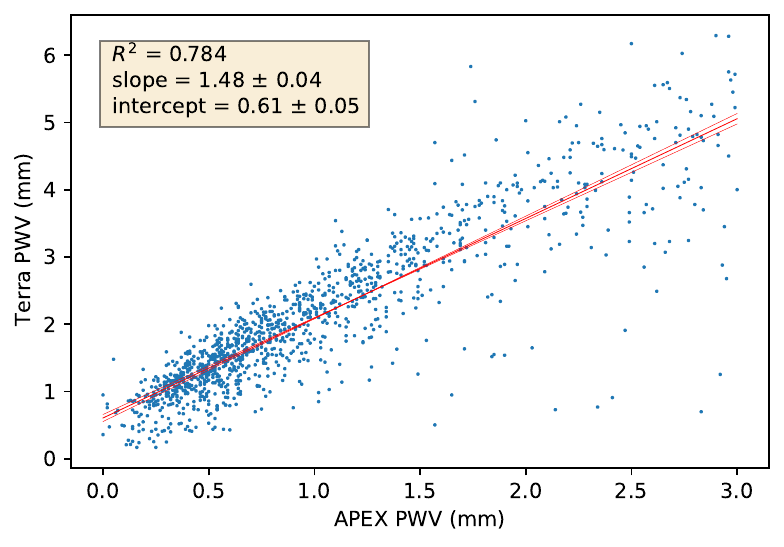}
    \caption{Regression analysis between the {\it Terra} and APEX measurements of PWV. Here we only use values for the driest conditions, $\text{PWV} < 3$~mm. We find that the slope $m=1.48$ and intercept $b=0.61$ can be treated as a constant versus time.}
    \label{fig:terra_regression}
\end{figure}

We find a significant time dependence whenever comparing the uncorrected {\it Aqua} and {\it Terra} PWV data, regardless of location. 
We then compare both MODIS data sets to the fiducial APEX radiometer data. 
The right four panels of Fig.~\ref{fig:pwv_recal} report the fit, time-dependent values for $m(t)$ and $b(t)$ used for re-calibration of the MODIS PWV data, assuming a linear form.  The error bars represent the $95\%$ confidence interval for each value. As the data reported only include up to the beginning of August 2020, and since APEX was largely shut from mid-March through late August, we exclude 2020 from our analysis. The data point for the first seven months of 2020, shown in blue, is only plotted for completeness. 

In the case of {\it Aqua}, a significant time dependence versus the APEX radiometer data is seen in the values for the slope $m(t)$ (upper left panel of the right four in Fig.~\ref{fig:pwv_recal}). 
This is supported by the higher value we find for the linear regression coefficient of determination $R^2$.  This value quantifies the improvement in fit versus a time-independent model, and is closest to $R^2 = 1$ when the variance after fit model subtraction is much smaller than the variance after subtraction of the mean data value.
We also note that the intercept (upper right panel of the right four, Fig.~\ref{fig:pwv_recal}) has slightly decreased, implying that {\it Aqua} now reports systematically higher PWV values than it did in its first several years. To correct for this, we re-calibrate the {\it Aqua} data by applying  Equation \ref{eq:correction} with $m(t) = 0.0237\cdot t-46.1863$, $b(t) = -0.0169\cdot t + 34.3632$, where $t$ is measured in years.

In the case of {\it Terra}, we do not see any significant variation versus time in the slope we inferred ($m(t)$, lower left panel of the right four in Fig.~\ref{fig:pwv_recal}). The apparent evolution in the $b(t)$ values is likely due to the relative sparsity of {\it Terra} measurements between 2014 and 2018 (note the larger error bars; the number of measurements drops to below 80 per year for the APEX site, compared to more than 250 for {\it Aqua} in each of these years). We exclude data from 2018 in particular as only 7 valid values of PWV were available for the bin including APEX. After finding no significant time evolution in the {\it Terra} data (vs that from APEX), we exclude any time dependence and fit the entire data set (see Figure \ref{fig:terra_regression}). 

\begin{figure}
    \centering
    \includegraphics[clip,trim=5mm 2mm 14mm 12mm,width=0.99\columnwidth]{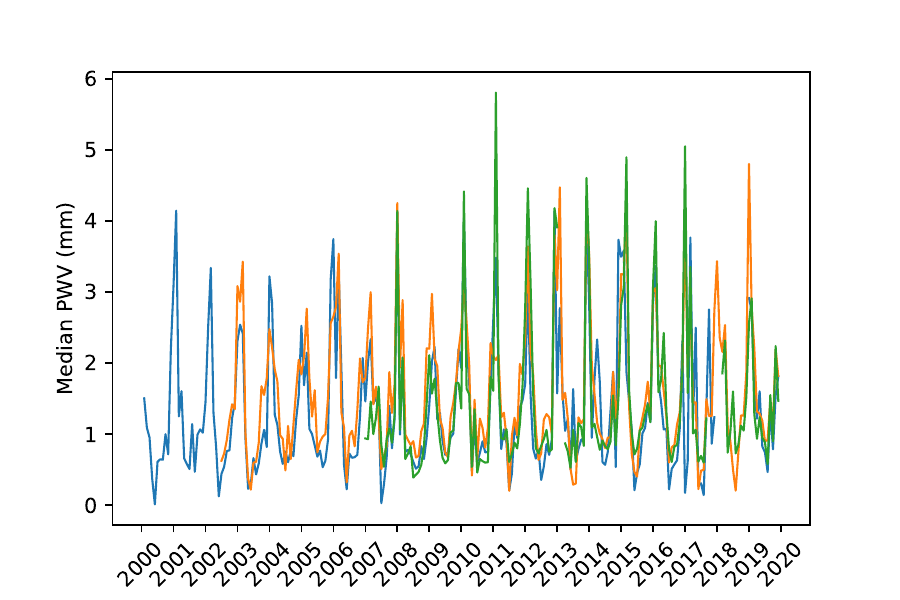}
    \caption{Using the re-calibration relation we inferred, the mean PWV vs time of {\it Terra} (blue) and {\it Aqua} (orange) are shown in comparison to the APEX radiometer data (green).}
    \label{fig:apex_terra_aqua_timeseries}
\end{figure} 

\begin{figure}
    \centering
    \includegraphics[clip,trim=0mm 0mm 10mm 0mm,width=0.99\columnwidth]{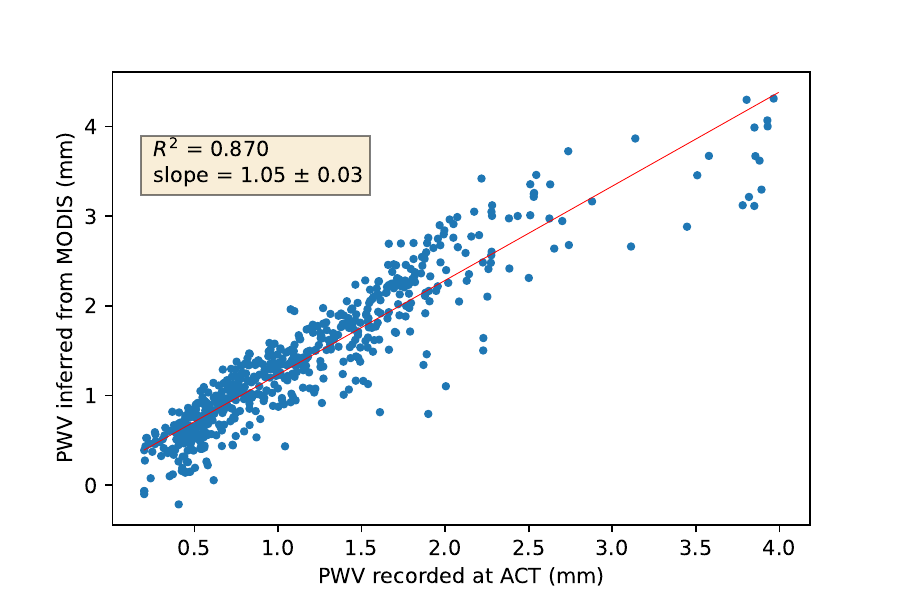}
    \caption{Comparison of the UdeC-UCSC radiometer located at the ACT telescope site with the recalibrated MODIS {\it Terra} and {\it Aqua} PWV values. The fit slope of 1.05 implies that on average the recalibrated MODIS PWV values are within 5\% of the PWV inferred from the UdeC-UCSC radiometer values.
    }\label{fig:UdeC-UCSC_radiometer}
\end{figure}

After applying these re-calibration corrections, we find the values for PWV versus time for the three instruments are in good agreement (see Figure \ref{fig:apex_terra_aqua_timeseries}), apparently justifying our approach to re-calibration.  
\change{Since the values of PWV inferred from MODIS measurements should not depend on the locations of {\it Aqua} and {\it Terra} above the Earth, we hypothesize that our inferred corrections for the 1~km resolution {\it Aqua} and {\it Terra} data could be applied to that from any dry location across the globe, as long as the value for $\text{PWV}$ remains in the range for which the calibration is valid. Further study and comparison to sites from around the globe will required to verify this hypothesis.  
We also verified the corrections we infer for MODIS with the UdeC-UCSC radiometer located at the ACT telescope site near Cerro Toco, and found the two to agree to within 5\% (see Figure \ref{fig:UdeC-UCSC_radiometer}.
Again, we average the radiometer measurements within the five minute interval corresponding to that of the MODIS data.}

After calibration, we use the MODIS data for the maps and time series analyses presented in the following sections.

\section{Results}\label{sec:results}

\begin{figure*}
    \centering
    \includegraphics[clip,trim=0mm 2mm 20mm 0mm,width=0.495\textwidth]{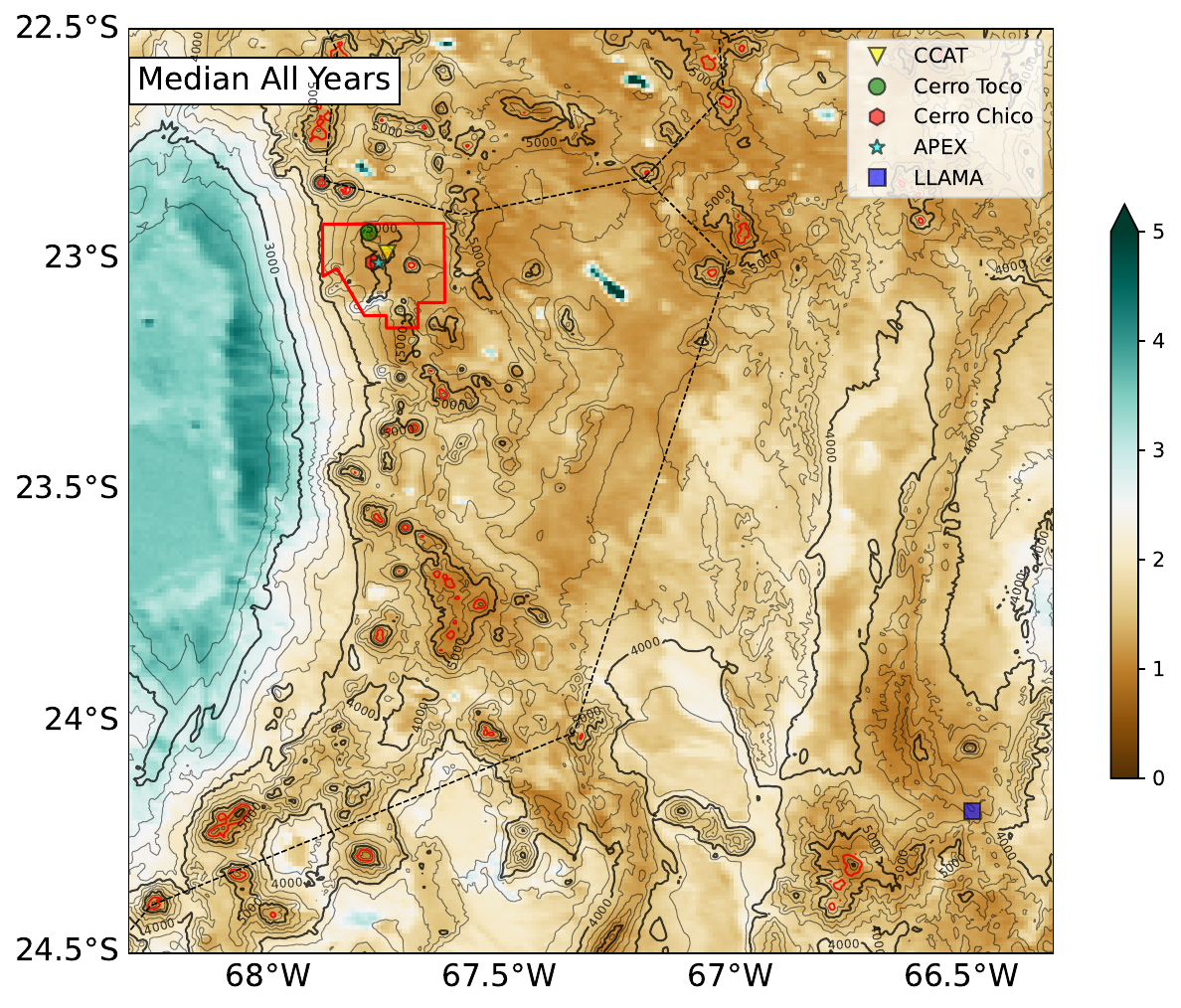}
    \includegraphics[clip,trim=20mm 2mm 0mm 0mm,width=0.495\textwidth]{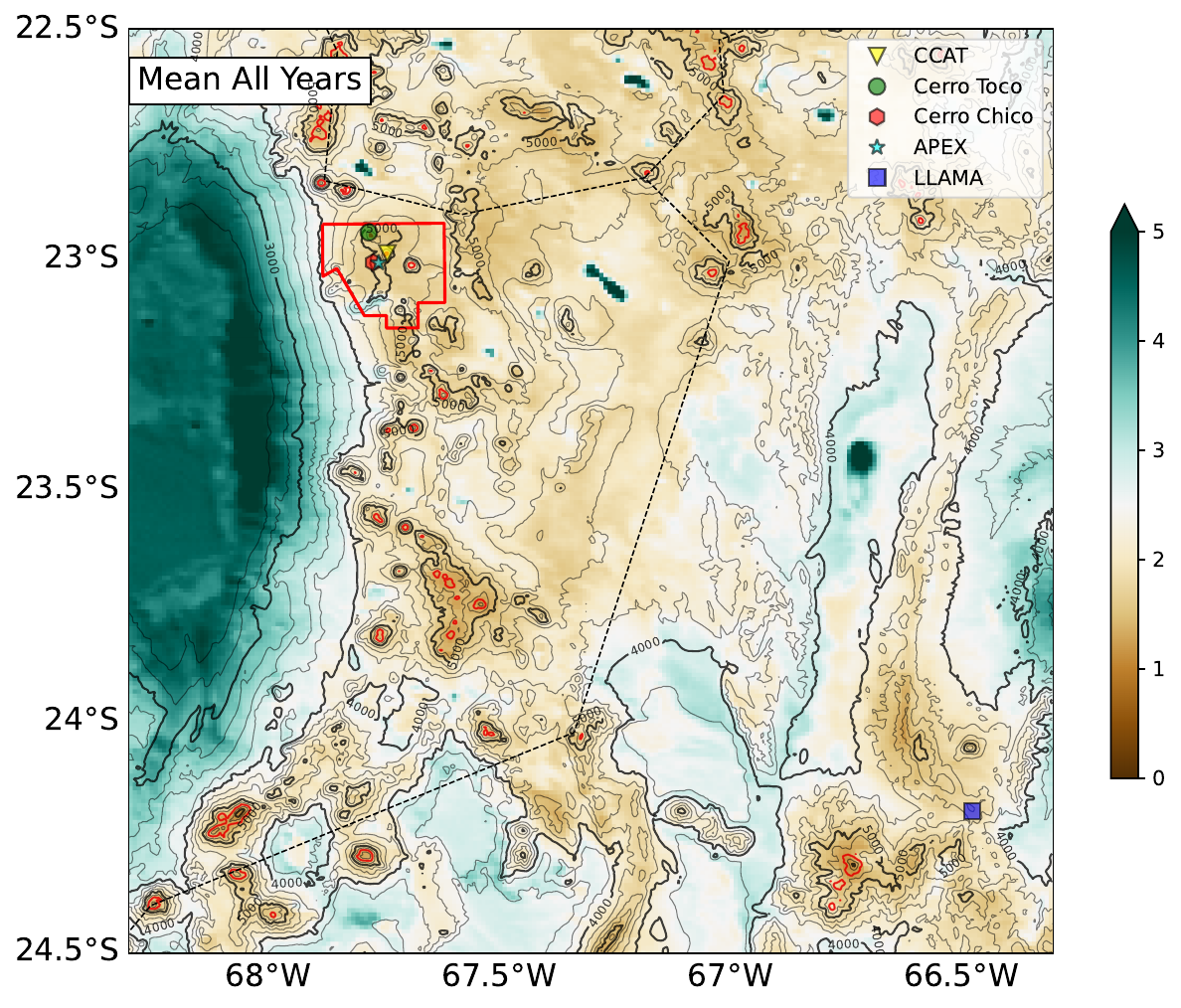}
    \caption{Median (left panel) and mean (right panel) daytime PWV values in millimeters for all available years of MODIS {\it Terra} and {\it Aqua} NIR measurements. The mean PWV values are systematically higher than the median values due to the skewed, non-Gaussian distribution intrinsic to the data.
    The color scale is the same in both maps, and deep green values can exceed 5~mm.
    The borders, AAP+ALMA concession boundary, markers, and elevation contours are the same as in Figure \ref{fig:topo}, with the addition of red contours to denote locations with elevations higher than 5500 meters a.s.l.
    }\label{fig:pwv_median_mean}
\end{figure*}

\subsection{Yearly and seasonal Averages}\label{sec:scaling}

After re-calibration, we map the yearly median and mean values, averaged over all years, in Fig.~\ref{fig:pwv_median_mean}, while in Appendix \ref{sec:appendix:pwv}, we show in Figure \ref{fig:pwv_seasonal_median} the median seasonal values for the entire region.  
For consistency with other studies \citep[e.g.][]{paine_scott_2017_438726}, we define the four seasons as the three integral months beginning with the one in which the season commences (i.e. `Austral summer' is treated as December-January-February, `autumn' is March-April-May, `winter' is June-July-August, and `spring' is September-October-November).
The median values in each map are consistently lower due to the skewed nature of the PWV distribution.  As is clear in the upper left panel of Fig.~\ref{fig:pwv_recal}, the distribution in measured PWV values is non-Gaussian and exhibits a long tail of high PWV values. 
We also report in Appendix \ref{sec:appendix:pwv} the monthly and the yearly median PWV (Figures \ref{fig:pwv_median_monthly}, \ref{fig:pwv_median_yearly_2000-2011}, and \ref{fig:pwv_median_yearly_2012-2020}).
We note that several features can be readily identified from the PWV maps: lakes and more humid areas such as the Salar de Atacama stand out as the wettest zones, while the highest volcanic peaks show up clearly as the driest.
\change{We also note that the mean value is likely biased high due to the non-Gaussian distribution of PWV values.}

\begin{figure}
    \centering
    \includegraphics[clip,trim=0mm 2mm 0mm 0mm,width=\columnwidth]{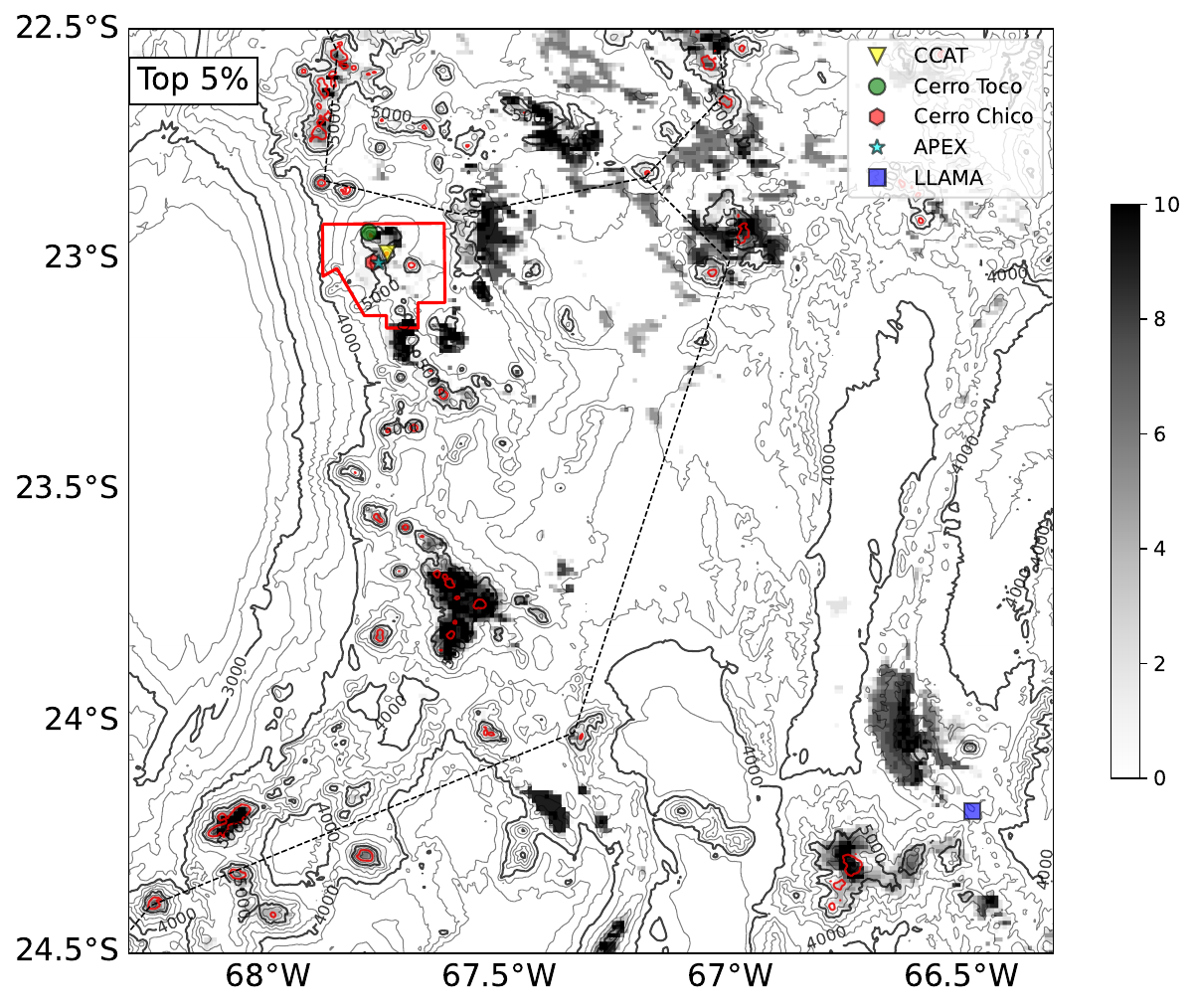}
    \caption{Number of PWV deciles where a given bin exhibits a PWV value in the best 5\% when compared to the whole region.  Bins with a value of 10 (closest to black) are in the best 5\% for all the deciles.
    Labels, markers, and overlays are the same as in Figure \ref{fig:pwv_median_mean}.
    }\label{fig:pwv_top5cdf}
\end{figure}

The seasonal values reported in Figure \ref{fig:pwv_seasonal_median} confirm that the median PWV is best (lowest) across the entire region during the Austral winter (i.e. July--September), while it is worst (highest) during the Austral summer, which coincides with the altiplanic winter (particularly January--March). 
We also confirm that the highest peaks in the region tend to show the lowest PWV.
This can be seen more clearly in Figure \ref{fig:pwv_top5cdf}, where we identify the locations having the best (driest) 5\% of PWV for each decile (percentiles of the distribution in steps of 10\%) of PWV conditions. To compute the values in this map, we calculate for each location the 10 deciles (10, 20, ...90, 100 percentiles) from the full distribution of daily values. Then, for each decile, we rank the corresponding decile value among all the locations in the whole region. The driest 5\% locations get a value of 1 in the map. We do that for each decile, and keep adding. Locations with a map value equal to 10 belong to the driest 5\% in all the deciles and hence, are best in all conditions.

\subsection{Seasonality and temporal trends near APEX}\label{sec:results:seasonality}

\begin{figure}
    \centering
    \includegraphics[clip,trim=8mm 5mm 15mm 12mm,width=0.99\columnwidth]{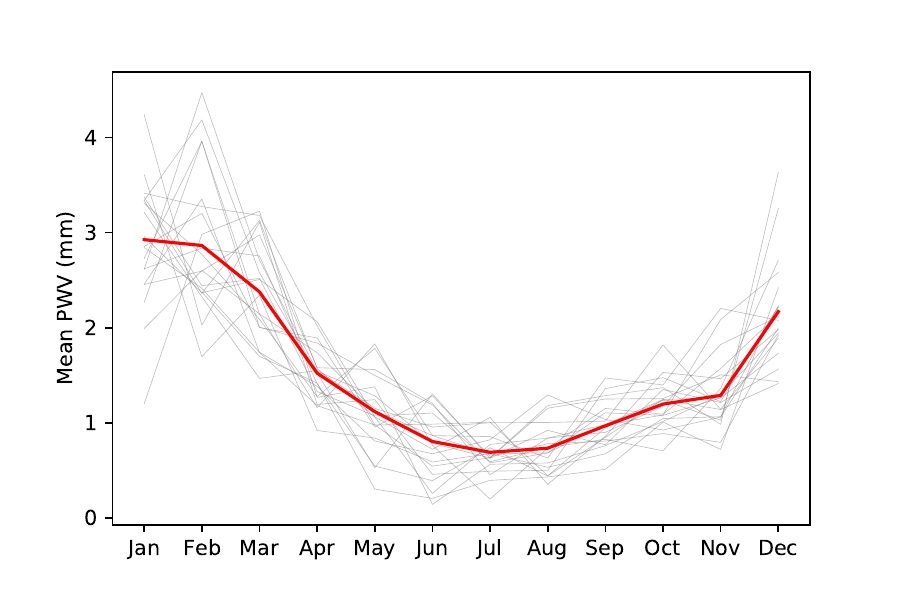}
    \includegraphics[clip,trim=8mm 5mm 15mm 12mm,width=0.99\columnwidth]{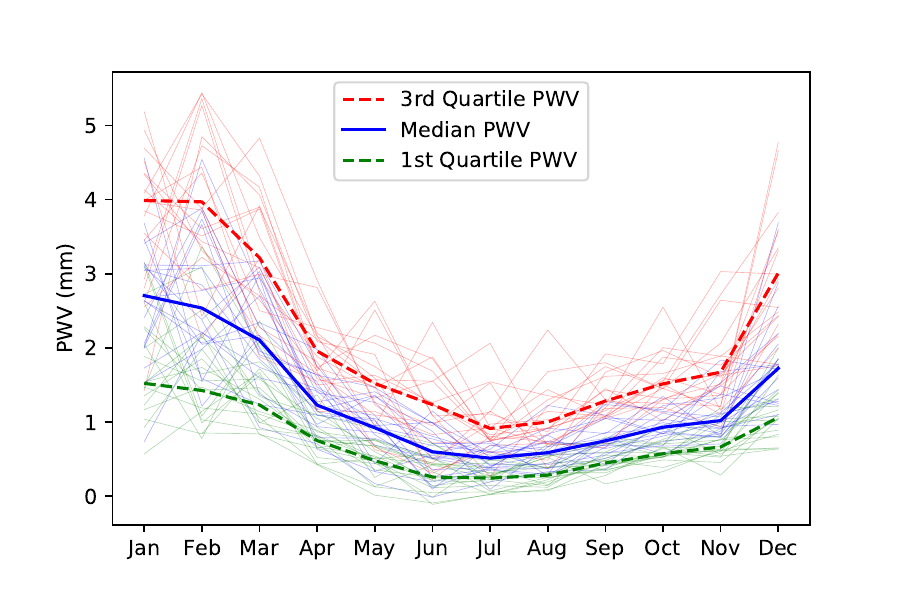}
    \caption{
    {\bf Upper:} Comparison between the mean year (red) and every year in the period 2000-2019 for the $1~\rm km \times 1~km$ cell \change{of MODIS satellite data} containing the APEX site.
    {\bf Lower:} Comparison between the first, second, and third quartiles for every year in the period 2000-2019 for the same cell \change{of MODIS data} as the upper panel.
    }
    \label{fig:comparison_year}
\end{figure}

\begin{figure}
    \centering
        \includegraphics[clip,trim=0mm 12mm 8mm 11mm,width=0.99\columnwidth]{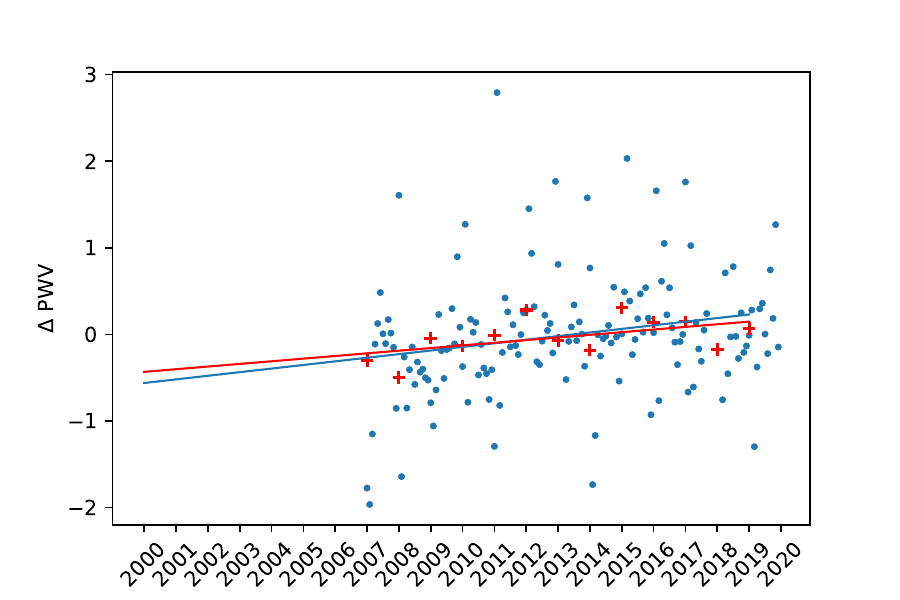}
    \includegraphics[clip,trim=0mm 0mm 8mm 10mm,width=0.99\columnwidth]{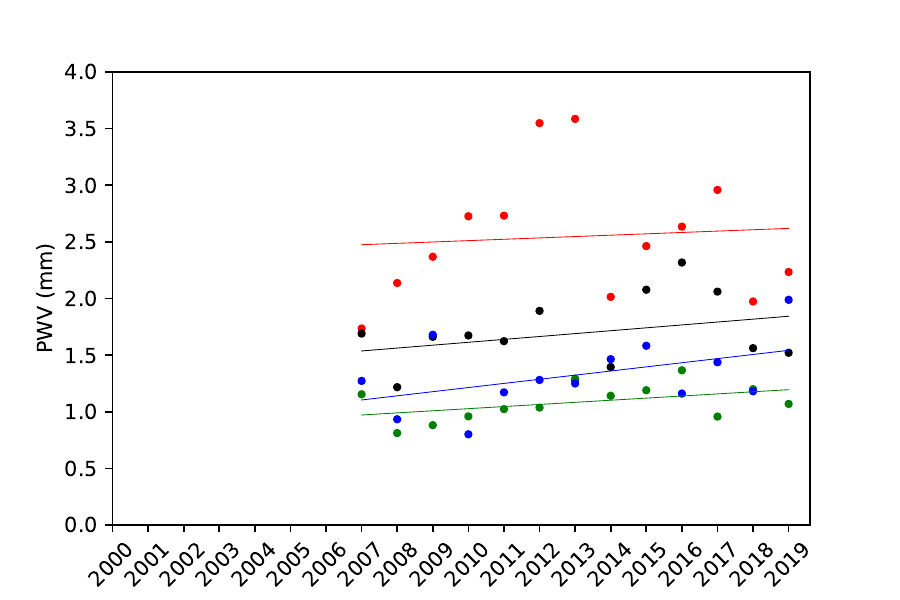}
    \caption{
    {\bf Upper:} Trend analysis in the deseasonalized monthly time series (blue dots) and the yearly mean time series (red crosses) in PWV using the radiometer data for the APEX site.
    {\bf Lower:} Seasonal trend analysis for radiometer data from the APEX site.  From highest to lowest average PWV, the curves are summer (red), autumn (black), spring (blue), and winter (green). The values of the fit slopes and the $R^2$ values for the fits are reported in Table~\ref{tab:fits}.
    }
    \label{fig:median_regression_APEX_Rad}
\end{figure}

\begin{figure}
    \centering
    \includegraphics[clip,trim=0mm 12mm 8mm 12mm,width=0.99\columnwidth]{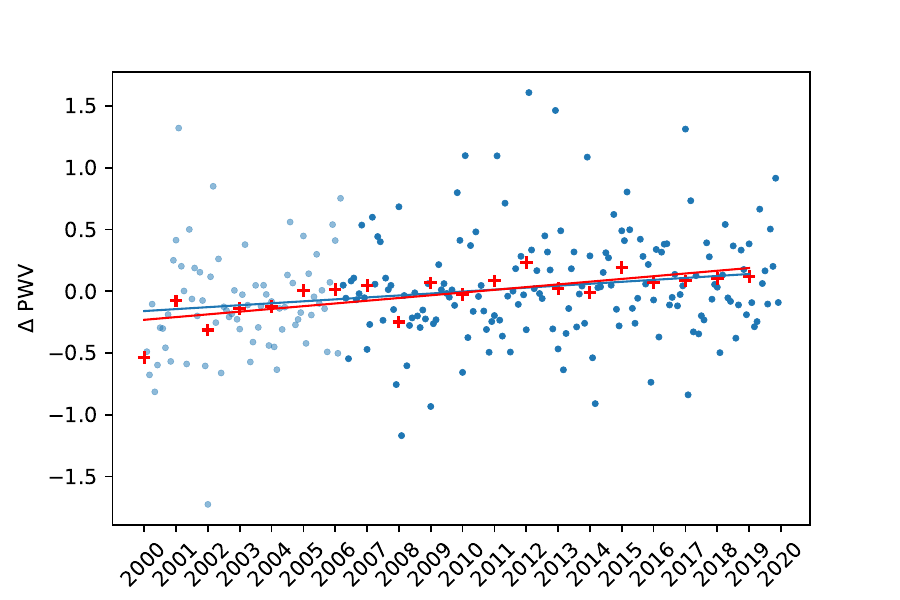}
    \includegraphics[clip,trim=0mm 10mm 8mm 10mm,width=0.99\columnwidth]{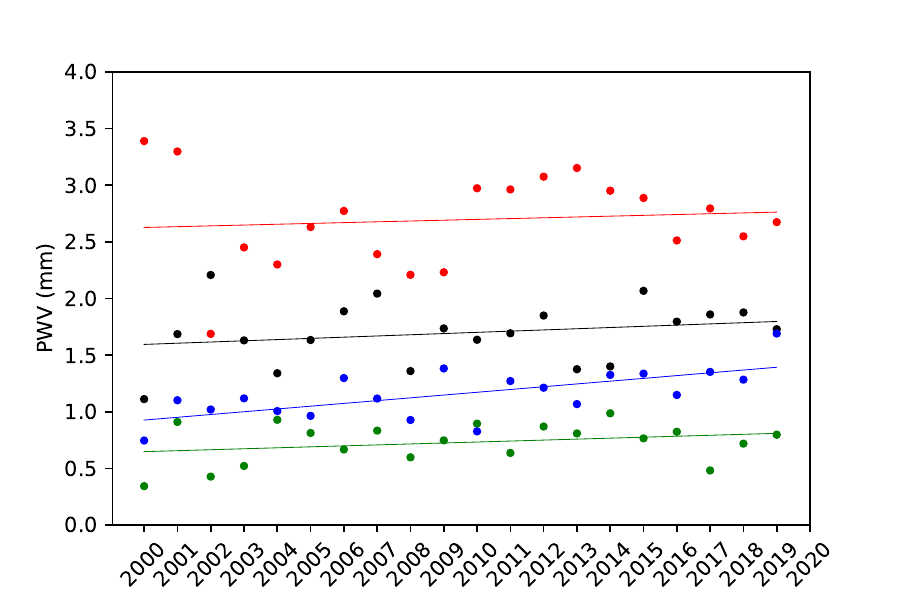}
     \includegraphics[clip,trim=0mm 0mm 8mm 12mm,width=0.99\columnwidth]{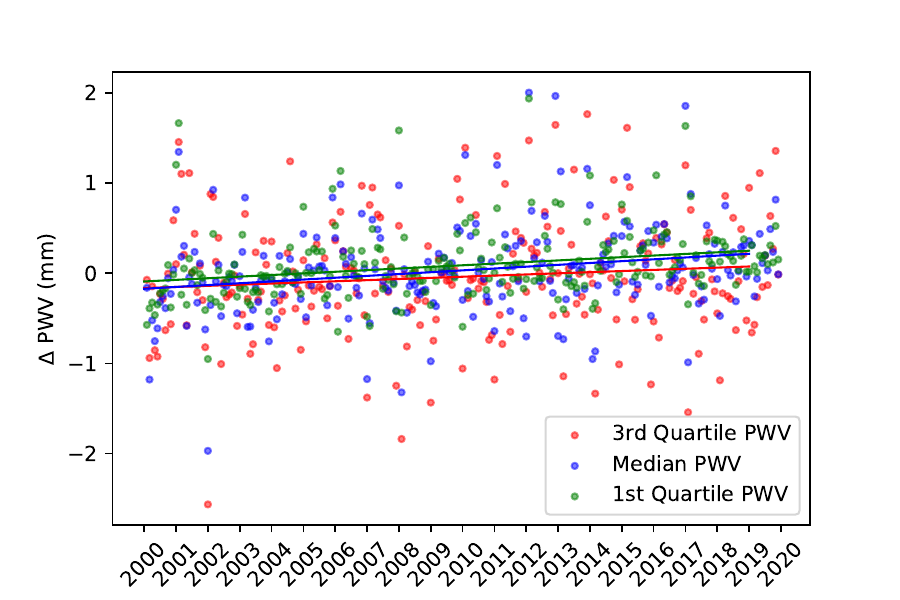}    
    \caption{
    {\bf Upper:} Trend analysis in the deseasonalized monthly time series (blue dots) and the yearly mean time series (red crosses) in PWV using the re-calibrated MODIS data for the $1~\rm km \times 1~km$ cell containing the APEX site. 
    {\bf Middle:} Seasonal trend analysis in seasons at the same location using the re-calibrated MODIS data. As in Fig.~\ref{fig:median_regression_APEX_Rad}, the curves going rom highest to lowest average PWV are summer (red), autumn (black), spring (blue), and winter (green).
    {\bf Lower:} Trend analysis for the first, second, and third quartiles \change{(i.e. the 25th, 50th, and 75th percentiles).}
    The values of the fit slopes and the $R^2$ values for the fits are reported in Table~\ref{tab:fits}.
    }
    \label{fig:median_regression_APEX_MODIS}
\end{figure}

We report here on the variations over the last two decades, focusing on the APEX site and then generalizing to other locations. 
The results for APEX are displayed in Figures \ref{fig:comparison_year}, \ref{fig:median_regression_APEX_Rad} and \ref{fig:median_regression_APEX_MODIS}.  
\change{In general, our results for the seasonality agree with those found by \cite{Cortes2020}, while we arrive a different interpretation for the temporal results, discussed later in this section.}
Figure \ref{fig:comparison_year} shows the seasonal variability at APEX, and it confirms that January and February are generally the worst months to observe, while July and August offer the best conditions.  

In Figure \ref{fig:median_regression_APEX_Rad}, we show our trend analysis in the monthly deseasonalised PWV mean and yearly mean (upper panel) of the APEX radiometer data (2007-2019), where we have used the \texttt{statsmodel} package in Python \citep{seabold2010statsmodels}. In order to deseasonalize the monthly time series data, we subtract the average seasonal cycle from each year of data, e.g. we subtract the average January value from each individual January value, the average February value from each individual February value, etc. The fitting results are reported in Table \ref{tab:fits}, with error bars corresponding to $\approx95\%$ confidence intervals (roughly 2-$\sigma$). The lower panel separates the trend analysis in seasons. There, it is clear again that winter offers the best observations, followed by spring, autumn, and summer (worst conditions). Although all the seasons show \change{increasing PWV over time, only the spring trend has a slope inconsistent with zero at more than 2-$\sigma$.} Note that the APEX radiometer data is the longest observational dataset in this region.    

\begin{table}
  \centering
  \begin{tabular}{lccc}
    \hline\hline
    \noalign{\smallskip}
    Conditions & Data set & Change & $R^2$  \\
     &  & [mm/year] &   \\
    \noalign{\smallskip}
    \hline
    \noalign{\smallskip}
    Deseasonalized & APEX  & $0.04\pm0.03$ & 0.049 \\
    Yearly  & APEX  & $0.03\pm0.03$ & 0.271 \\
    \noalign{\smallskip}
    \hline
    \noalign{\smallskip}
    Summer  & APEX  & $0.01\pm0.10$ & 0.01 \\
    Autumn  & APEX  & $0.03\pm0.05$ & 0.09 \\
    Winter  & APEX  & $0.02\pm0.02$ & 0.21 \\
    Spring  & APEX  & $0.04\pm0.05$ & 0.21 \\
    \noalign{\smallskip}
    \hline
    \noalign{\smallskip}
    Deseas. & MODIS  & $0.02\pm0.01$ & 0.044 \\
    Yearly  & MODIS  & $0.02\pm0.01$ & 0.523 \\    \noalign{\smallskip}
    \hline
    \noalign{\smallskip}
    Summer  & MODIS & $0.01\pm0.03$ & 0.01 \\
    Autumn  & MODIS & $0.01\pm0.02$ & 0.05 \\    
    Winter  & MODIS & $0.01\pm0.01$ & 0.08 \\
    Spring  & MODIS & $0.02\pm0.01$ & 0.44 \\
    \noalign{\smallskip}
    \hline
    \noalign{\smallskip}
    1st Quartile  & MODIS & $0.02\pm0.01$ & 0.074 \\
    2nd Quartile  & MODIS & $0.02\pm0.01$ & 0.055 \\    
    3rd Quartile  & MODIS & $0.01\pm0.01$ & 0.013 \\
    \noalign{\smallskip}
    \hline    \noalign{\medskip}
  \end{tabular}
  \caption{Fit slopes and $R^2$ values for the fit relations shown in Figures \ref{fig:median_regression_APEX_Rad} and \ref{fig:median_regression_APEX_MODIS}.  
  }\label{tab:fits}
\end{table}

In Figure \ref{fig:median_regression_APEX_MODIS}, we show the 20-year trend in the re-calibrated MODIS data for the square kilometer cell containing the APEX site. Both the APEX radiometer and MODIS satellite based trend analyses coincide, and show a slight increase in PWV over the last 20 years. In the lower panel, we show that the trend is similar in the monthly median and in the monthly top quartile (\change{25th percentile}) while it does not change significantly in the worst conditions. The fitting results for the MODIS trend data are also reported in Table \ref{tab:fits}.
We note that since APEX does not provide PWV measurements from before May 2006, the uncertainty in our re-calibration relations could increase for times prior to that.

As noted in e.g. \cite{BOHM2020103192} and \cite{Cortes2020}, variability in PWV over time may depend on phenomena such as ENSO.
Periods of higher Oceanic Niño Index (ONI), which is one way to trace the strength of ENSO, are known to correlate with more rain at both lower and higher latitudes along the western coast of South America \citep{GlantzRamirez2020}.
We do not see a clear correlation between the ONI and the deseasonalized PWV or the 1-year boxcar smoothed time series. The cause of interannual variability likely depends on the season, as confirmed in the middle panel of Figure \ref{fig:median_regression_APEX_MODIS}, which hints at differing levels of variability over time for each season. During austral summer, variability depends on factors such as the altiplanic winter, the period around February when wet air from the Amazon region traverses the Andes and precipitates predominantly in Northern Chile \citep[e.g.]{Gan2005, Canedo-Rosso2019}.  Moreover, variability in the Andes precipitation also depend on variability in the Antarctic Ozone and in the Southern Annular Mode, which is strengthening with climate change \citep{Feron2020}.
In a future study, we will more fully explore the relation between PWV in the Atacama Desert and the South American Monsoon (SAM), ENSO, and variations in the ozone layer.

As the temperature increases with climate change and heat waves become more and more frequent \citep{Feron2019}, the total amount of water vapor in the atmosphere is predicted to increase.  The severity of rains will also increase \citep{Asadieh2015}, but the change in the patterns of water vapor and rain fall will vary with location. 
According to {\it Climate Change 2021: The Physical Science Basis} \citep{IPCC2021},  the mean annual precipitation is predicted to decrease everywhere in the Atacama desert in the 33 multi-model trend. Seasonally, most models agree that the austral summer will become drier over time (i.e. the altiplanic winter will be less intense) but the winter season, which is best for observations, could become wetter in the northern Atacama.  These seasonal predictions are corroborated by \cite{CABRE201635}, who use the Fifth-Generation Penn State/NCAR Mesoscale Model (MM5) specifically to look at precipitation and temperature trends in South America, and also by private communication with the authors of \cite{Feron2020}. It is worth noting that the amount of consecutive dry (precipitation-free) days is predicted to decrease in the northern Atacama.  See the projected Maps in the IPCC WGI Interactive Atlas. \footnote{\url{https://interactive-atlas.ipcc.ch/}.}
\change{In section \ref{sec:results:spatial}, we also perform our own comparisons using more recent climate models.}

\begin{figure}
    \centering
    \includegraphics[clip,trim=0mm 2mm 2mm 2mm,width=0.99\columnwidth]{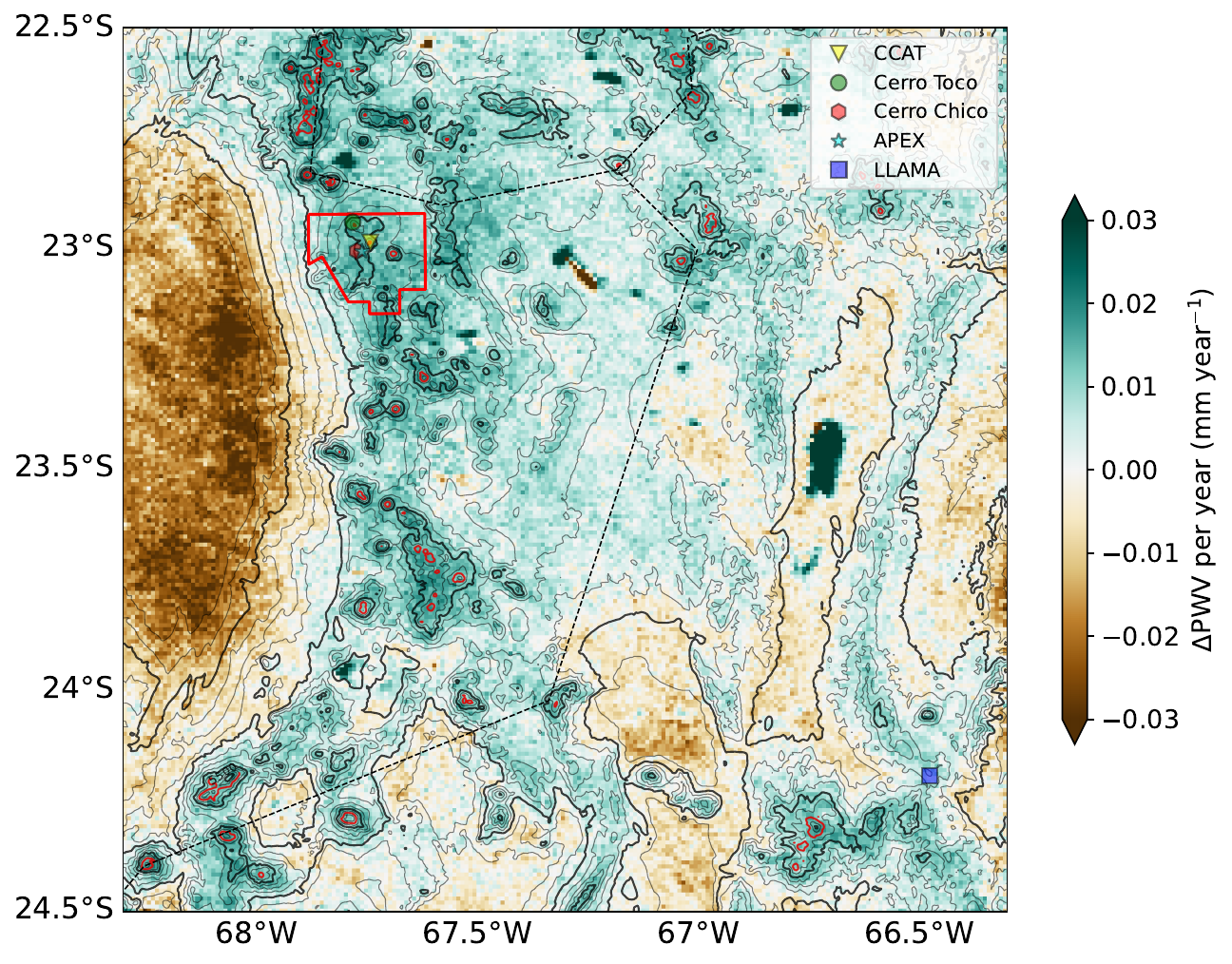}   
    \caption{20-year long term change in the \change{median PWV}, in units of mm year$^{-1}$, calculated from the deseasonalized time series.
    The borders, AAP+ALMA boundary, markers, and elevation contours are the same as in Figure \ref{fig:topo}.
    }
    \label{fig:trend_area}
\end{figure}

\subsection{Comparison to other sites}\label{sec:results:spatial}

\begin{figure}
    \centering
    \includegraphics[clip,trim=6mm 5mm 15mm 12mm,width=0.99\columnwidth]{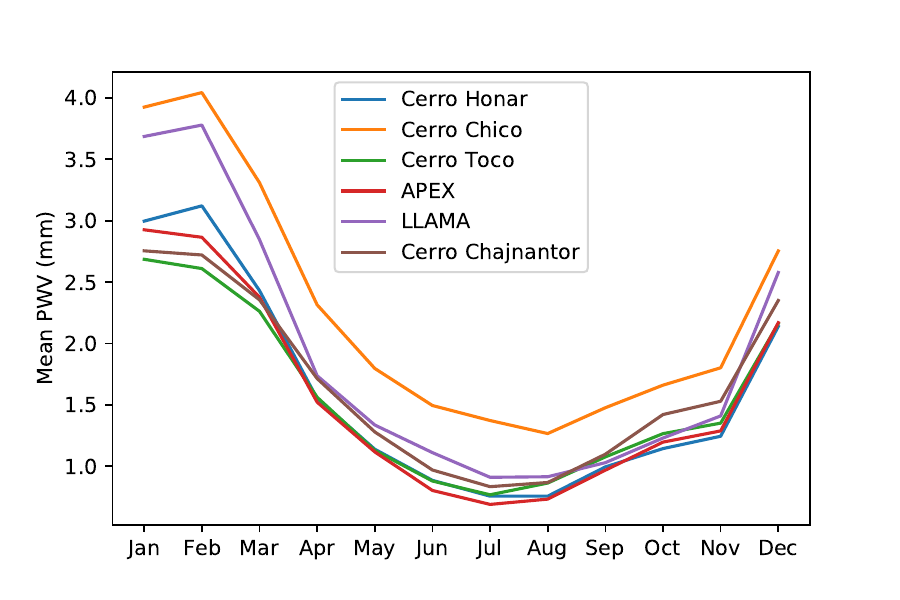}
    \caption{Comparison of seasonality across several sites selected in and around the AAP.  In general, it appears that Cerros Toco and Honar have similarly low PWV as those available at Cerro Chajnantor, while Cerro Chico has a higher mean PWV than the others.  \change{We note that each of these curves is for the $1\times1~\rm km^2$ cell that contains the site of interest, and may not be well centered on the site.  Each cell may also contain a mix of data for a range of elevations, as the topography changes rapidly throughout the region.}
    }
    \label{fig:sites_comparison}
\end{figure}

One of the primary motivations of this study is to inform site selection for future submillimeter observatories such as the Atacama Large Aperture Submillimeter Telescope Project \citep[AtLAST;][]{Klaassen2019, Klaassen2020}\footnote{\url{https://atlast-telescope.org/}}.

Figure \ref{fig:sites_comparison} shows the average seasonal cycle for several existing and planned facilities that are sited in locations featuring low PWV, including the entire AAP region and ALMA concession. Overall, it appears that the site near Cerro Toco as well as Cerro Chico and Cerro Honar enjoy similarly low PWV values as those available around Cerro Chajnantor. 

Given the trend seen for the APEX site in Figures  \ref{fig:median_regression_APEX_MODIS} and \ref{fig:median_regression_APEX_Rad}, we are interested in whether and how the conditions for the region have shifted over the last 20 years. The rate of \change{change in the median} PWV, in units of mm year$^{-1}$, is shown in Figure \ref{fig:trend_area}.\footnote{With Figure \ref{fig:trend_area} in mind, the same trend is apparent in the yearly median data in Figures \ref{fig:pwv_median_yearly_2000-2011} and \ref{fig:pwv_median_yearly_2012-2020} in the appendix.}   
Generally speaking, there has been a slight increase in the PWV recorded values for the entire region, following the same overall trend as that the APEX location. 
This represents a maximum increase in the \change{median} PWV of about $+0.5$~mm over the past 20 years, but the trend, which is common across all $220\times220$ pixels, is statistically more significant than that for the APEX site alone.  We note that, had we excluded the time variability for the MODIS {\it Aqua} data, the trend inferred without re-calibration would have been systematically stronger.  \change{Further, the trend for regions with a PWV higher than 3~mm, we our recalibration is an extrapolation, may not be robust and we urge the reader to treat this with caution; such higher PWV regions are generally lower than 4000~m a.s.l. and not under consideration for mm/submm astronomy.}

Our findings that the mean and median precipitable water vapor has increased over the past two decades (Figures \ref{fig:median_regression_APEX_Rad} and \ref{fig:median_regression_APEX_MODIS}) appear to be in contrast to those of \cite{Cantalloube2020}, who report that the number of days with low water vapor has increased in recent years for the region surrounding Paranal.  
As noted in \cite{Cantalloube2020}, however, the \change{long term portions of their study, for both ERA and CMIP6, are generally limited to the low spatial resolution of $\gtrsim$31~km,} and their radiometer-based measurements only cover the last 5 years, which is too short to draw strong statistical conclusions.
Most importantly, the trends in precipitation might in fact be opposite with the region surrounding Paranal becoming drier over time, especially during the winter season, while the northern Atacama could become wetter.
Ultimately, our findings, while indicating the trend is significant at $\approx 95\%$ confidence (i.e. $\approx $2-$\sigma$), will require more robust confirmation.

\begin{figure}
    \centering
    \includegraphics[clip,trim=0mm 0mm 0mm 0mm,width=0.99\columnwidth]{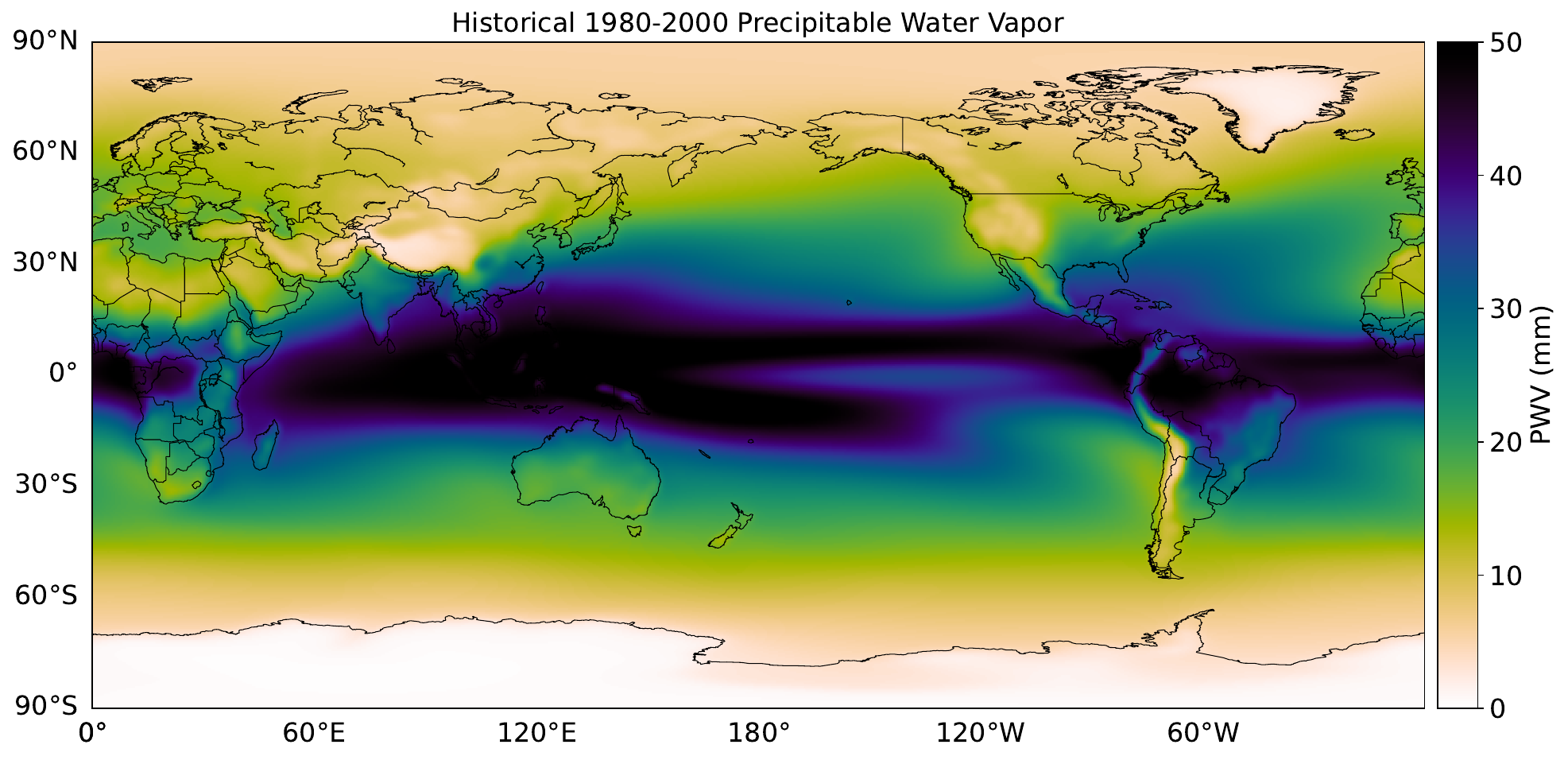}
    \includegraphics[clip,trim=0mm 0mm 0mm 0mm,width=0.99\columnwidth]{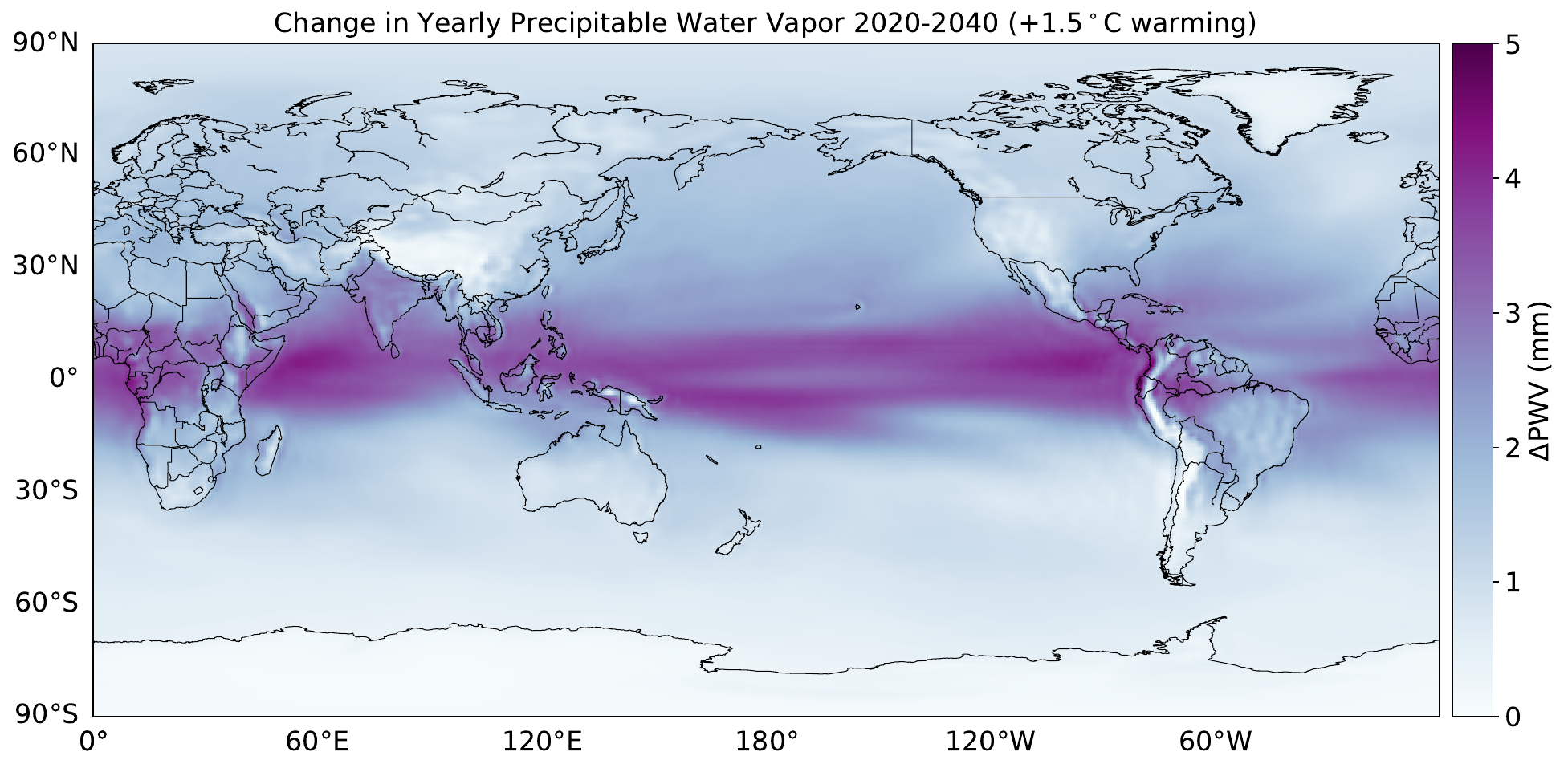}
    \includegraphics[clip,trim=0mm 0mm 0mm 0mm,width=0.99\columnwidth]{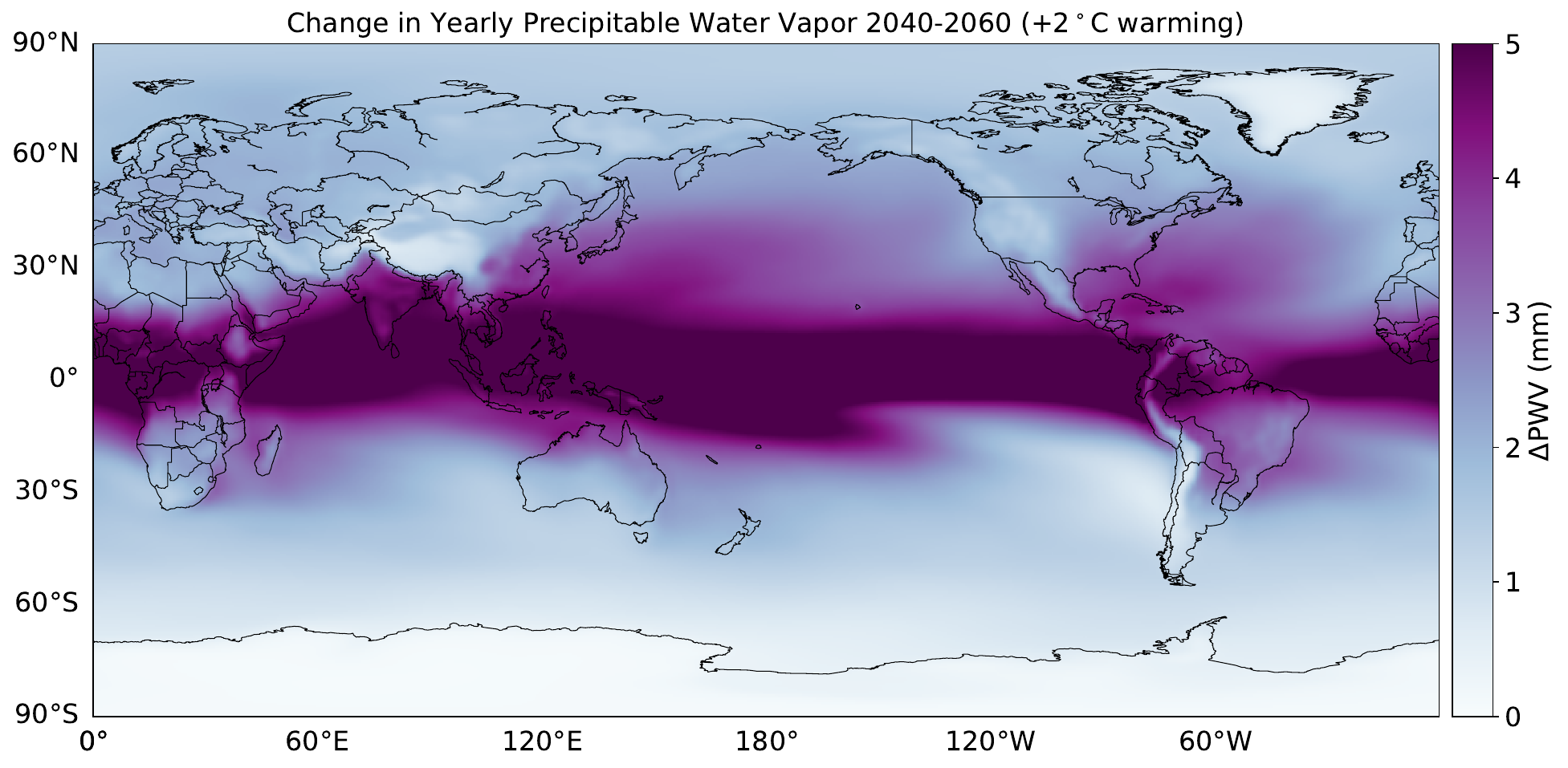}
    \includegraphics[clip,trim=0mm 0mm 0mm 0mm,width=0.99\columnwidth]{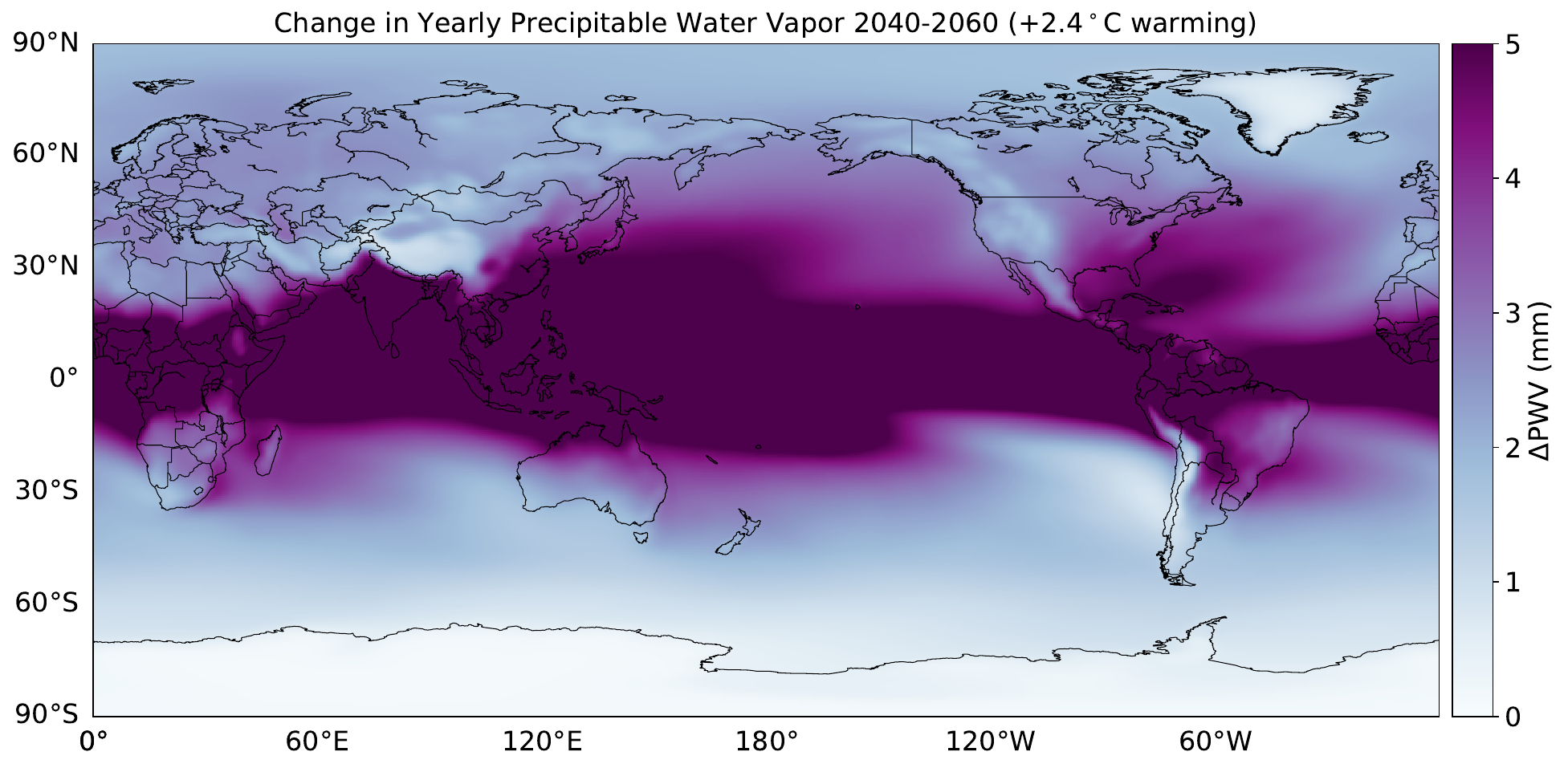}
    \caption{
    {\bf First panel:} Map showing the historical yearly average PWV for 1980-2000 using the values from the CMIP6 dataset \citep{CMIP6}. 
    {\bf Second panel:} Best-cast scenario showing the 2020-2040 CMIP6 SSP3-7.0 prediction for the PWV minus the historical average for 1980-2000, assuming warming with respect to pre-industrial values of $+1.5^\circ$C.
    {\bf Third panel:} Median case showing the 2040-2060 CMIP6 SSP2-4.5 prediction for the PWV minus the historical average for 1980-2000, assuming warming with respect to pre-industrial values of $+2.0^\circ$C.
    {\bf Fourth panel:} Worst-case scenario showing the 2040-2060 CMIP6 SSP5-8.5 prediction for the PWV minus the historical average for 1980-2000, assuming warming with respect to pre-industrial values of $+2.4^\circ$C.
    }\label{fig:cmip6_pwv}
\end{figure}

The observed increase in PWV over time \change{for the region} may be attributed to either inter-decadal variability or climate change, and a longer baseline of time will be required before we understand the nature of the time dependence.
\change{To examine this trend further, we performed our own analysis of the CMIP6 data,\footnote{The data can be obtained at \url{https://esgf-node.llnl.gov/search/cmip6/}.} showing the results in Figure~\ref{fig:cmip6_pwv}.  
For this we compare three Shared Socioeconomic Pathways (SSPs) corresponding to warming with respect to pre-industrial values of $+1.5^\circ$C by 2020-2040 (optimistic case, SSP3-7.0), $+2.0^\circ$C by 2040-2060 (median, SSP2-4.5), and $+2.4^\circ$C by 2040-2060 (worst case, SSP5-8.5).
Note that as of 2020, the value for the temperature increase was $+1.2^\circ$C, and according to the Intergovernmental Panel on Climate Change (IPCC) Sixth Assessment Report \cite{IPCC2021} will likely reach +1.5~C$^\circ$ by the end of this decade.\footnote{The IPCC report is available at \url{https://www.ipcc.ch/report/sixth-assessment-report-working-group-i/}.} 
Also noted by the IPCC Sixth Assessment Report, the worst-case scenario we adopt here assumes that CO$_2$ emissions will continue to increase at the same rate it has for the past two decades.

The maps in Figure~\ref{fig:cmip6_pwv} illustrate the difference between the predicted PWV yearly mean values and the historical average for 1980-2000.  The predicted global trend is for the PWV to increase everywhere, which can be understood simply as warmer air having the ability to hold more moisture.  
We also note that both the Atacama desert and Antarctica are predicted to see the lowest increases in PWV, though even a small increase is non-negligible for submillimeter astronomy. 
We emphasize that, at the resolution of CMIP6, in no case does the $\Delta$PWV decrease significantly for any region of interest.  

Importantly, while the CMIP6 results are low spatial resolution, we identify a roughly $\Delta$PWV = +0.3~mm increase in the average PWV for the Atacama when examining the CMIP6 data for the 20 years under consideration, and is consistent with the $\Delta$PWV trend found in the MODIS data.
This is found by comparing the historical data in CMIP6 covering the period 2001-2014 and the model forecasts for 2015-2020 with the historical data from 1980-2000.  
Further, while we do not show it here, the global increase in precipitable water vapor does not imply an increase in precipitation, which is expected to decrease for much of the world, following changes in the global water cycle and atmospheric cells.
}

\section{Discussion and Conclusions}

In this work, we report on relatively high ($1/110^\circ$, $\approx 1$~km) resolution satellite studies of the precipitable water vapor above the Atacama Astronomy Park and surrounding regions for a period spanning nearly two decades. \change{In comparison, many previous studies have relied on much lower spatial resolution data, such as ERA5, which yields $0.25^\circ$ resolution for atmospheric data (e.g. many of the results in \cite{Cantalloube2020}), or on ground-based measurements made from specific sites stationed in the field \citep[e.g.][]{Otarola2019,Cantalloube2020}, and may not be as applicable for the entire region.} 
As noted above, the work presented relies on data taken in the near infrared that can only probe the daytime PWV values.  \change{Given the typical diurnal variations in PWV, our results should generally be regarded as an upper limit on the nighttime PWV values.}

From the maps in Figures \ref{fig:pwv_median_mean} \& \ref{fig:pwv_median_mean} as well as in Appendix \ref{sec:appendix:pwv}, one can identify several promising locations in terms of having exceptionally low PWV.  We note this is a single parameter, of course, and does not account for active or dormant volcanoes, the accessibility of the site (e.g. existing roads, power, data transport), or wind conditions. 

A full in-situ characterization will be required before a site for AtLAST (or any other future facility) can be selected.
However, our results confirm that out of the prospective submillimeter astronomical sites considered, those exhibiting the driest conditions are generally located at the some of highest accessible altitudes.  These include Cerro Chajnantor (CCAT site, 5640~m a.s.l), Cerro Toco (5604~m a.s.l.), Cerros de Honar (a broad series of peaks $\sim$5400~m a.s.l.), and the LLAMA site (4800~m a.s.l.), and are shown in Figure \ref{fig:sites_comparison}.  We also note that several exceptionally dry regions at elevations $< 5500$~ m a.s.l. and not currently used for astronomy can be identified in Figure \ref{fig:pwv_top5cdf}.

\change{
Using the re-calibrated satellite data, we identify in this work a worrying but initially low (2-$\sigma$) significance trend for the region of interest.  In the context of the recently released IPCC Sixth Assessment report as well as the CMIP6 climate model predictions, we demonstrate that the predictions support this finding, showing it is both globally present and that it can be understood readily in terms of climate change.  Specifically, as the temperature increases, so will the capacity for the atmosphere to hold water.
For context, \cite{WANG20181} noted a related trend, that as the sea surface temperature increases, so will the PWV in the atmosphere. 
We therefore conclude that, as astronomers and as humans, we have a strong self-interest in minimizing the impacts of climate change.
}

Finally, we hypothesize that our re-calibration approach may be extensible to any site worldwide that exhibits low values of PWV \change{($\lesssim 3$~mm)} on average.  This condition typically describes any site of interest to the ground-based millimeter, submillimeter, and optical astronomical communities.
In addition to potentially saving substantial time, effort, and costs in the field by allowing one to pre-select prospective locations for further in-situ characterization, our approach may allow for further studies of the variability to establish seasonal variations and trends associated with climate change, particularly in areas where water and precipitation are scarce.  Future studies will address climate and interannual variability, will expand to other locations, and will include yet higher spatial resolution through interferometric satellite measurements \citep[e.g.][]{Mateus2017}.

\section*{Acknowledgments}
We thank the referee for the thorough and constructive comments that greatly improved the context, content, and robustness of the work presented.

This paper makes use of data from the NASA satellites MODIS {\it Terra} and {\it Aqua}, SRTM15+ data from the GEBCO Compilation Group (2020) GEBCO 2020 Grid, and ground-based measurements from APEX.
PGT was supported by the European Southern Observatory's (ESO) 2020 Summer Research Programme\footnote{\url{http://eso.org/sci/meetings/2020/summerresearch.html}}, which in turn is supported by ESO's Directorate for Science.
TM and CdB are supported in part by the European Union’s Horizon 2020 research and innovation programme under grant agreement No 951815.

We thank the developers of many Python libraries, including \texttt{Matplotlib} (including the \texttt{Basemap} package), \texttt{numpy}, and \texttt{statsmodels}. 

\bibliographystyle{aa}
\bibliography{bibliography}
\appendix

\section{Seasonal, monthly, and year variability in the precipitable water vapor}\label{sec:appendix:pwv}

\begin{figure*}
    \centering
    \includegraphics[clip,trim=0mm 2mm 20mm 0mm,width=0.495\textwidth]{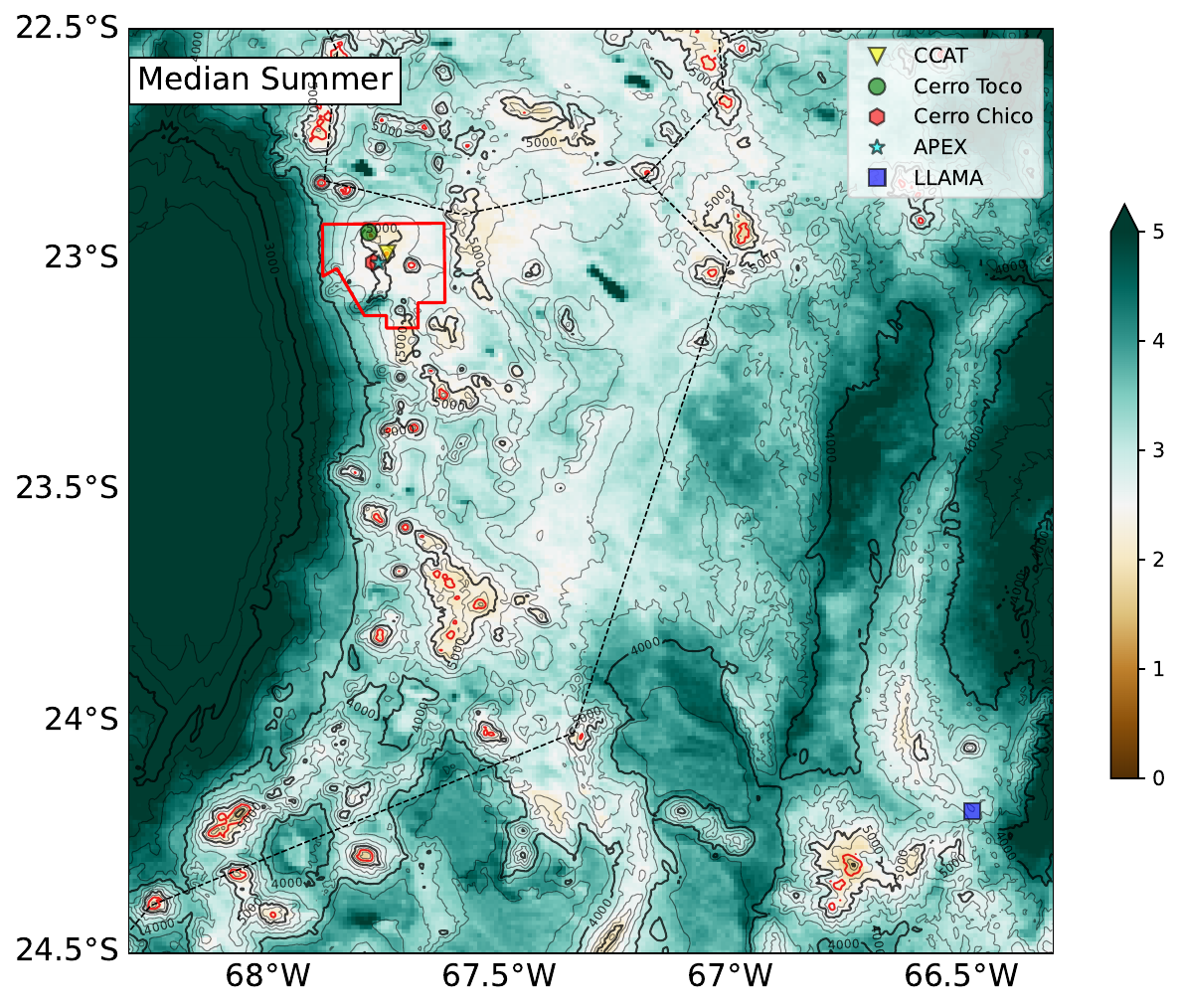}
    \includegraphics[clip,trim=20mm 2mm 0mm 0mm,width=0.495\textwidth]{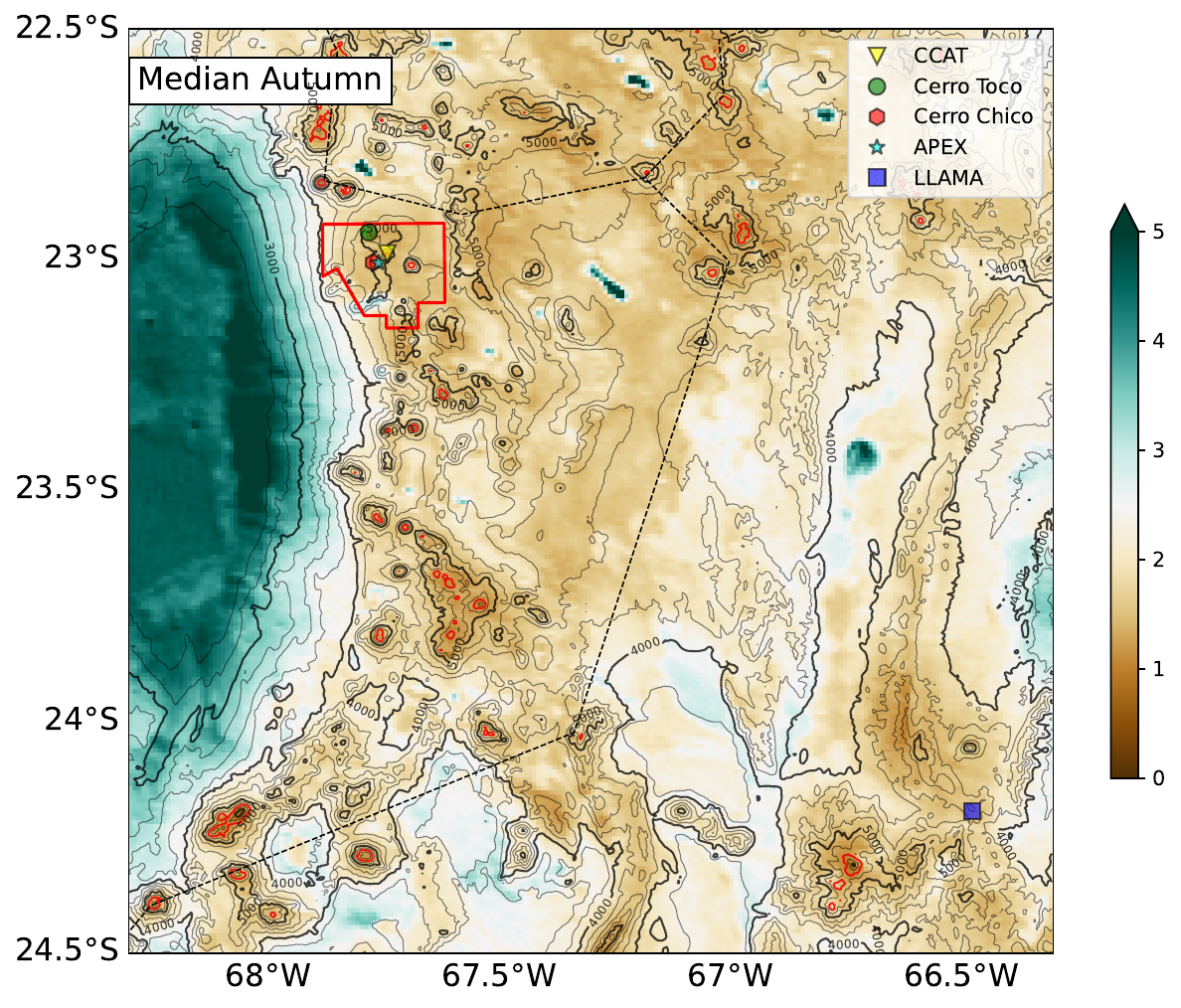}
    \includegraphics[clip,trim=0mm 2mm 20mm 0mm,width=0.495\textwidth]{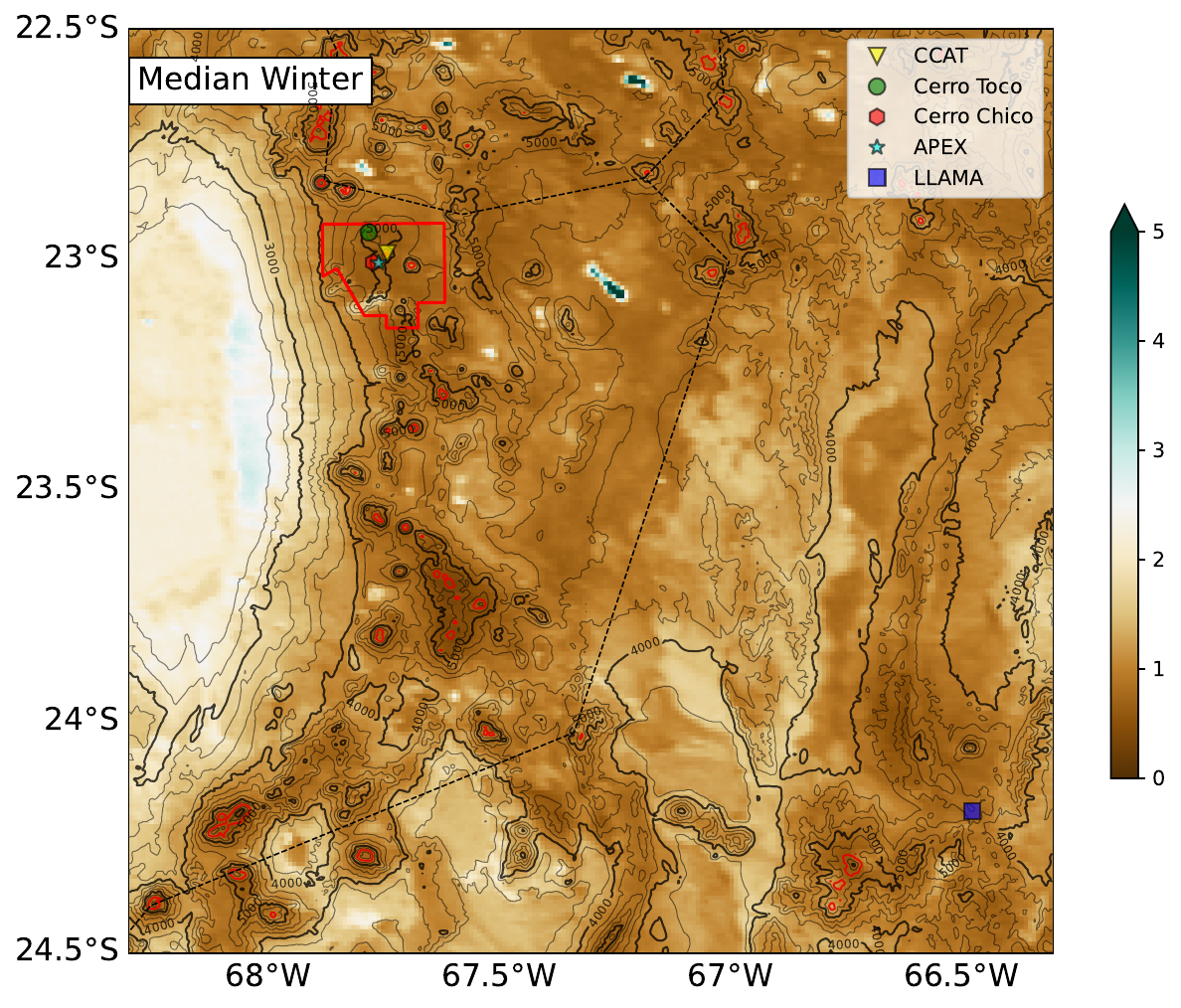}
    \includegraphics[clip,trim=20mm 2mm 0mm 0mm,width=0.495\textwidth]{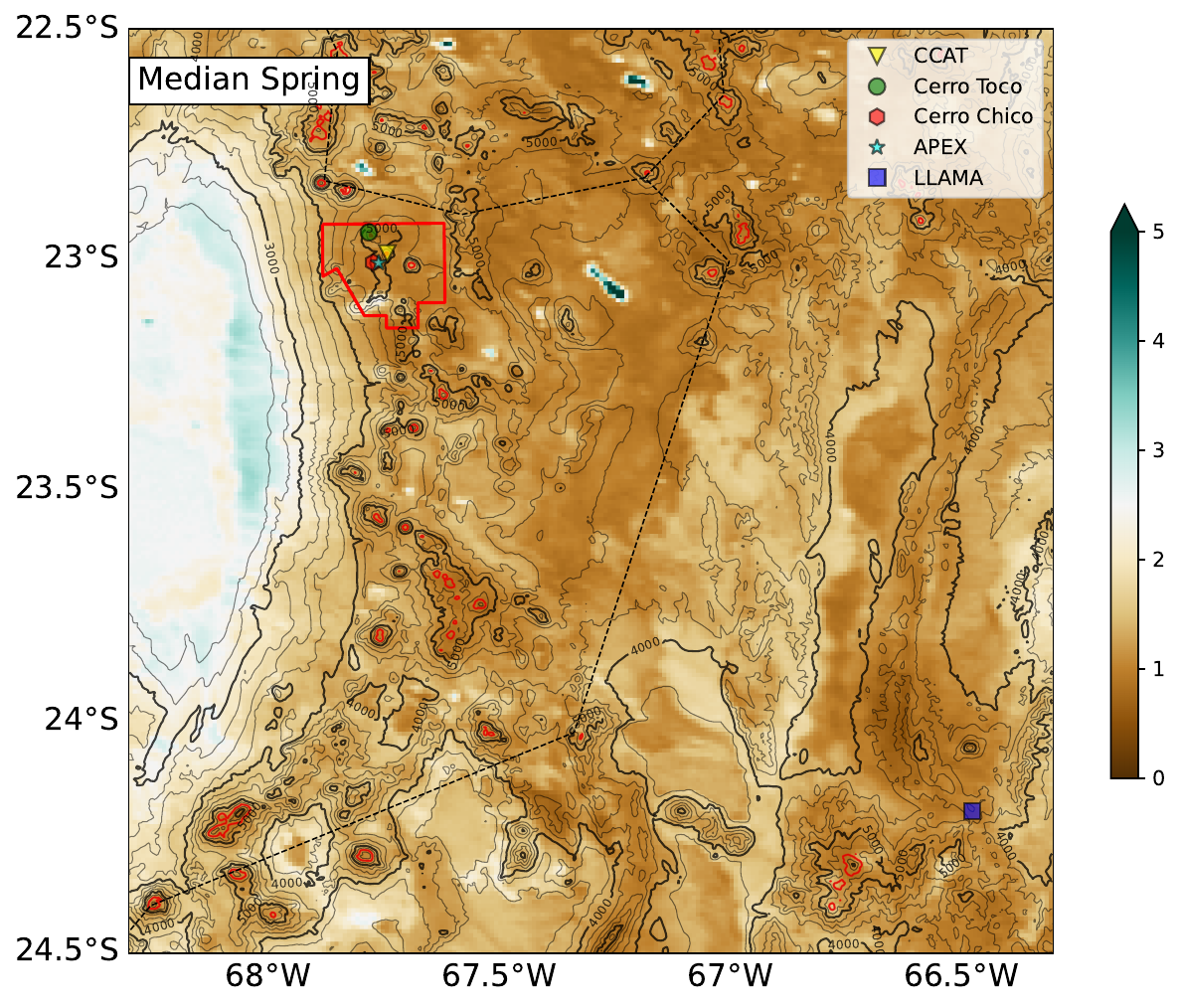}
    \caption{Seasonal median PWV values. The panels show summer, autumn, winter, and spring, where we have defined the seasons for consistency with \cite{paine_scott_2017_438726} as the three integral months beginning with the one in which the season commences (e.g. `Austral summer' is treated as December-January-February, `autumn' is March-April-May, `winter' is June-July-August, and `spring' is September-October-November). 
    The borders, AAP+ALMA boundary, site markers, and elevation contours are the same as in Figure \ref{fig:pwv_median_mean}. 
    }\label{fig:pwv_seasonal_median}
\end{figure*}

We show the monthly maps of the median PWV values in Figure \ref{fig:pwv_median_monthly}.  

{
\begin{figure*}
    \centering
    \includegraphics[clip,trim=0mm 2mm 21mm 2.5mm,height=0.31\textwidth]{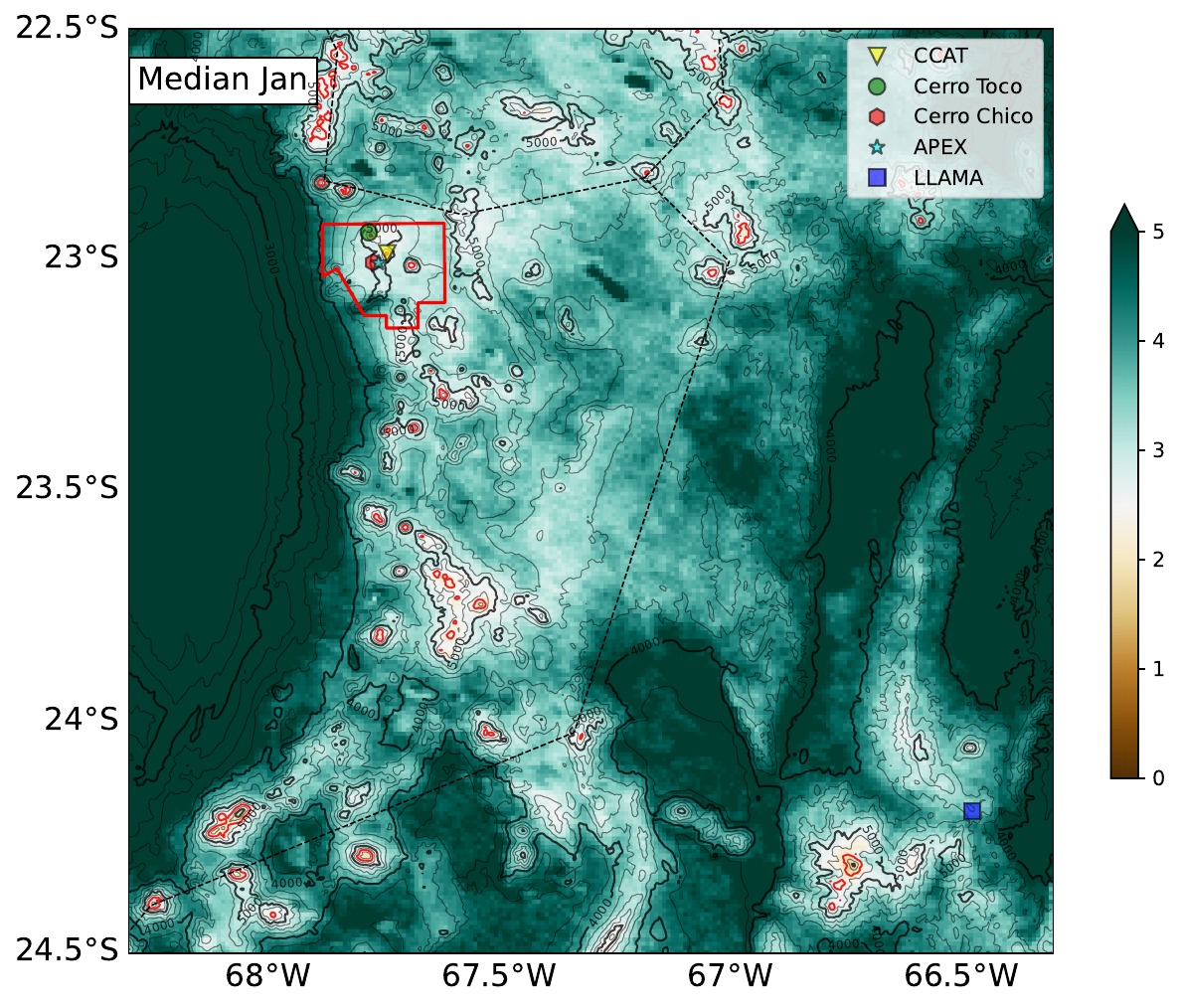}
    \includegraphics[clip,trim=20.4mm 2mm 21mm 2.5mm,height=0.31\textwidth]{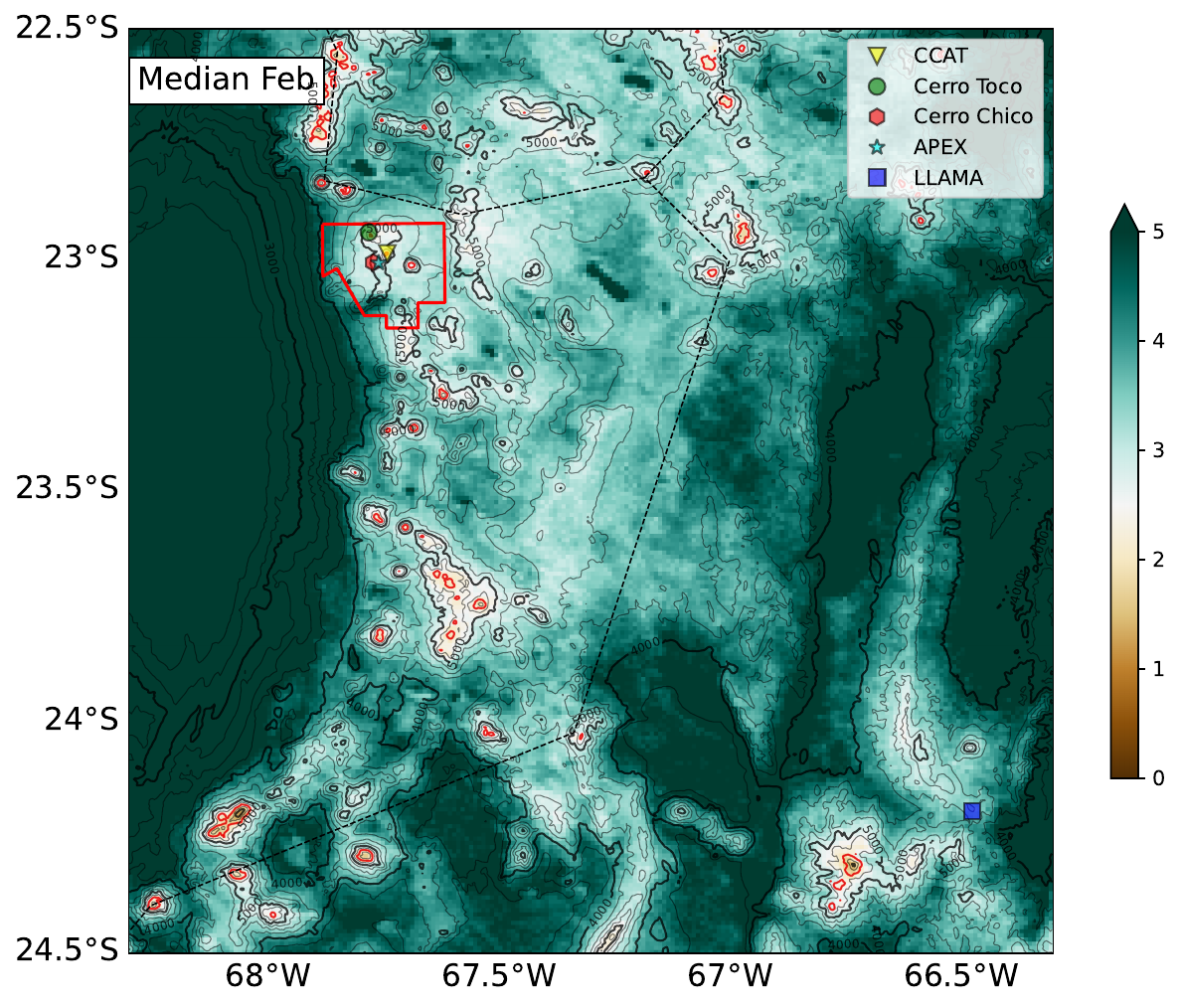}
    \includegraphics[clip,trim=20.4mm 2mm 0mm 2.5mm,height=0.31\textwidth]{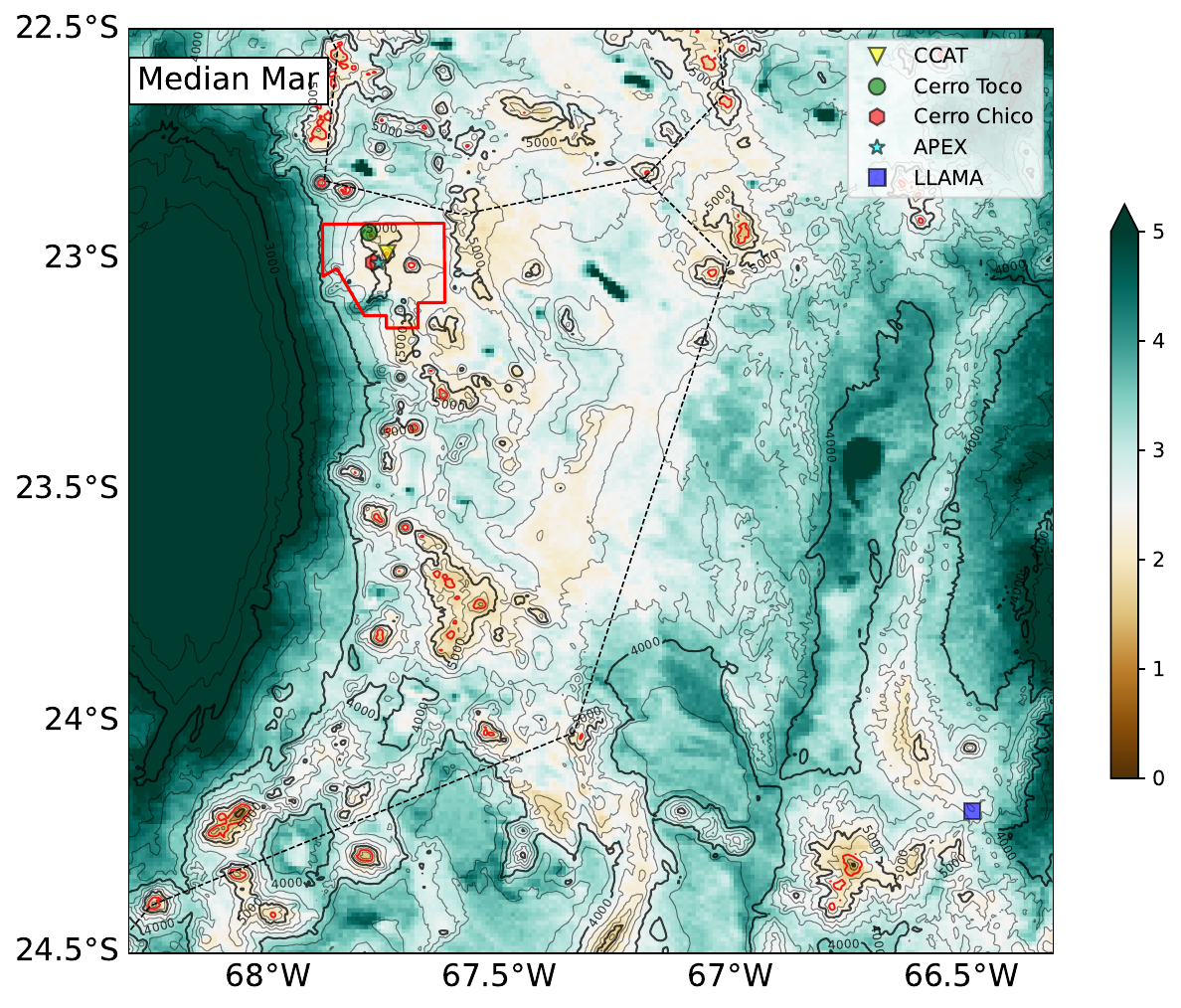}\\
    \includegraphics[clip,trim=0mm 2mm 21mm 2.5mm,height=0.31\textwidth]{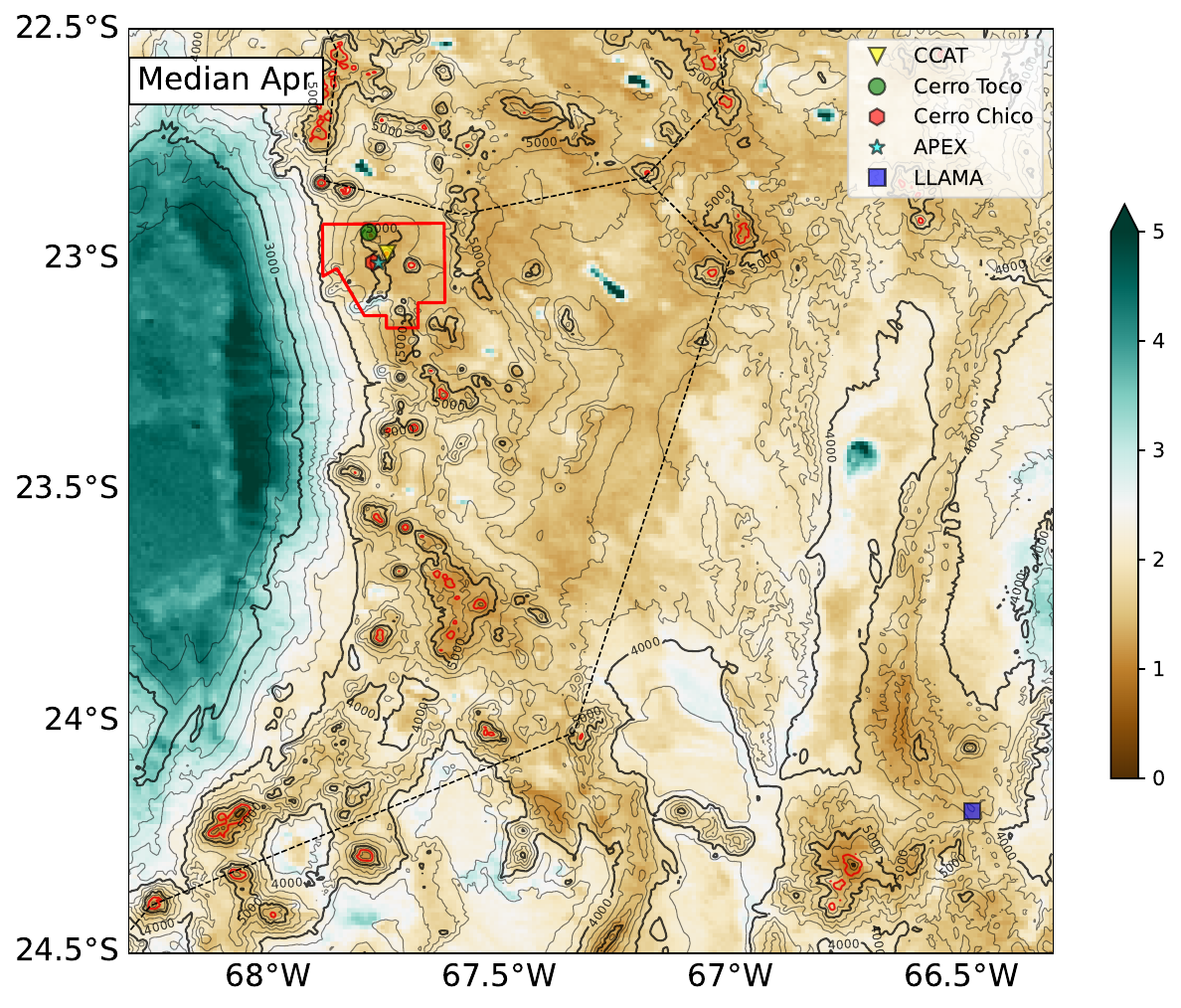}
    \includegraphics[clip,trim=20.4mm 2mm 21mm 2.5mm,height=0.31\textwidth]{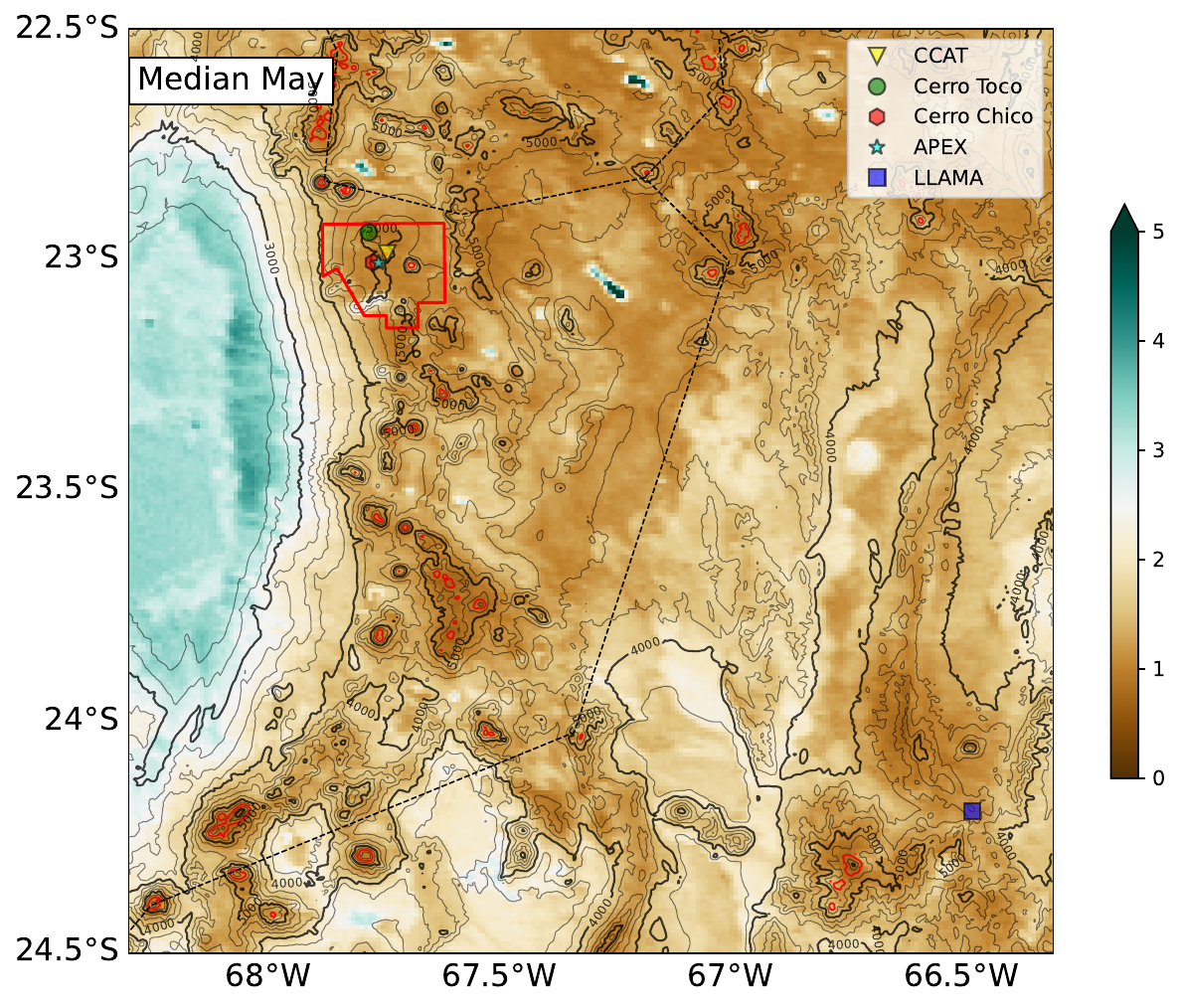}
    \includegraphics[clip,trim=20.4mm 2mm 0mm 2.5mm,height=0.31\textwidth]{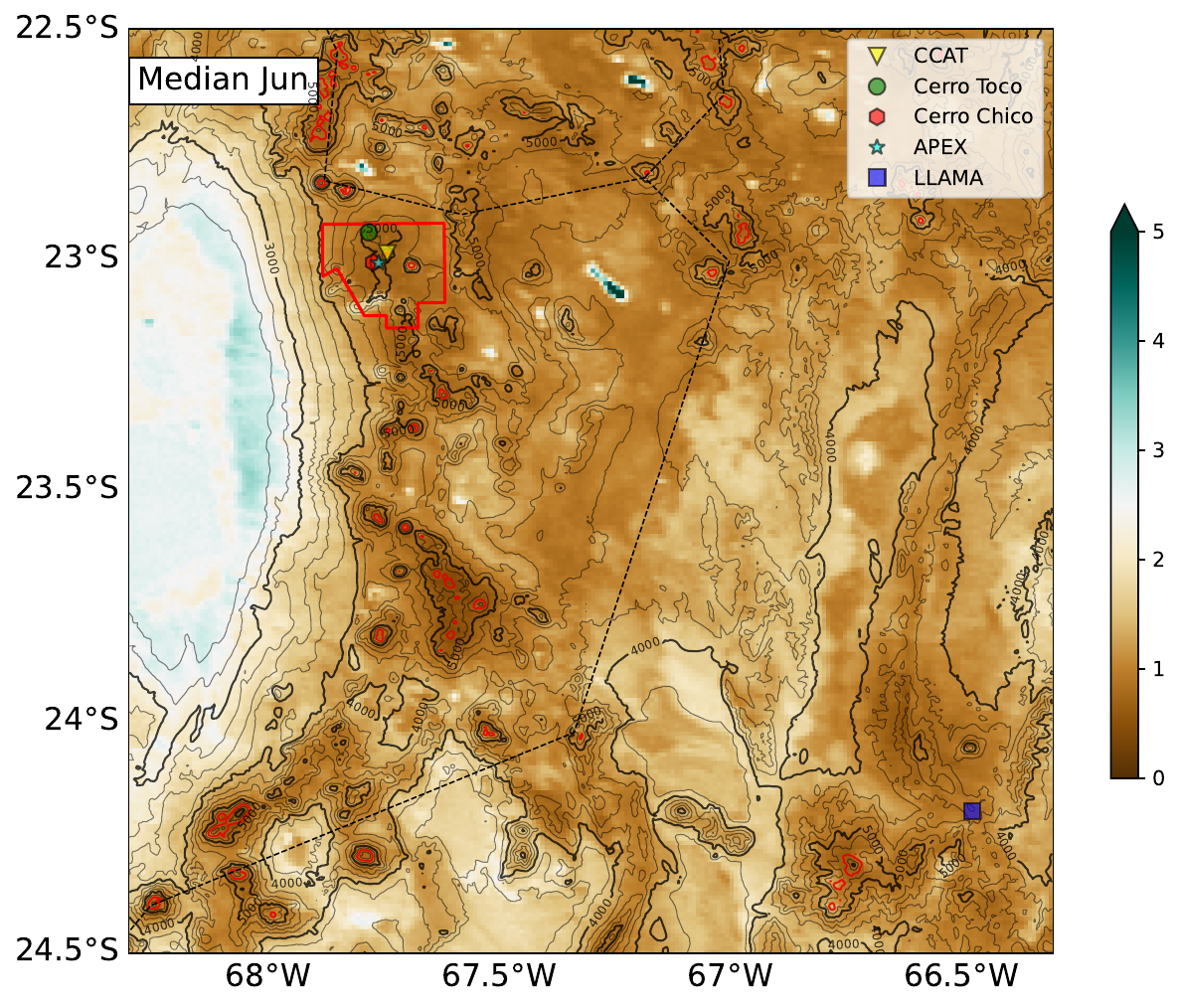}\\
    \includegraphics[clip,trim=0mm 2mm 21mm 2.5mm,height=0.31\textwidth]{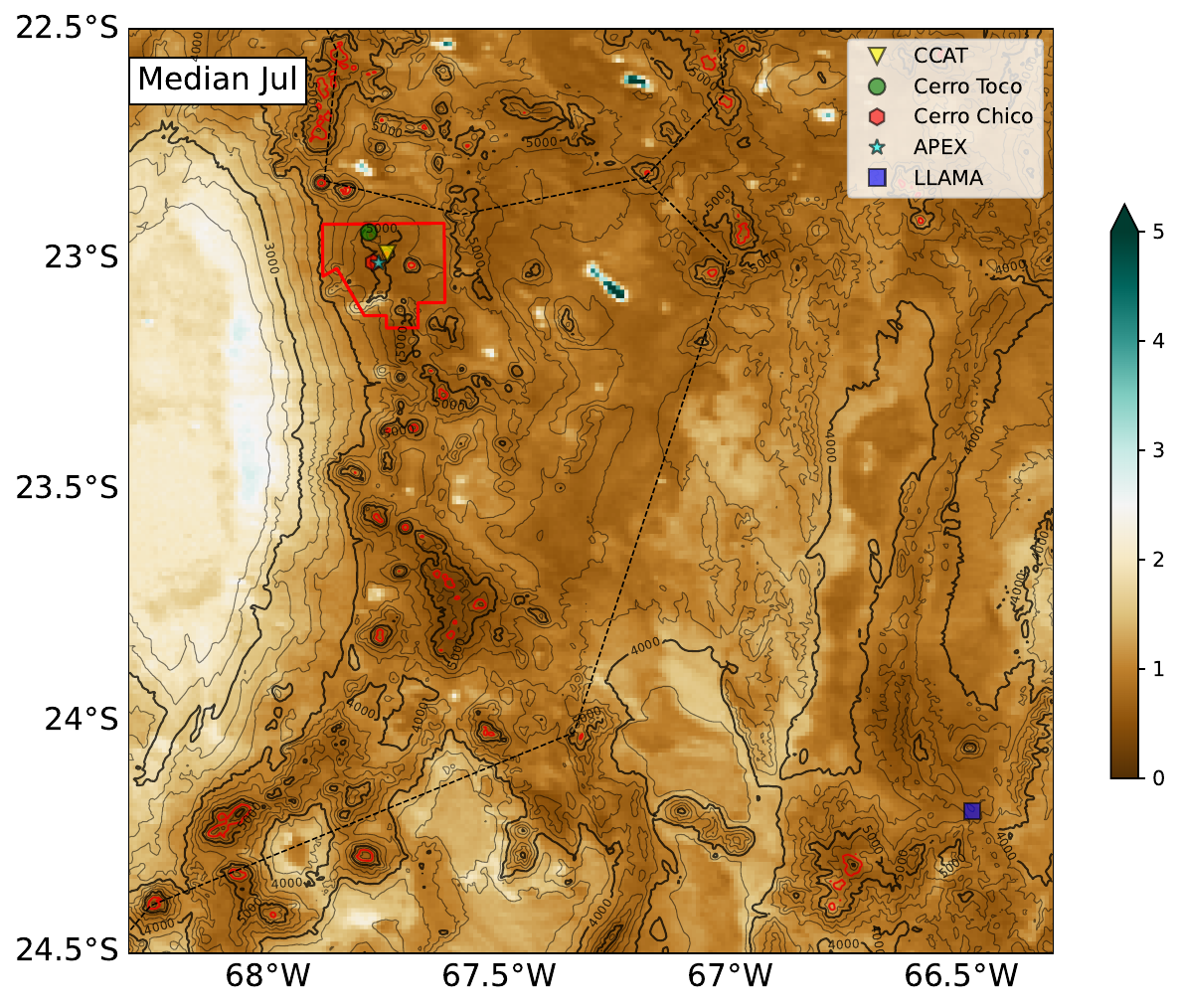}
    \includegraphics[clip,trim=20.4mm 2mm 21mm 2.5mm,height=0.31\textwidth]{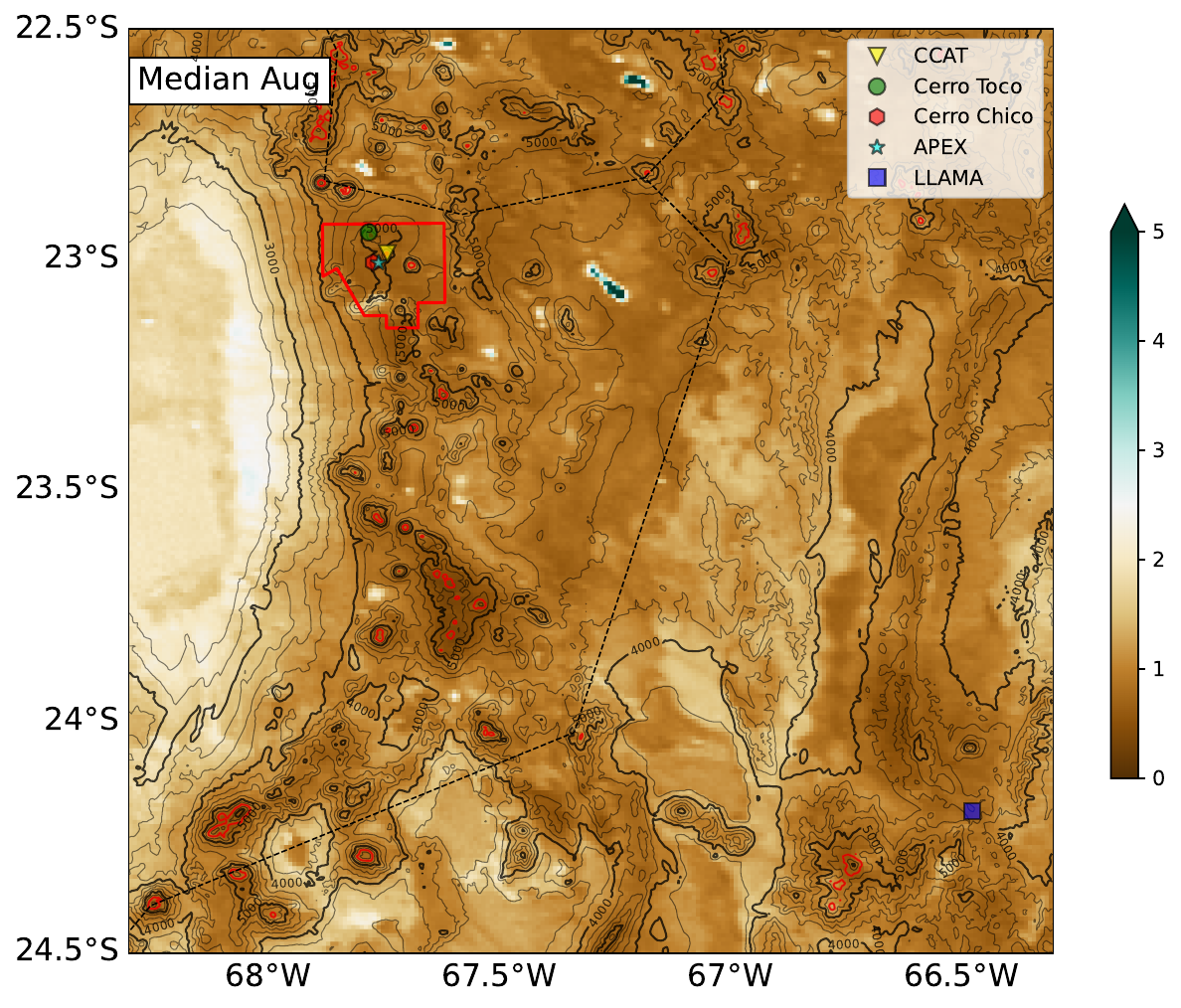}
    \includegraphics[clip,trim=20.4mm 2mm 0mm 2.5mm,height=0.31\textwidth]{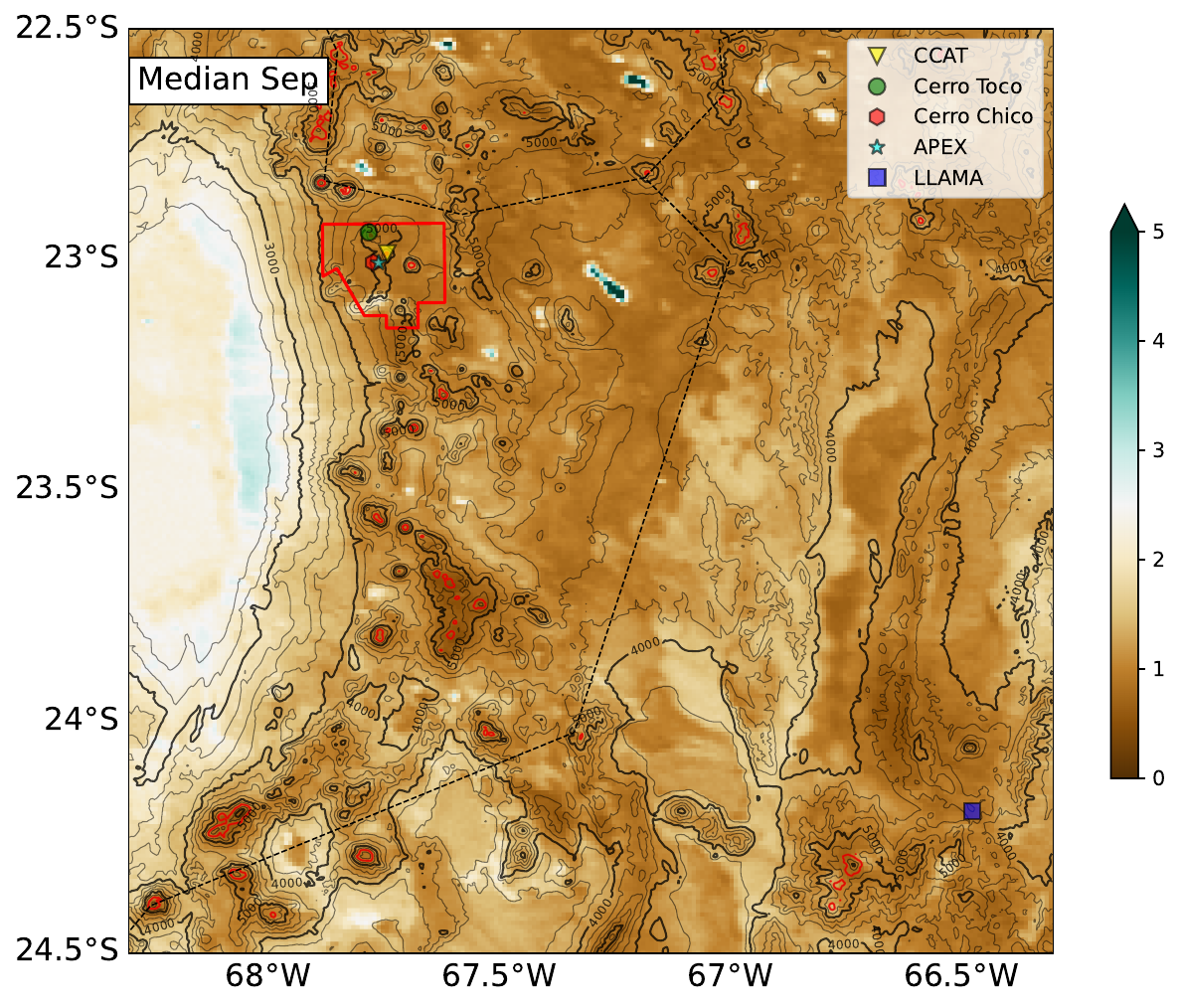}\\
    \includegraphics[clip,trim=0mm 2mm 21mm 2.5mm,height=0.31\textwidth]{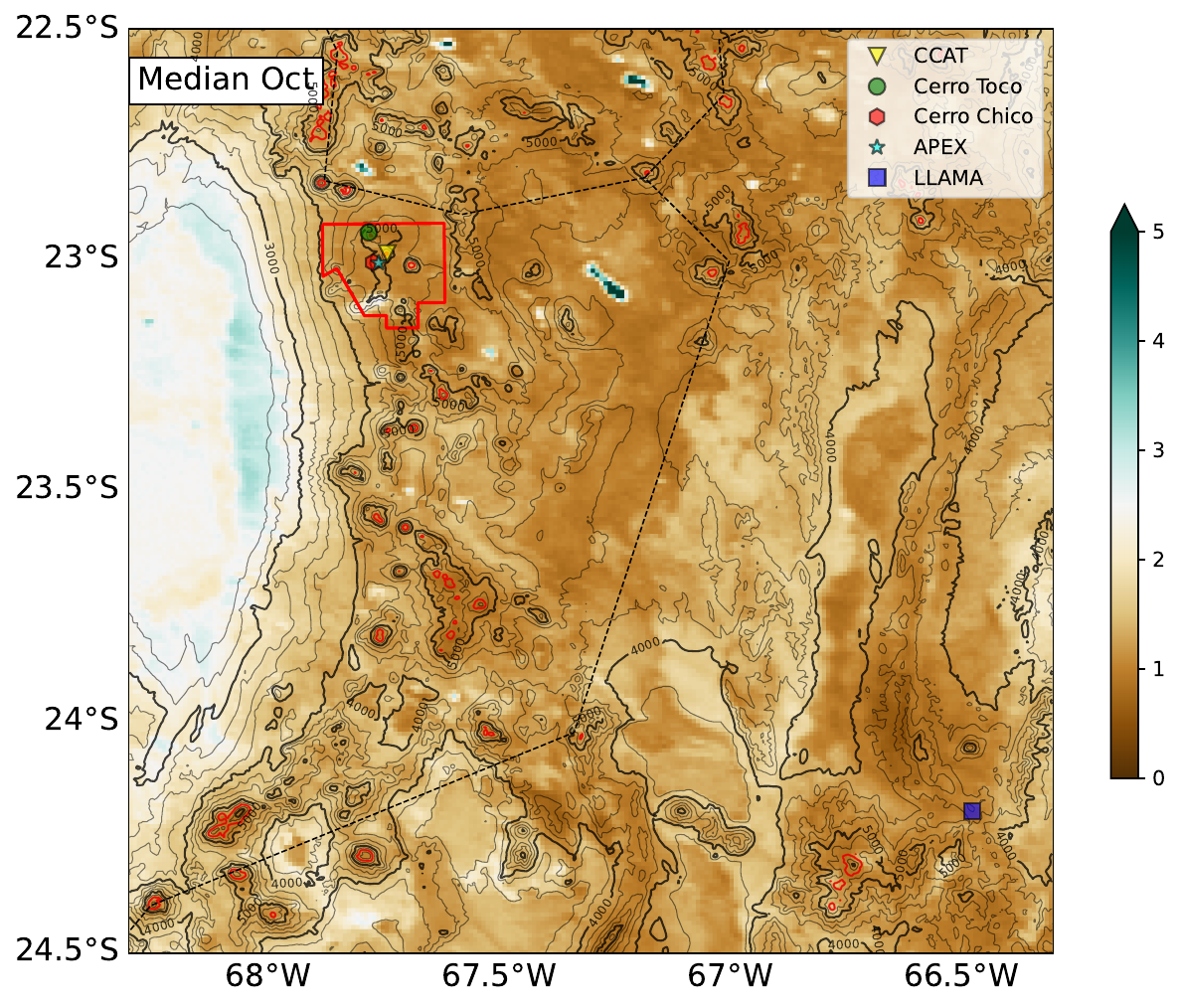}
    \includegraphics[clip,trim=20.4mm 2mm 21mm 2.5mm,height=0.31\textwidth]{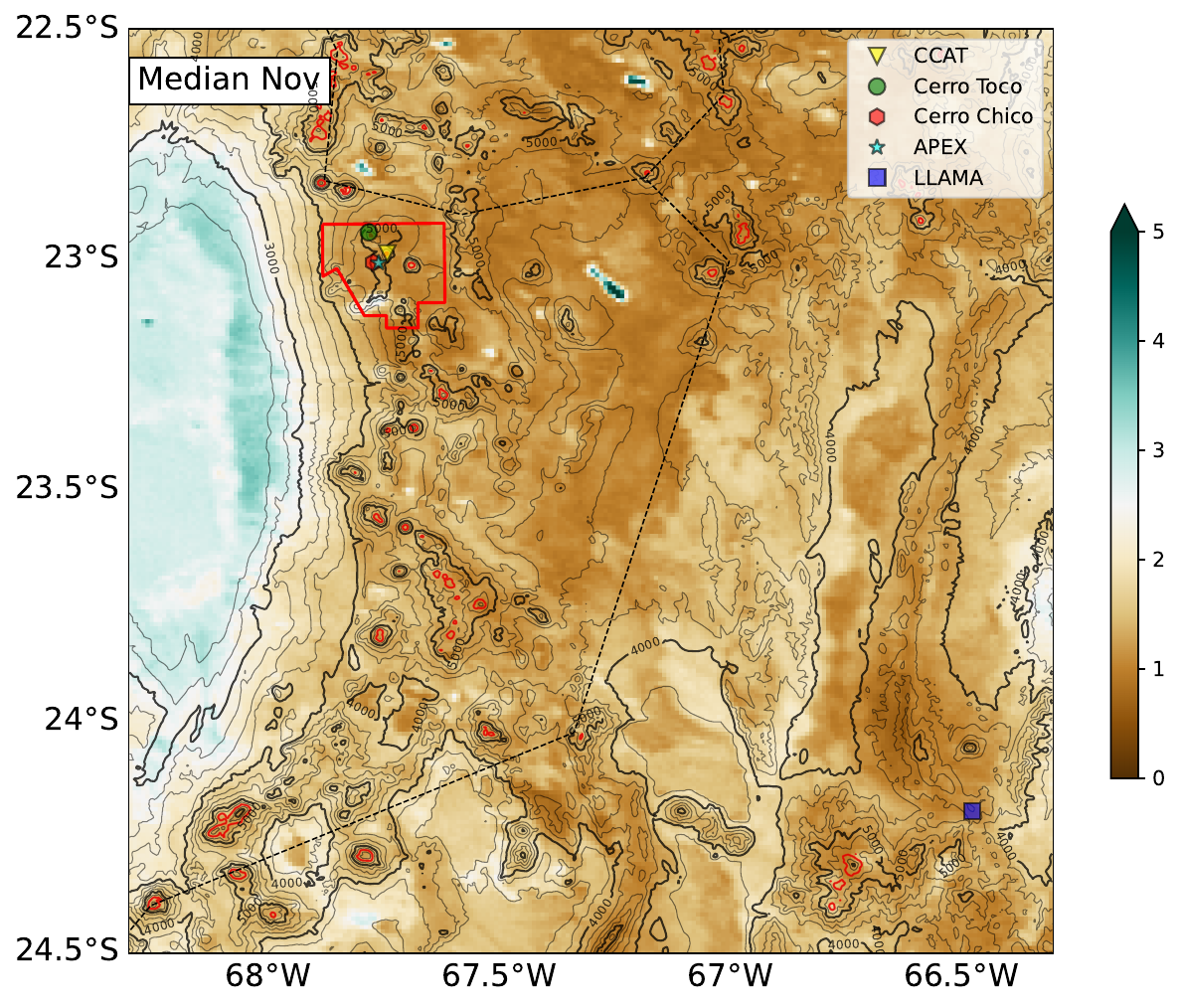}
    \includegraphics[clip,trim=20.4mm 2mm 0mm 2.5mm,height=0.31\textwidth]{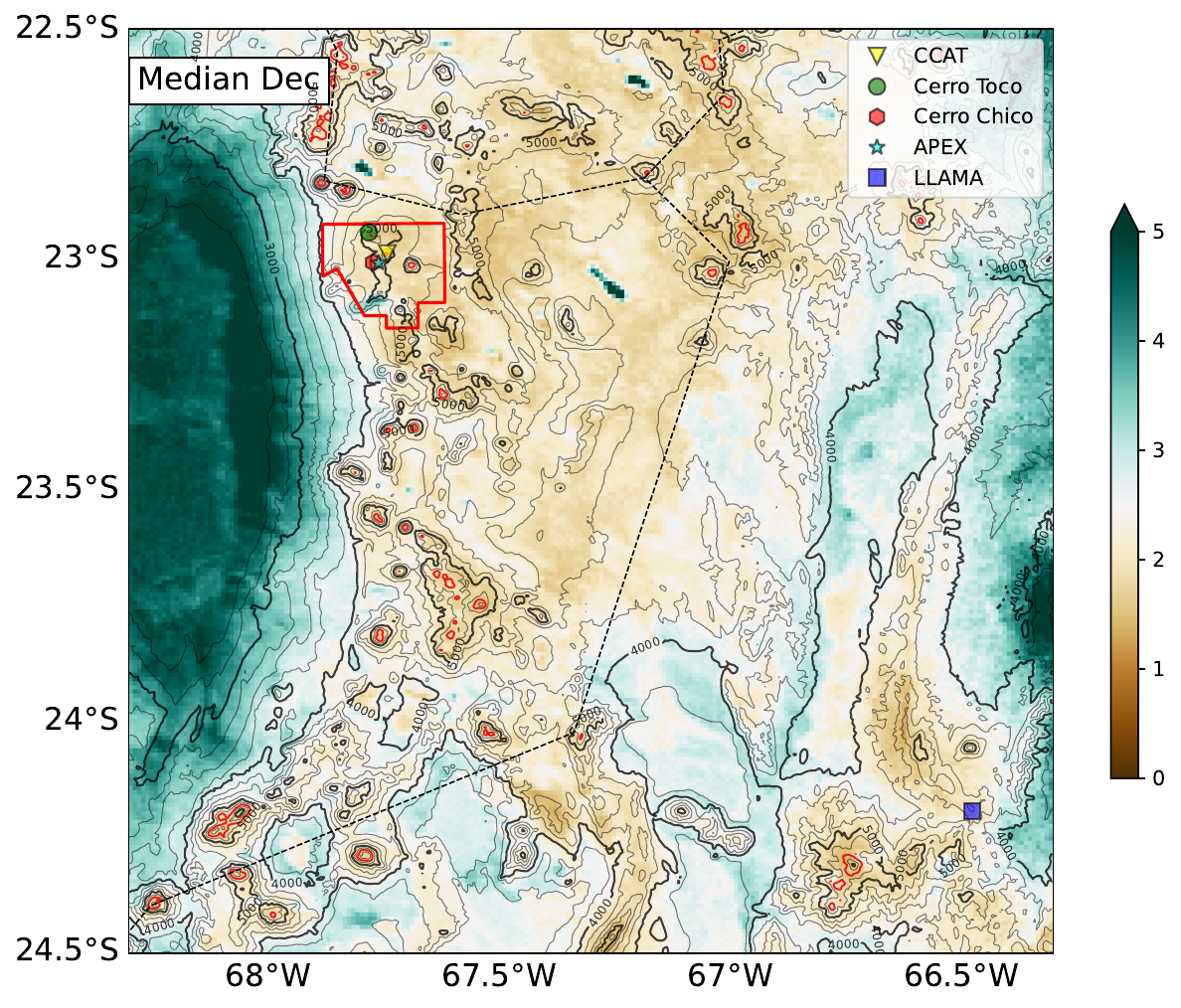}
    \caption{Median monthly daytime PWV values across all years. The borders, AAP+ALMA boundary, site location markers, and elevation contours are the same as in Figure \ref{fig:pwv_median_mean}.
    }\label{fig:pwv_median_monthly}
\end{figure*}

\begin{figure*}
    \centering
    \includegraphics[clip,trim=0mm 2mm 21mm 2.5mm,height=0.31\textwidth]{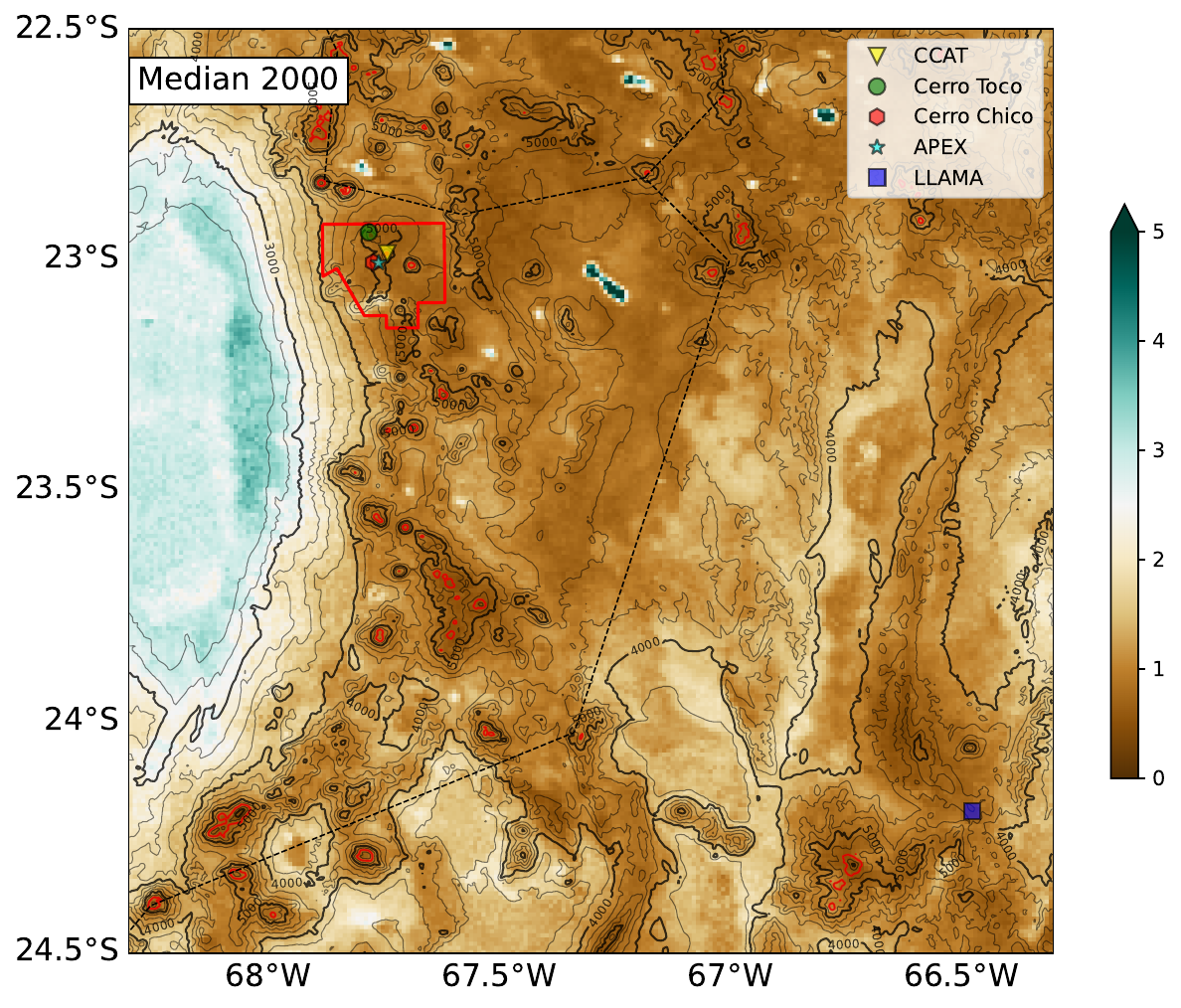}
    \includegraphics[clip,trim=20.4mm 2mm 21mm 2.5mm,height=0.31\textwidth]{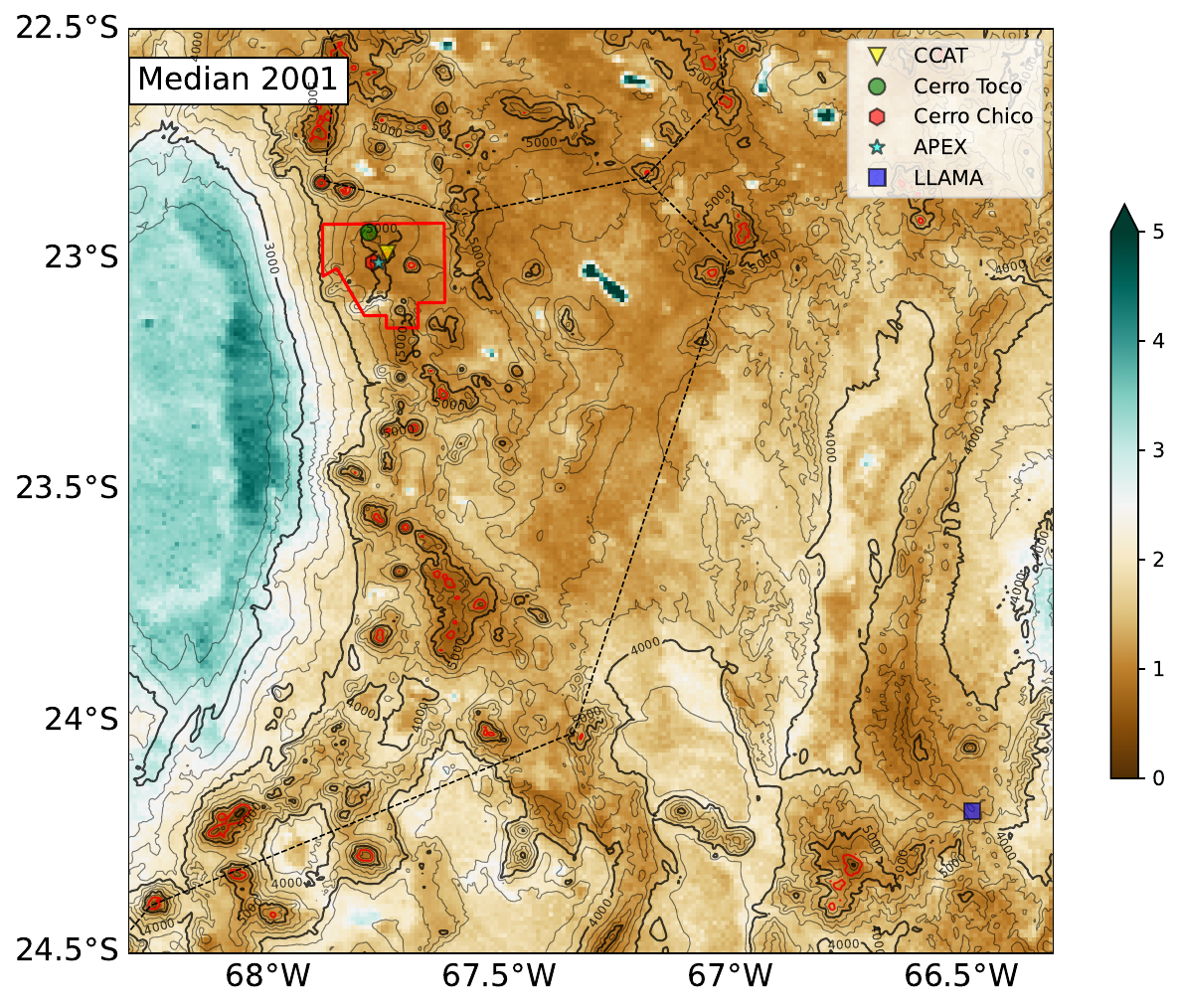}
    \includegraphics[clip,trim=20.4mm 2mm 0mm 2.5mm,height=0.31\textwidth]{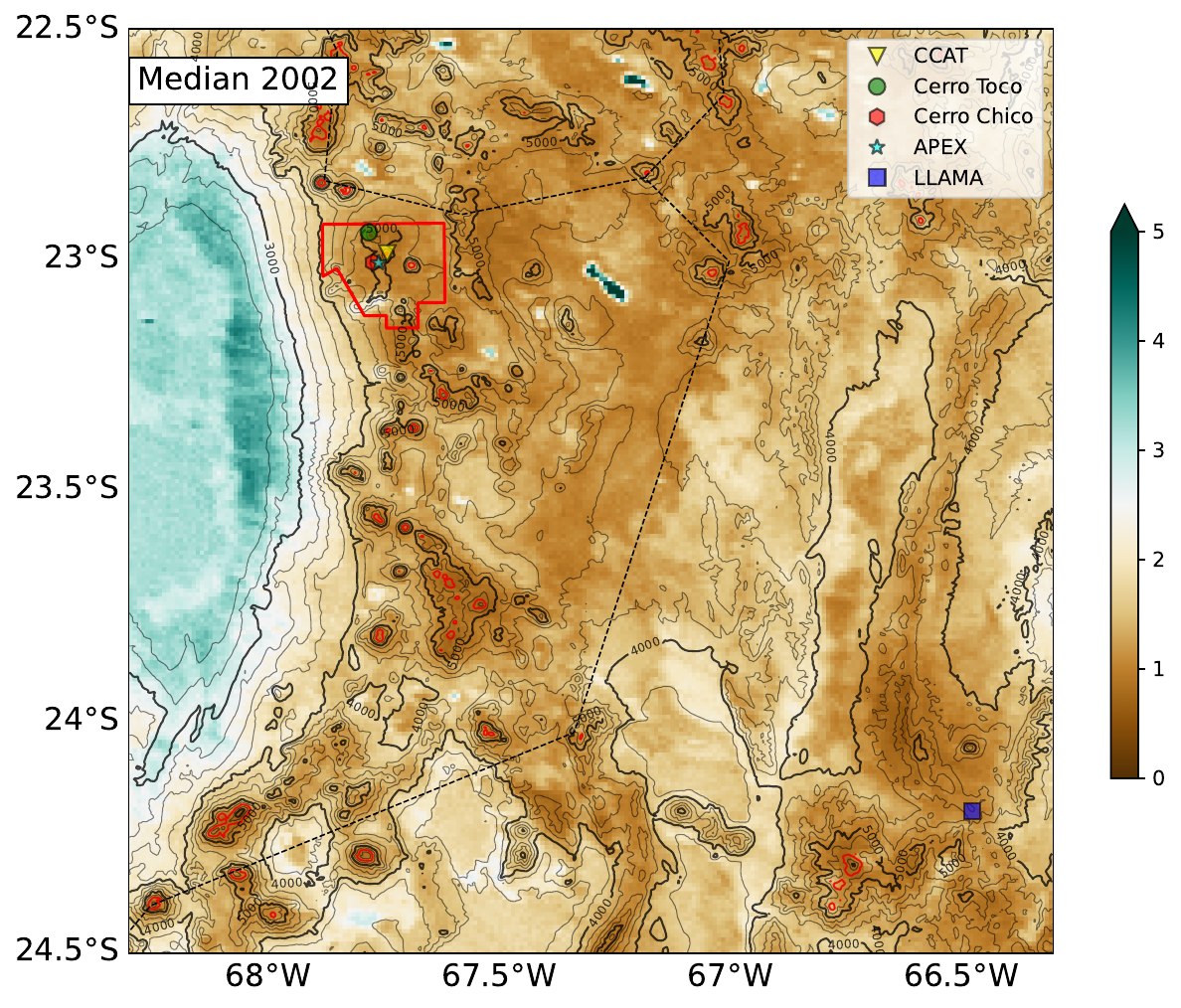}\\
    \includegraphics[clip,trim=0mm 2mm 21mm 2.5mm,height=0.31\textwidth]{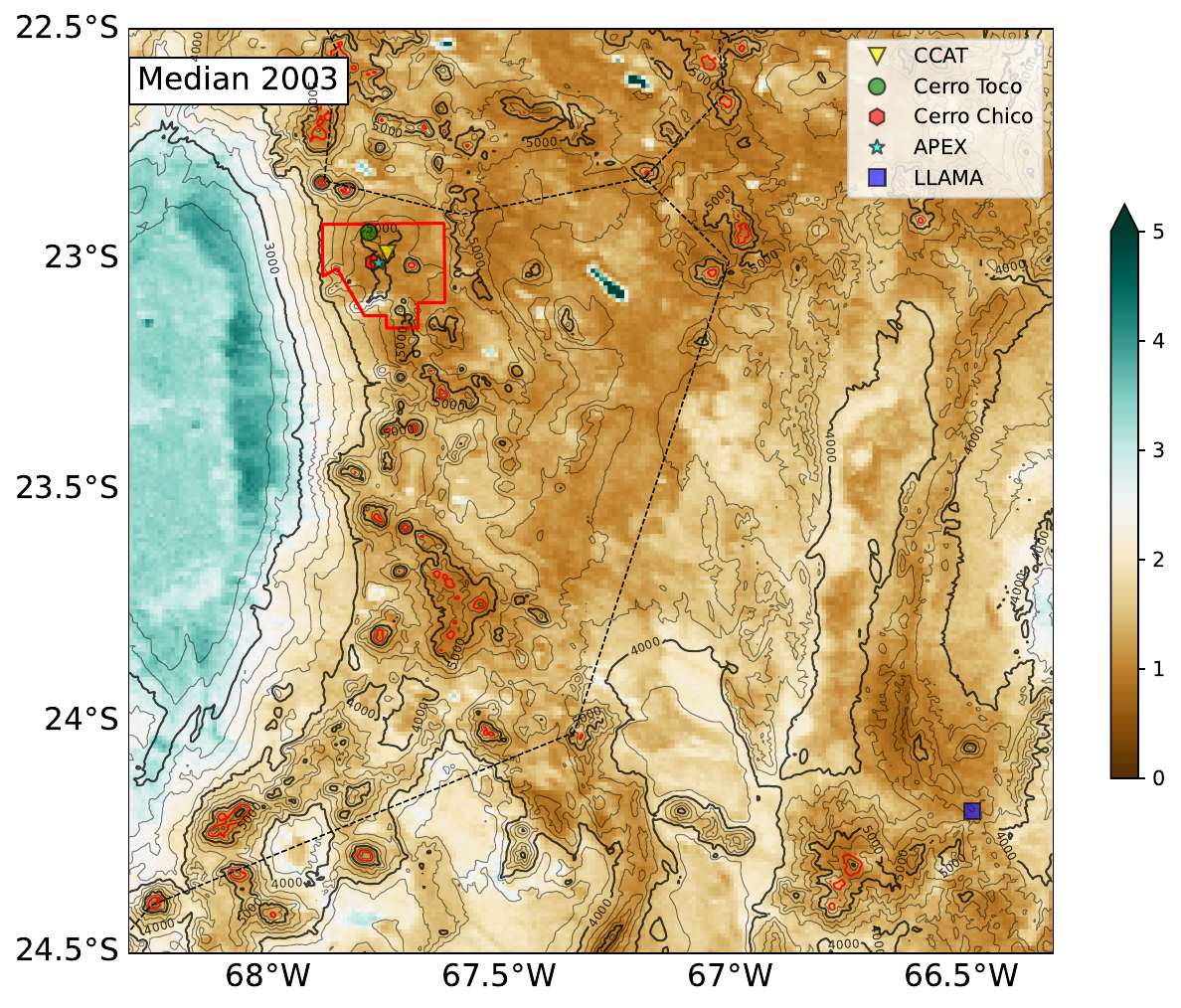}
    \includegraphics[clip,trim=20.4mm 2mm 21mm 2.5mm,height=0.31\textwidth]{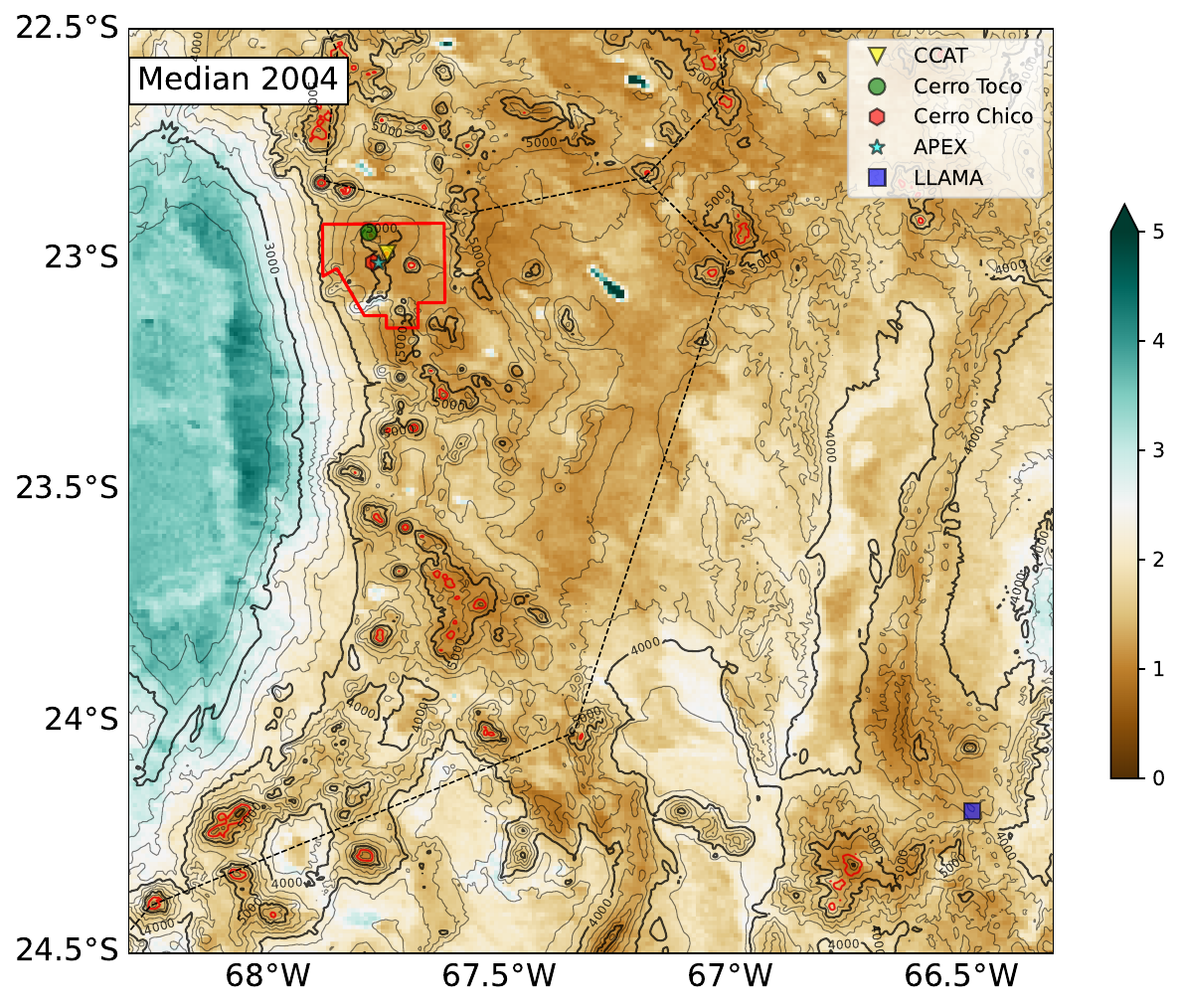}
    \includegraphics[clip,trim=20.4mm 2mm 0mm 2.5mm,height=0.31\textwidth]{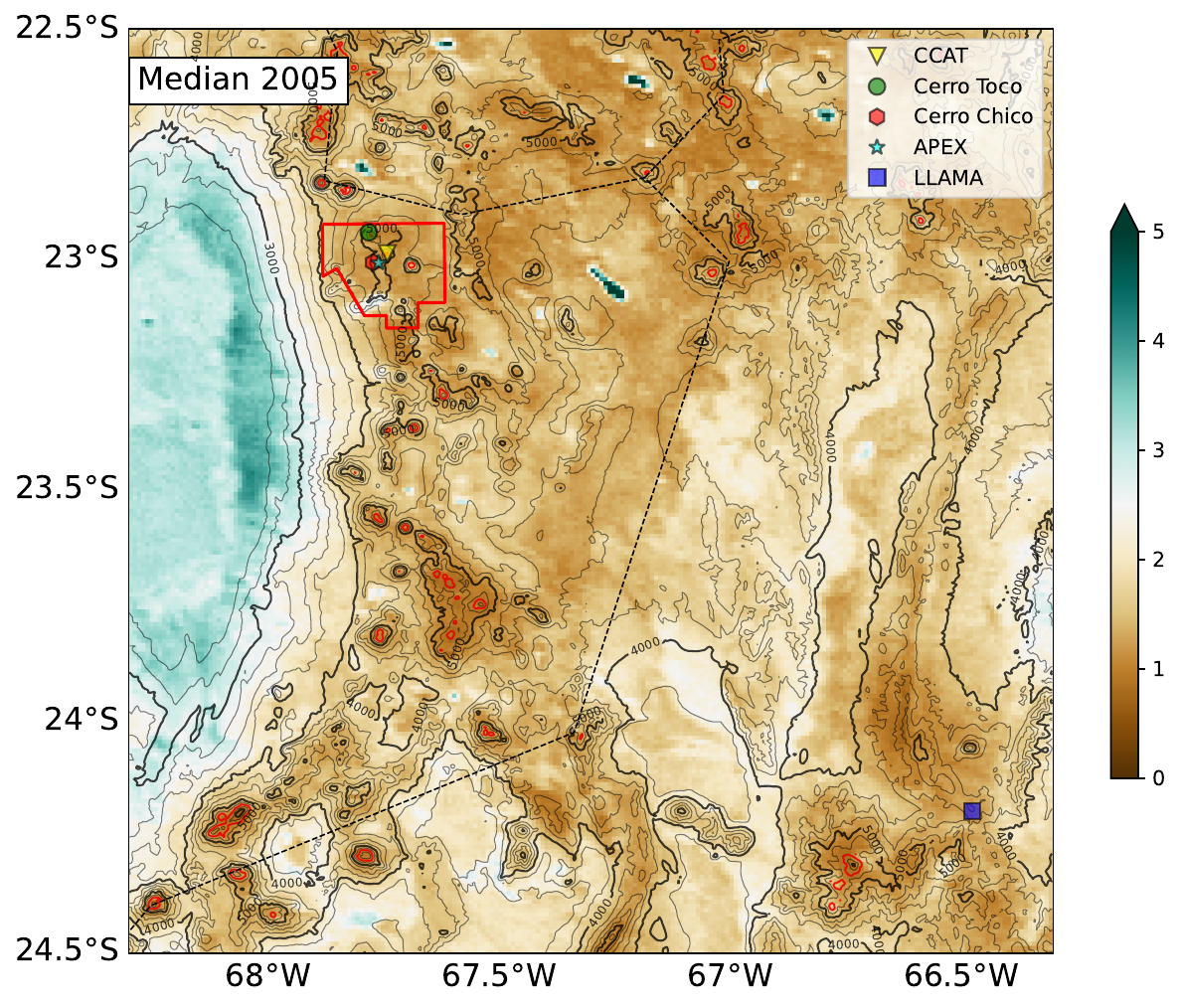}\\
    \includegraphics[clip,trim=0mm 2mm 21mm 2.5mm,height=0.31\textwidth]{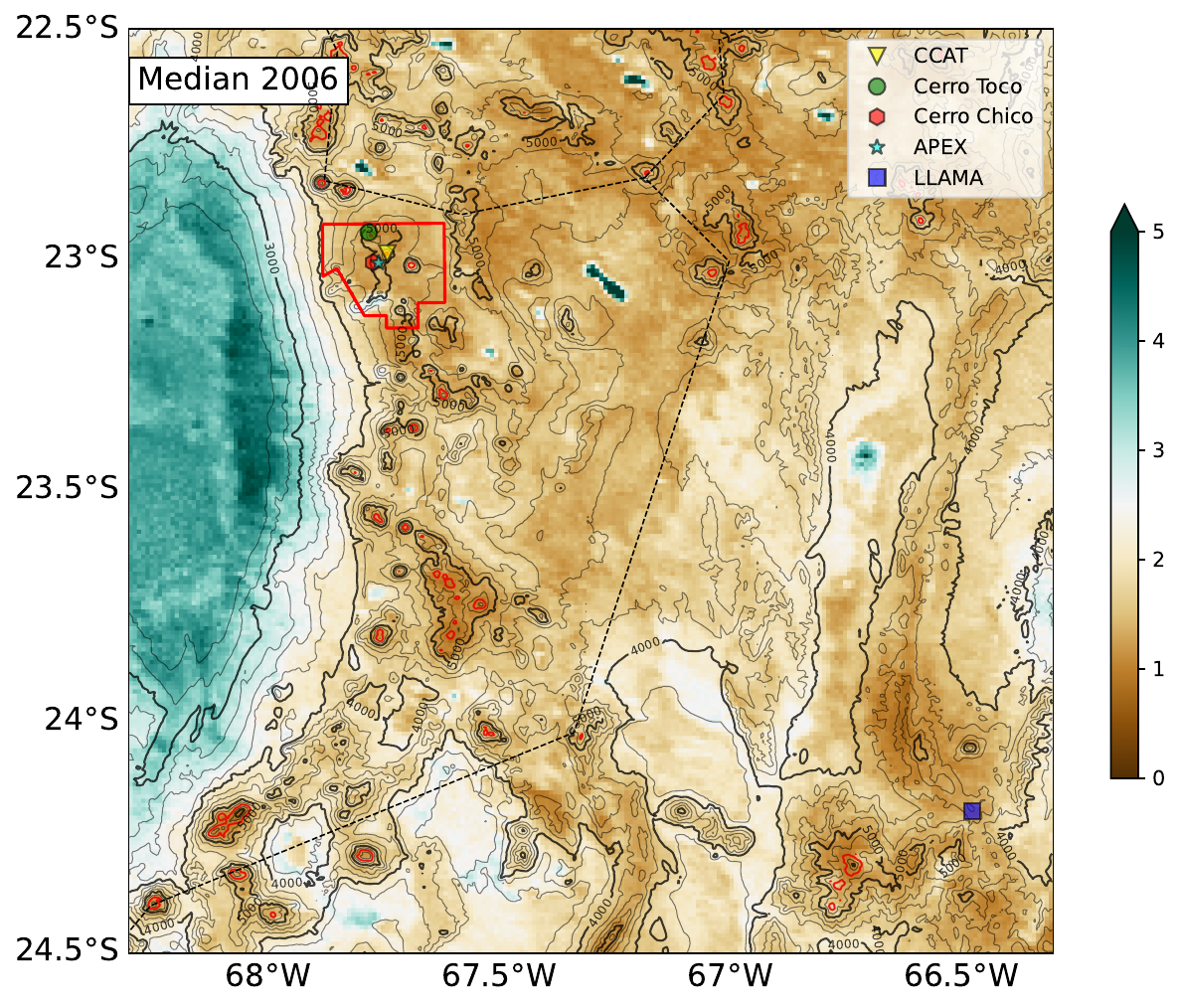}
    \includegraphics[clip,trim=20.4mm 2mm 21mm 2.5mm,height=0.31\textwidth]{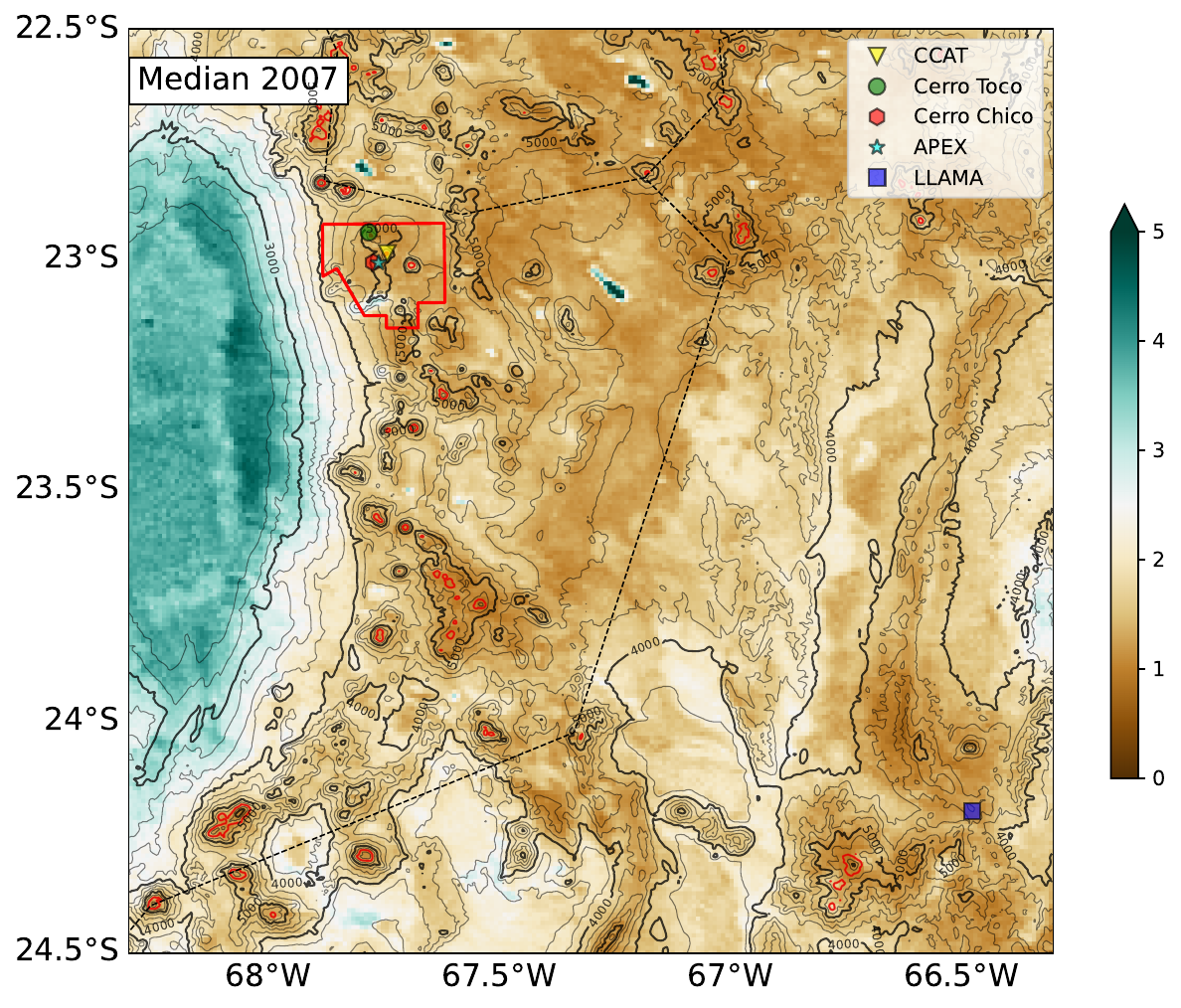}
    \includegraphics[clip,trim=20.4mm 2mm 0mm 2.5mm,height=0.31\textwidth]{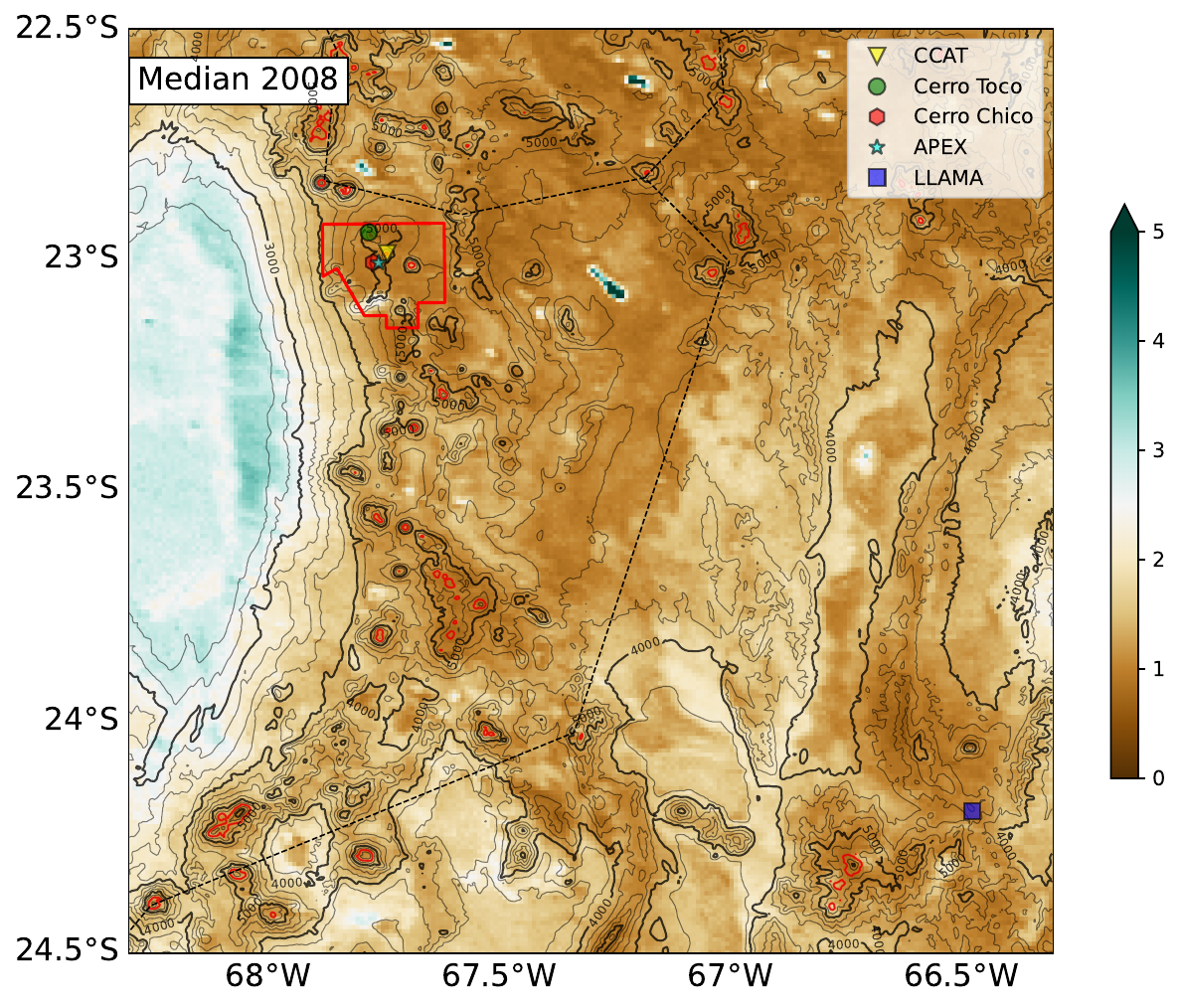}\\
    \includegraphics[clip,trim=0mm 2mm 21mm 2.5mm,height=0.31\textwidth]{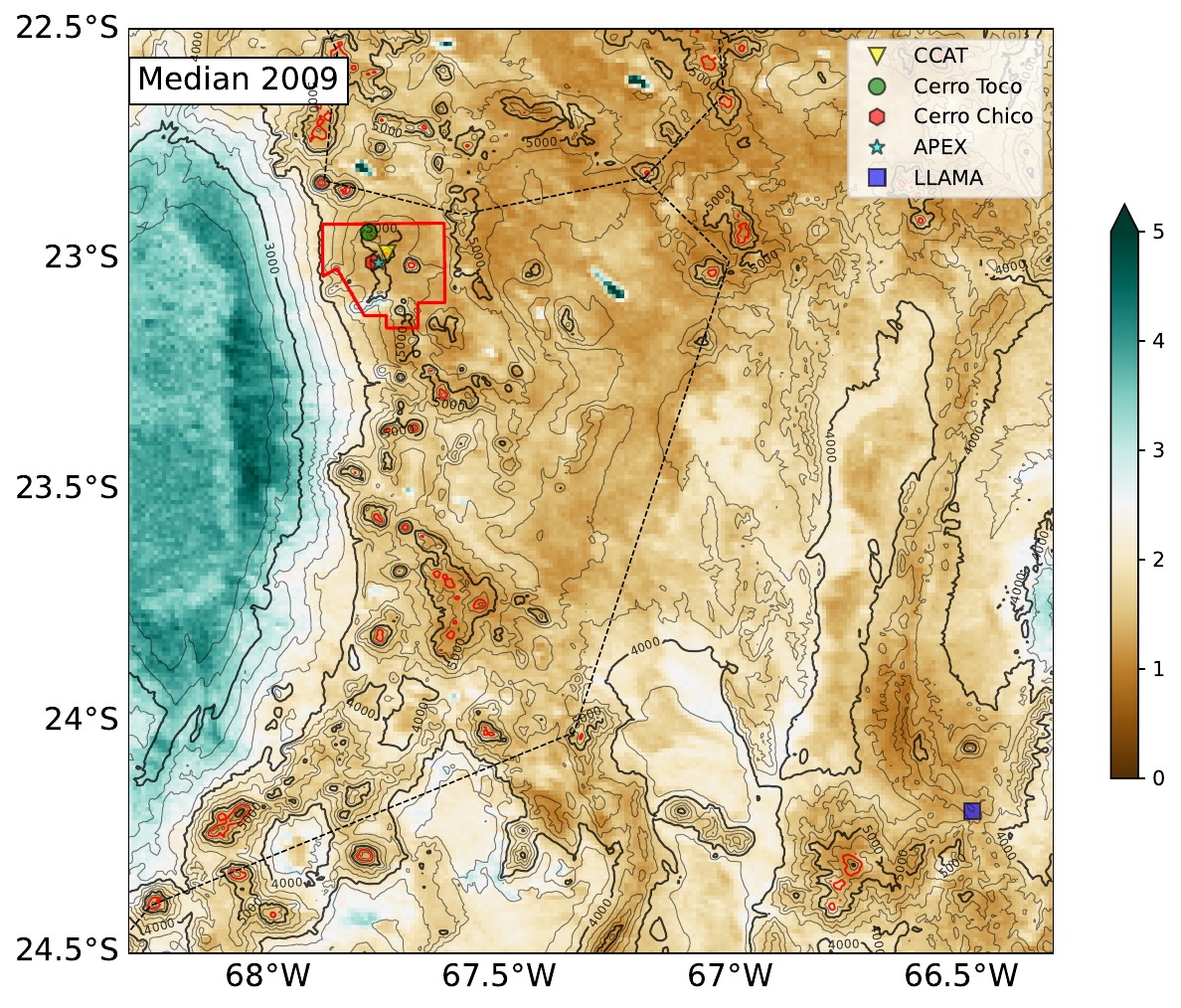}
    \includegraphics[clip,trim=20.4mm 2mm 21mm 2.5mm,height=0.31\textwidth]{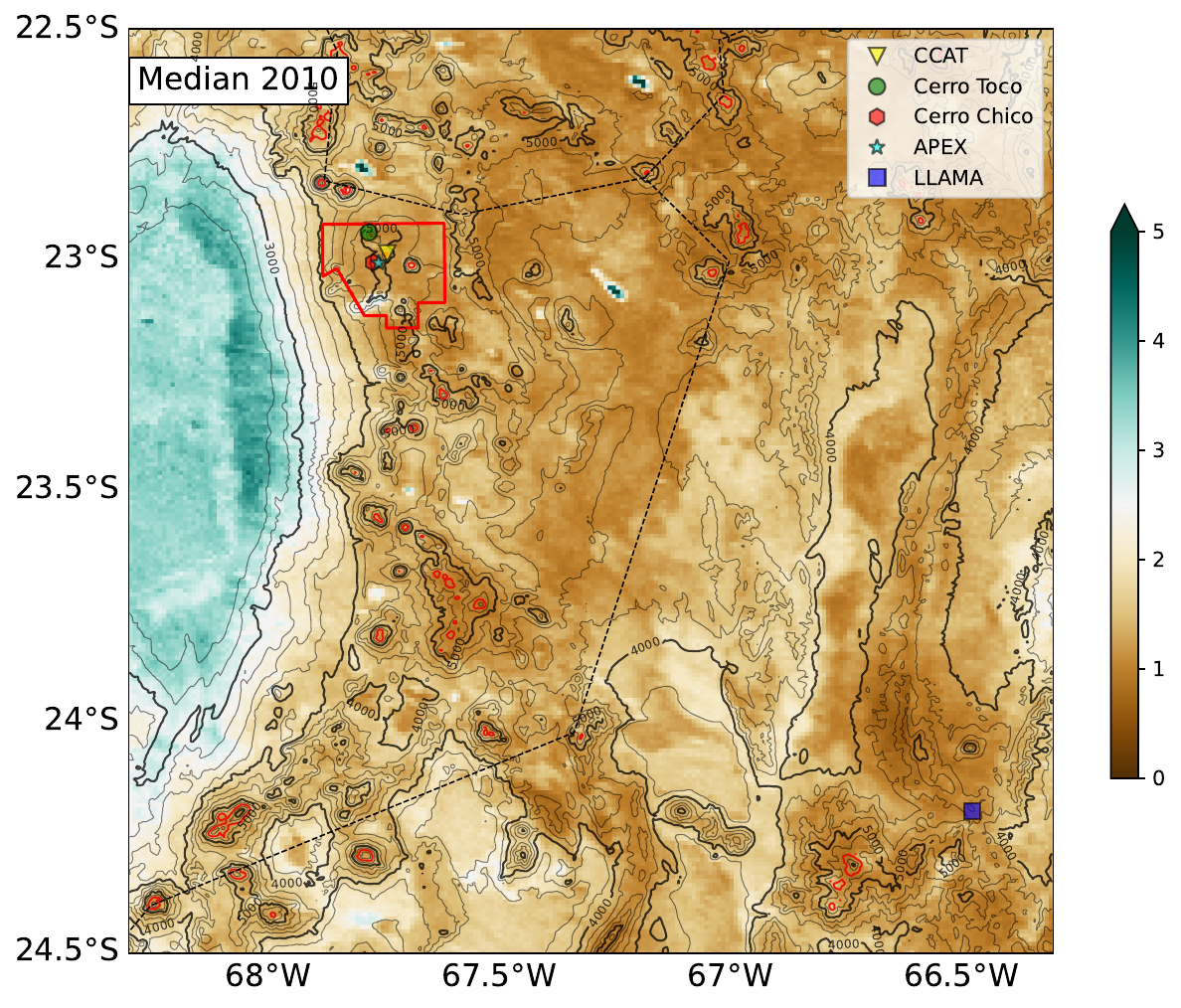}
    \includegraphics[clip,trim=20.4mm 2mm 0mm 2.5mm,height=0.31\textwidth]{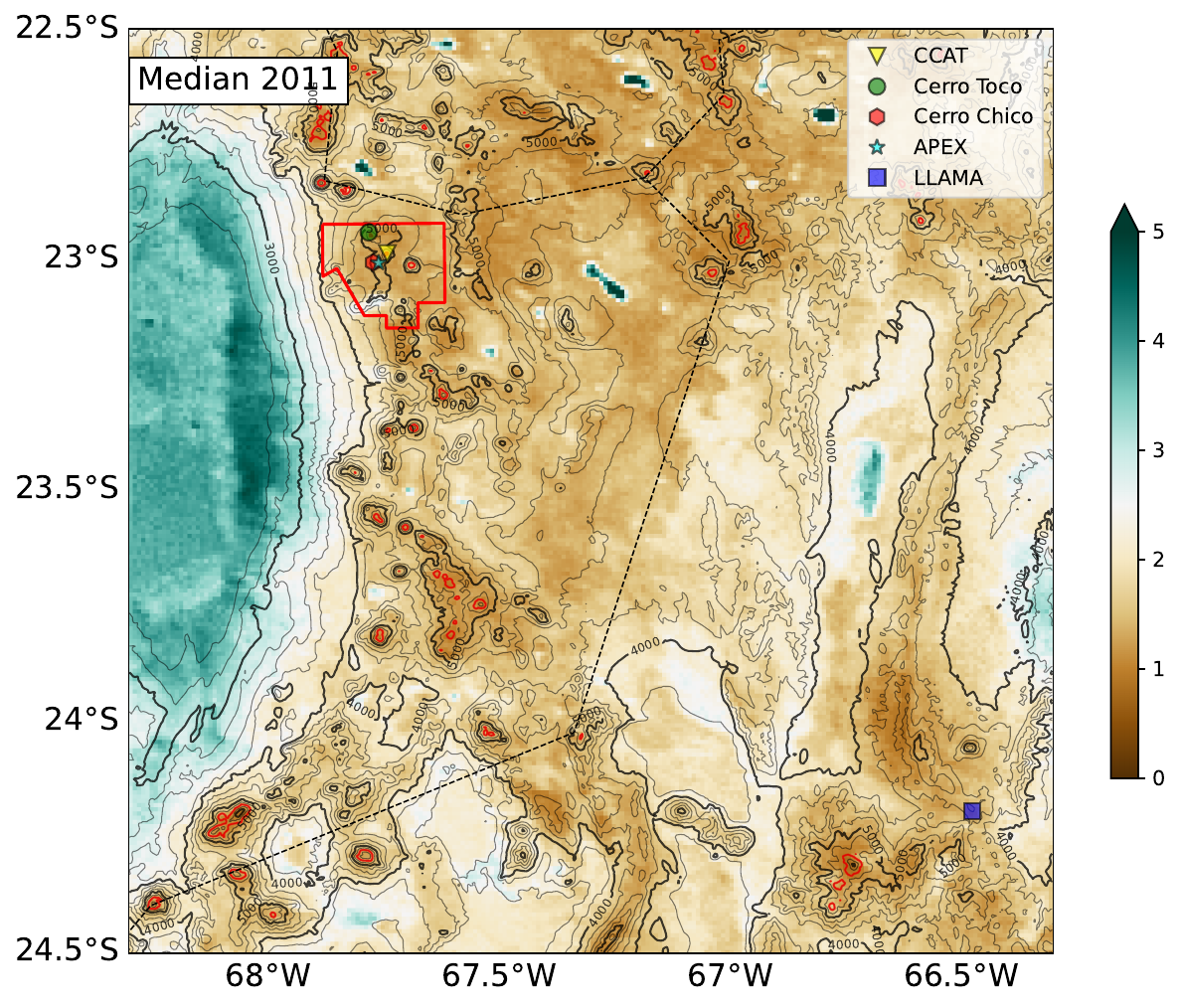}\\
 \caption{Median yearly daytime PWV values across years 2000-2011. The borders, AAP+ALMA boundary, site location markers, and elevation contours are the same as in Figure \ref{fig:pwv_median_mean}.
    }\label{fig:pwv_median_yearly_2000-2011}
\end{figure*}

\begin{figure*}
    \centering
    \includegraphics[clip,trim=0mm 2mm 21mm 2.5mm,height=0.31\textwidth]{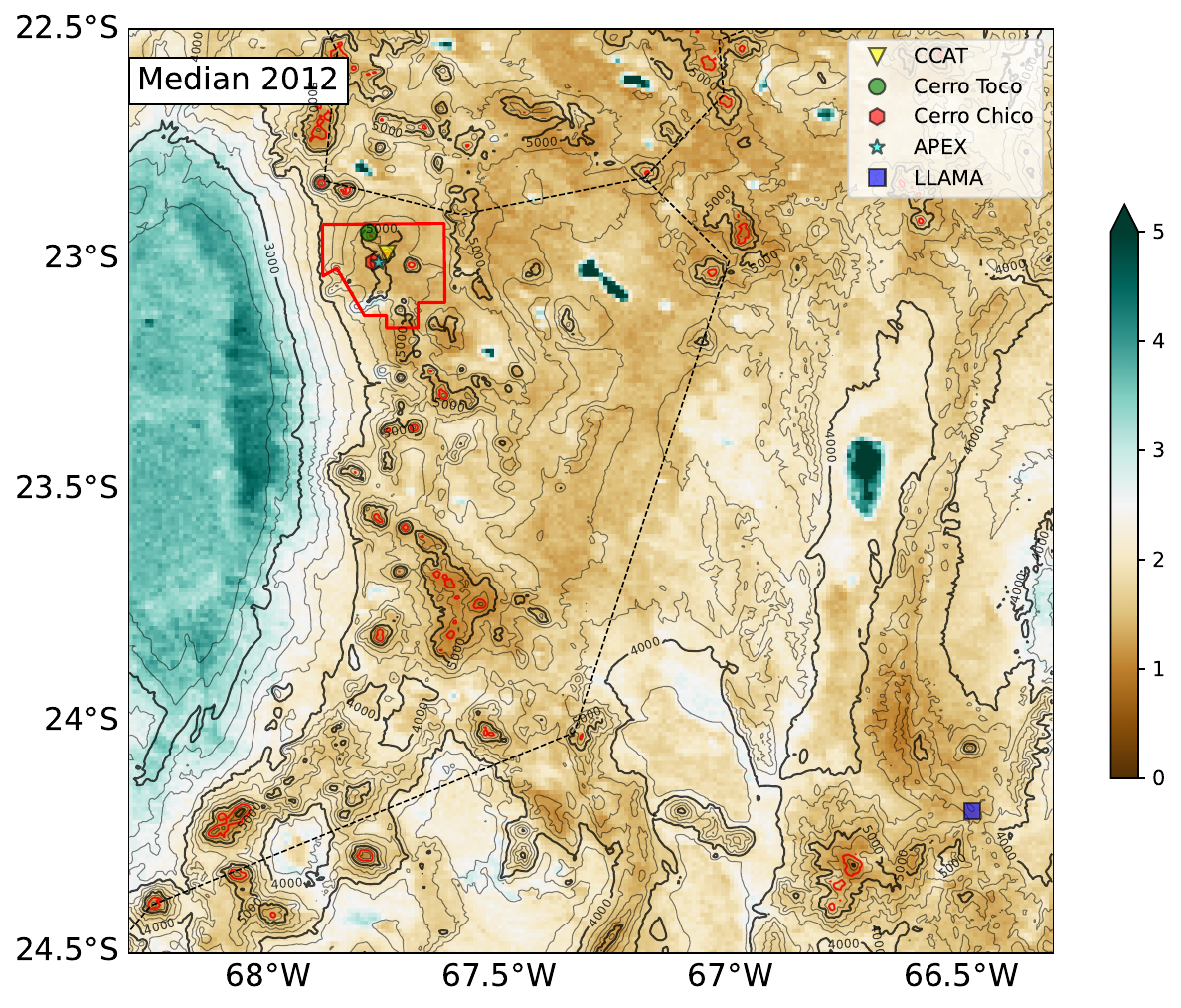}
    \includegraphics[clip,trim=20.4mm 2mm 21mm 2.5mm,height=0.31\textwidth]{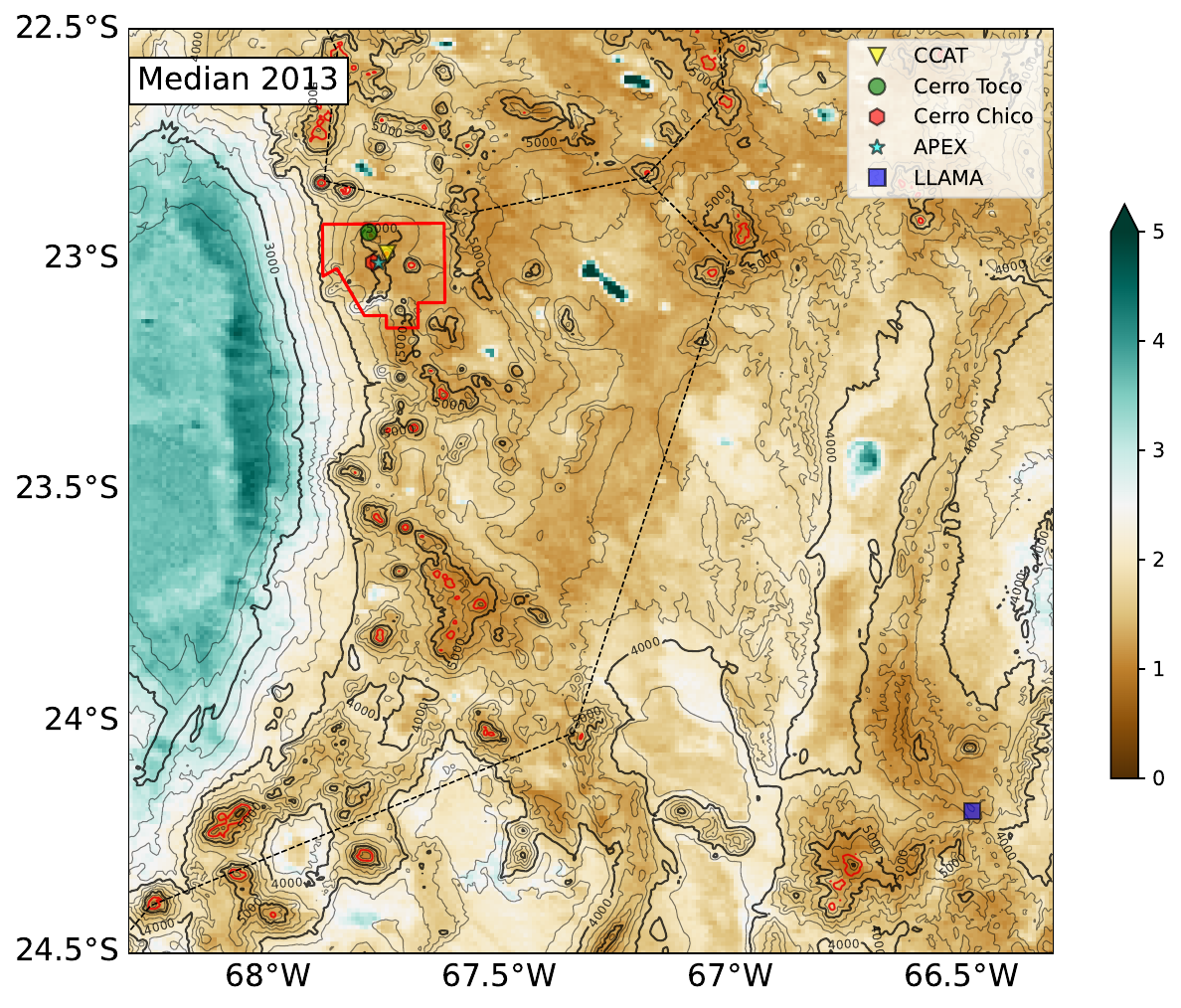}
    \includegraphics[clip,trim=20.4mm 2mm 0mm 2.5mm,height=0.31\textwidth]{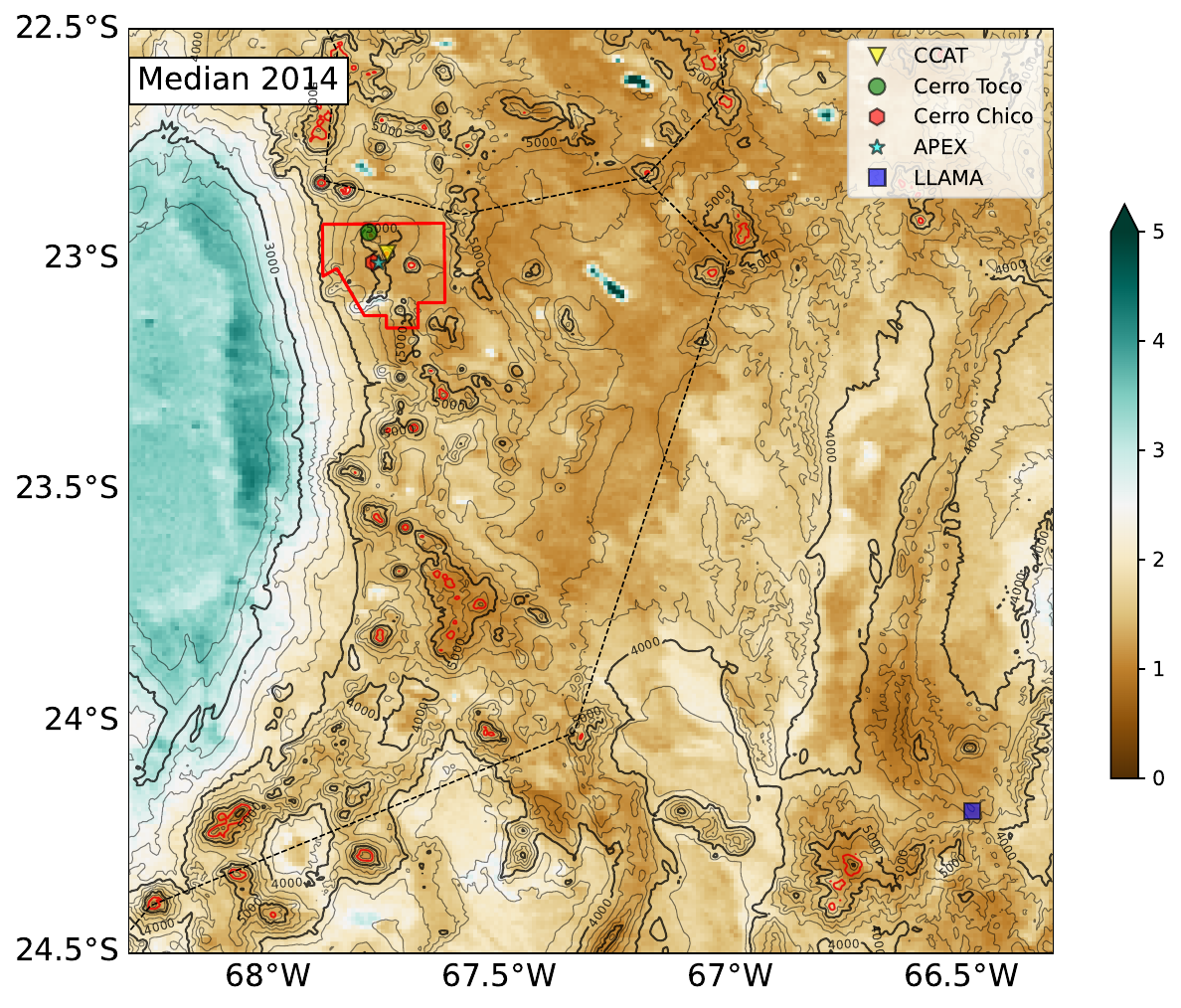}\\    
    \includegraphics[clip,trim=0mm 2mm 21mm 2.5mm,height=0.31\textwidth]{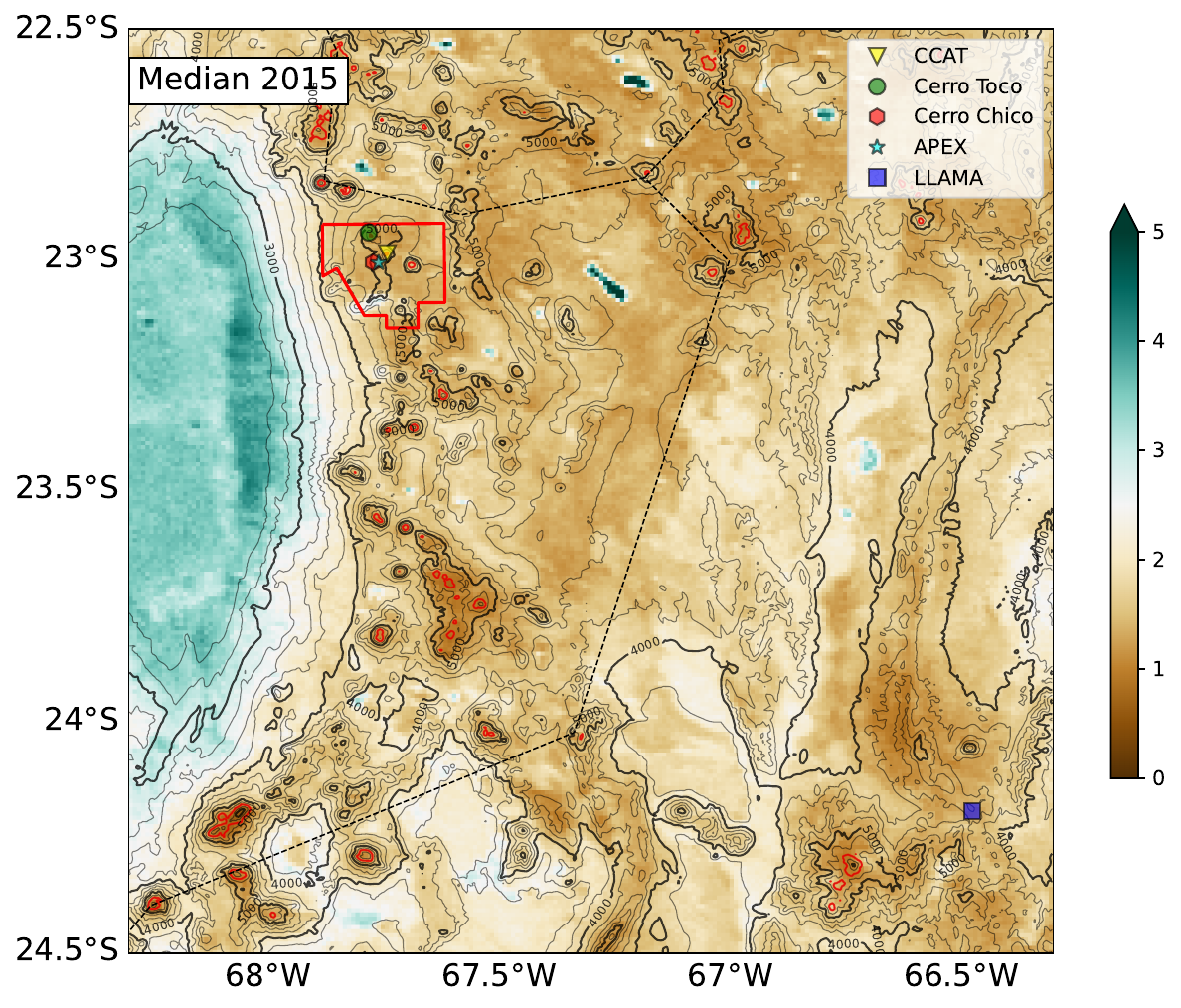}
    \includegraphics[clip,trim=20.4mm 2mm 21mm 2.5mm,height=0.31\textwidth]{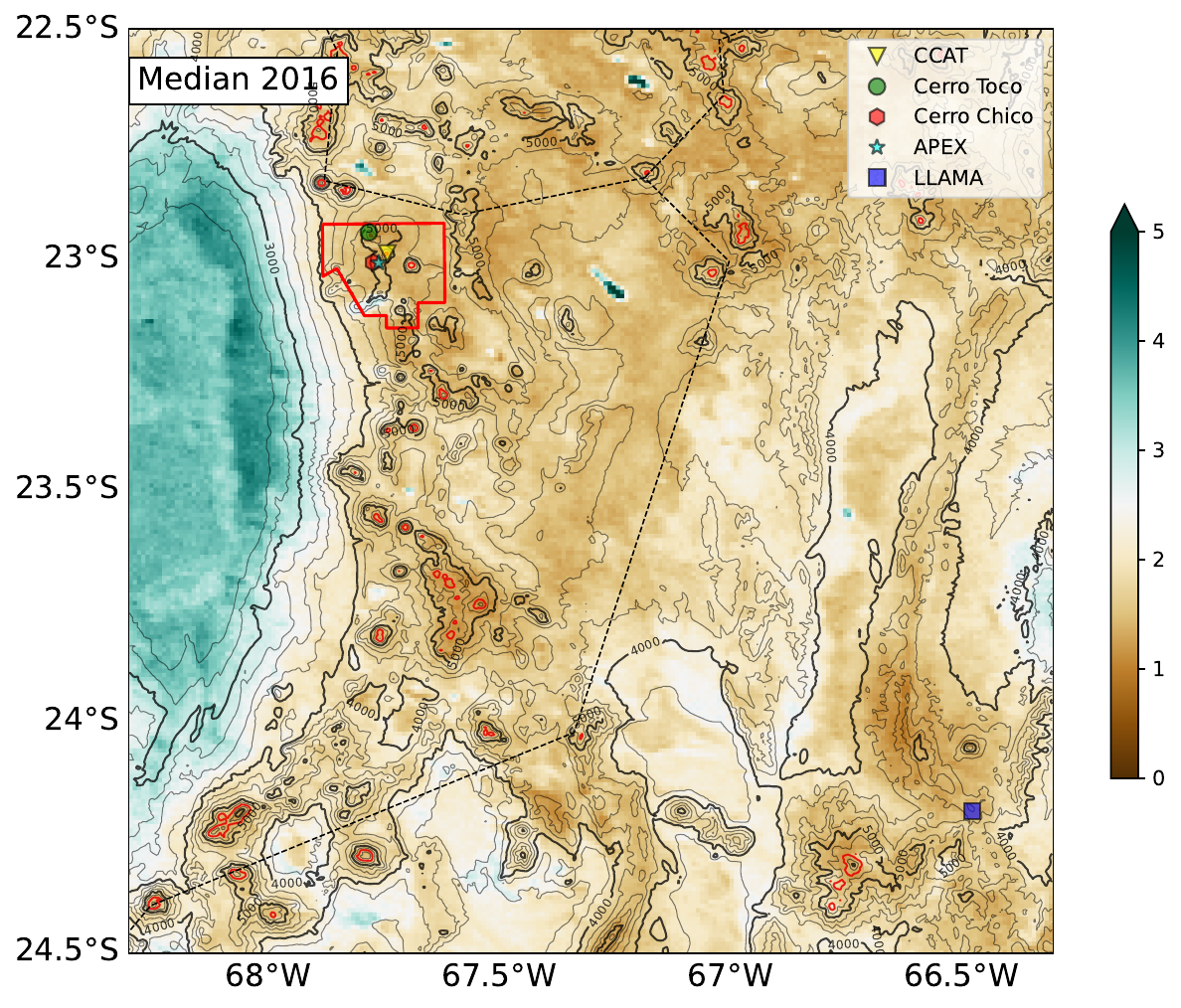}
    \includegraphics[clip,trim=20.4mm 2mm 0mm 2.5mm,height=0.31\textwidth]{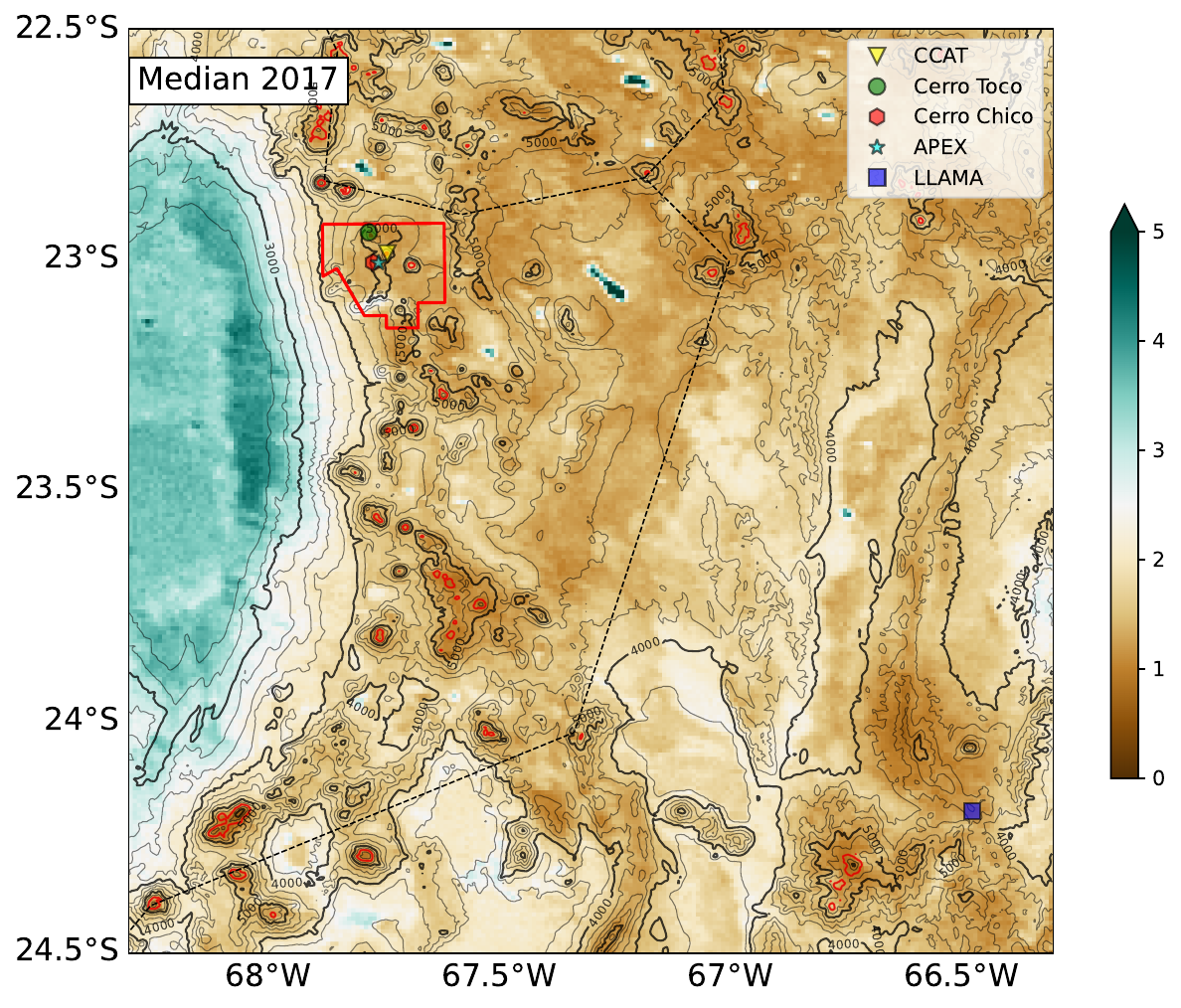}\\ 
    \includegraphics[clip,trim=0mm 2mm 21mm 2.5mm,height=0.31\textwidth]{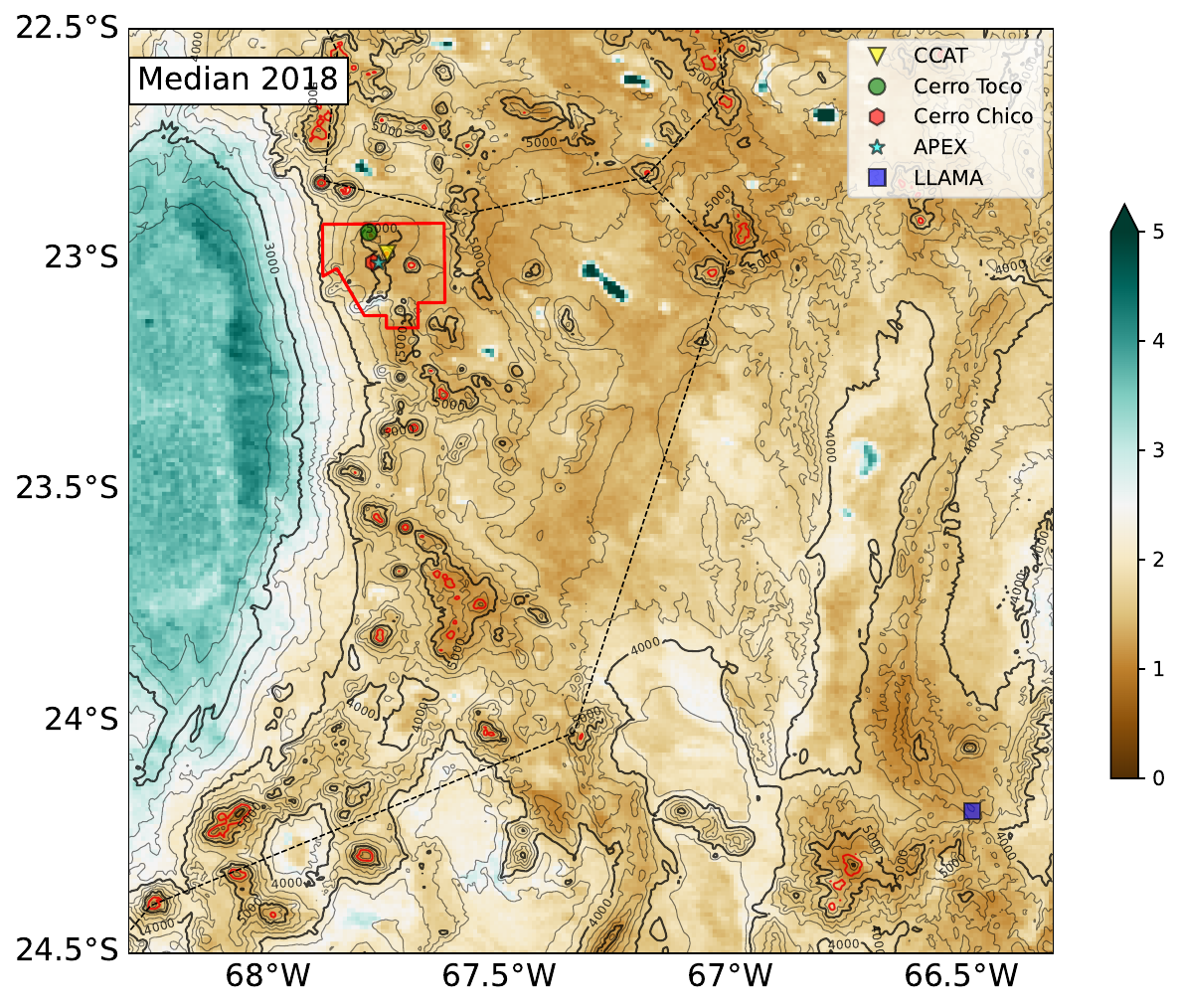}
    \includegraphics[clip,trim=20.4mm 2mm 21mm 2.5mm,height=0.31\textwidth]{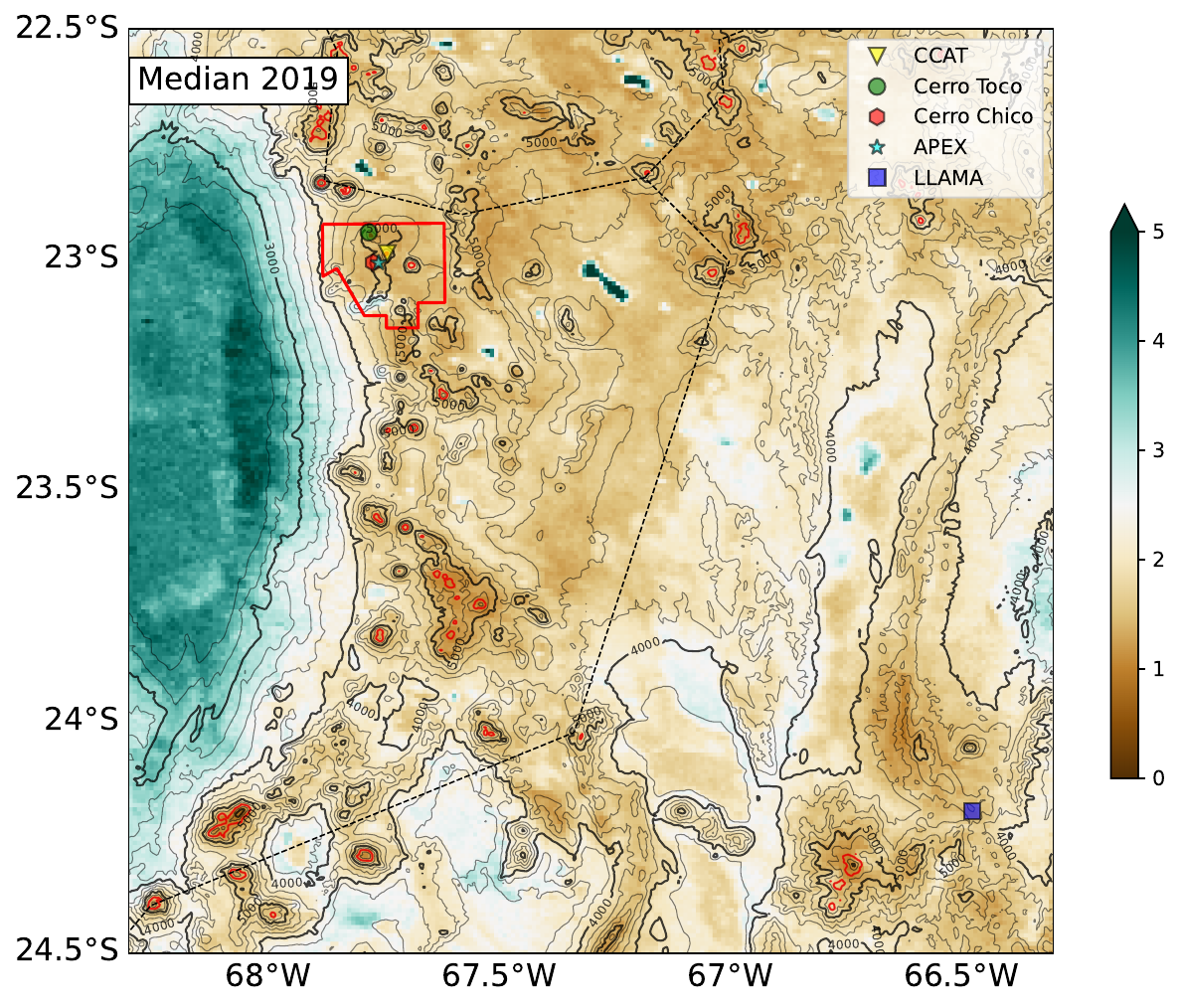}
    \includegraphics[clip,trim=20.4mm 2mm 0mm 2.5mm,height=0.31\textwidth]{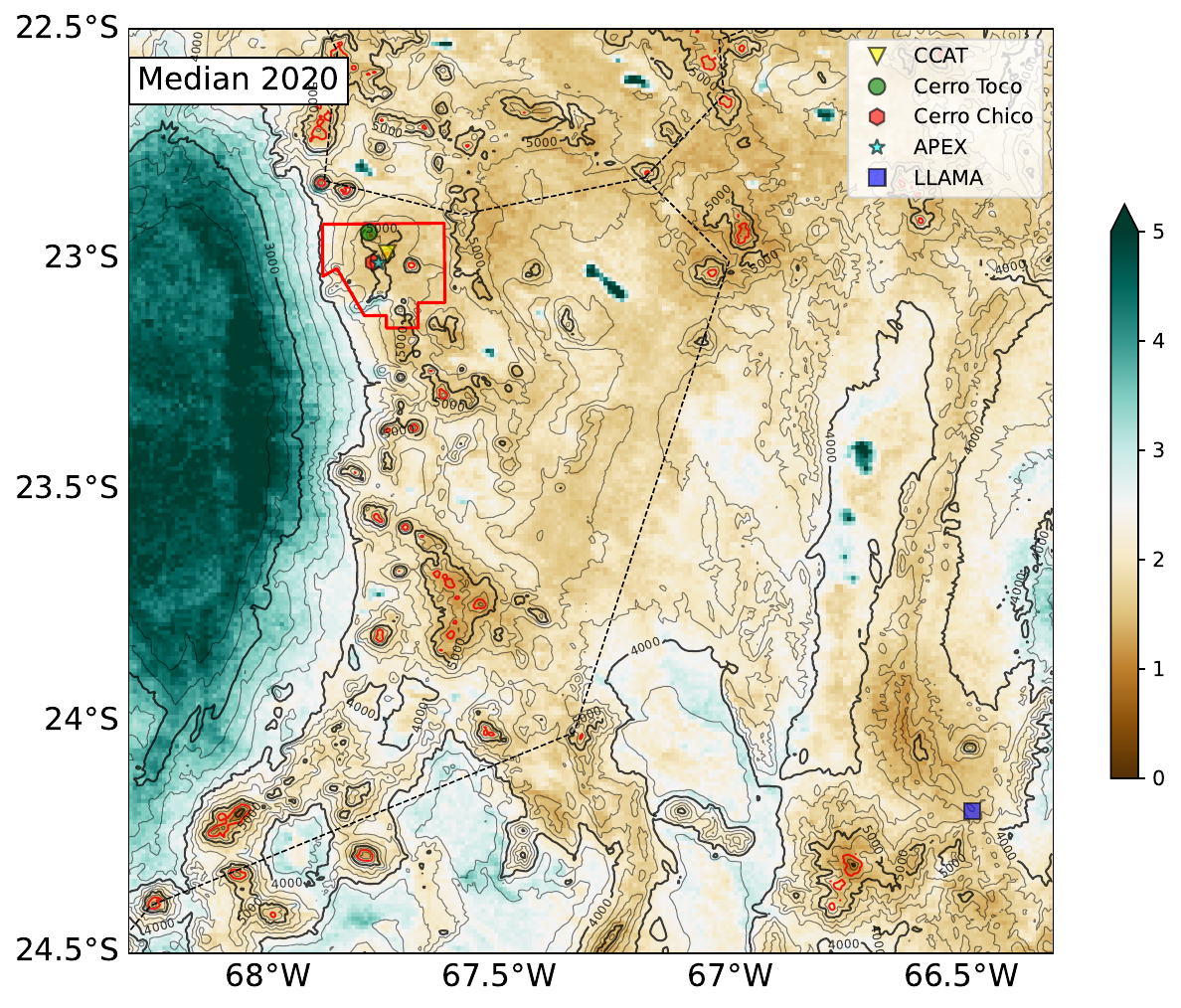} \caption{Median yearly daytime PWV values across years 2012-2020. The borders, AAP+ALMA boundary, site location markers, and elevation contours are the same as in Figure \ref{fig:pwv_median_mean}.
    }\label{fig:pwv_median_yearly_2012-2020}
\end{figure*}
}

\end{document}